\documentclass[fleqn,usenatbib]{mnras}

\usepackage{newtxtext,newtxmath}
\usepackage{float}

\usepackage[T1]{fontenc}

\DeclareRobustCommand{\VAN}[3]{#2}
\let\VANthebibliography\thebibliography
\def\thebibliography{\DeclareRobustCommand{\VAN}[3]{##3}\VANthebibliography}

\usepackage{graphicx}	
\usepackage{amsmath}	
\usepackage{tikz}       
\usepackage{xcolor}
\usepackage{hyperref}

\newcommand{\angstrom}{\mbox{\normalfont\AA}}
\definecolor{lime}{HTML}{A6CE39}
\DeclareRobustCommand{\orcidicon}{%
    \begin{tikzpicture}
    \draw[lime, fill=lime] (0,0) 
    circle [radius=0.16] 
    node[white] {{\fontfamily{qag}\selectfont \tiny ID}};
    \draw[white, fill=white] (-0.0625,0.095) 
    circle [radius=0.007];
    \end{tikzpicture}
    \hspace{-2mm}
}
\newcommand{\orchidJanka}{\href{https://orcid.org/0000-0002-0831-3330}{\orcidicon}}
\newcommand{\orchidJerkstrand}{\href{https://orcid.org/0000-0001-8005-4030}{\orcidicon}}
\newcommand{\orchidVanBaal}{\href{https://orcid.org/0009-0001-3767-942X}{\orcidicon}}
\newcommand{\orchidWongwathanarat}{\href{https://orcid.org/0000-0001-8400-8891}{\orcidicon}}

\title[3D nebular phase supernovae]{Diagnostics of 3D explosion asymmetries of stripped-envelope supernovae by nebular line profiles}
\author[B. van Baal et al.]{
Bart F. A. van Baal\orchidVanBaal,$^{1}$\thanks{E-mail: barteld.vbaal@astro.su.se}
Anders Jerkstrand\orchidJerkstrand,$^{1}$
Annop Wongwathanarat\orchidWongwathanarat$^{2}$
and Hans-Thomas Janka\orchidJanka$^{2}$
\\
$^{1}$The Oskar Klein Centre, Department of Astronomy, Stockholm University, AlbaNova, Se-10691 Stockholm, Sweden\\
$^{2}$Max Planck Institute for Astrophysics, Karl-Schwarzschild-Str 1, D-85748 Garching, Germany\\
}

\date{Accepted XXX. Received YYY; in original form ZZZ}

\pubyear{2024}

\begin{document}
\label{firstpage}
\pagerange{\pageref{firstpage}--\pageref{lastpage}}
\maketitle

\begin{abstract}  
Understanding the explosion mechanism and hydrodynamic evolution of core-collapse supernovae is a long-standing quest in astronomy. The asymmetries caused by the explosion are encoded into the line profiles which appear in the nebular phase of the SN evolution -- with particularly clean imprints in He star explosions. Here, we carry out nine different supernova simulations of He-core progenitors, exploding them in 3D with parametrically varied neutrino luminosities using the \texttt{Prometheus-HotB} code, hydrodynamically evolving the models to the homologeous phase. We then compute nebular phase spectra with the 3D NLTE spectral synthesis code \texttt{ExTraSS} (EXplosive TRAnsient Spectral Simulator). We study how line widths and shifts depend on progenitor mass, explosion energy, and viewing angle. We compare the predicted line profile properties against a large set of Type Ib observations, and discuss the degree to which current neutrino-driven explosions can match observationally inferred asymmetries. With self-consistent 3D modelling -- circumventing the difficulties of representing $^{56}$Ni mixing and clumping accurately in 1D models -- we find that neither low-mass He cores exploding with high energies nor high-mass cores exploding with low energies contribute to the Type Ib SN population. Models which have line profile widths in agreement with this population give sufficiently large centroid shifts for calcium emission lines. Calcium is more strongly affected by explosion asymmetries connected to the neutron star kicks than oxygen and magnesium. Lastly, we turn to the NIR spectra from our models to investigate the potential of using this regime to look for the presence of He in the nebular phase.
\end{abstract}

\begin{keywords}
supernovae: general -- stars: evolution -- stars: massive -- line: profiles --  methods: numerical
\end{keywords}



\section{Introduction}

Massive stars ($M_\text{ZAMS} \gtrsim 8 M_\odot$) will end their lives in a core-collapse supernova (CCSN) event \citep{heger2003massive}. The central compact object formed in these explosions can be either a neutron star or black hole, while the rest of the star is ejected into the interstellar medium, leading to enrichment of the chemical elements synthesized in both the hydrostatic and explosive burning \citep{woosley1995,arnett1996supernovae,woosley2002evolution,Limongi2003,ceverino2009role}. These ejecta expand over time and are radioactively powered, slowly transitioning from an initially optically thick diffusion phase into a more optically thin "nebular" phase wherein emission lines start appearing. Studying these emission lines allows for the inference of masses of specific elements as well as the distribution of these elements across the nebula \citep{jerkstrand2017spectra}.

Observational evidence suggests that significant asymmetries are common in CCSNe. Complex structures such as knots, filaments and plumes can be directly imaged in nearby remnants \citep[e.g.][]{hughes2000nucleosynthesis,fesen2006expansion,larsson2013morphology,larsson2023JWST,grefenstette2014asymmetries,grefenstette2017distribution,Mili2024}. Deviations from symmetry can often be inferred from line profiles both on small scales, linking to clumpiness \citep[e.g.][]{filippenko1989spectroscopic,spyromilio1994clumping} and on global scales, linking to large-scale ejecta asymmetries \citep[e.g.][]{mazzali2001nebular,modjaz2008double,taubenberger2009nebular}. Neutron stars have been observed to be moving at space velocities of up to $\sim1000\,\text{km}\,\text{s}^{-1}$\citep[e.g.][]{arzoumanian2002velocity,hobbs2005statistical}, indicating fundamental asymmetry in the explosion mechanism. Spectropolarimetry results \citep[e.g.][]{tanaka2012three,reilly2016spectropolarimetry,tinyanont2021infrared} indicate likewise. Detections from gamma-ray decay lines in SN 1987A \citep{teegarden1991gamma,boggs201544ti} and Cas A \citep{grefenstette2017distribution} also suggest asymmetry.

Asymmetries in the ejecta can be effectively studied in the nebular phase, when the largely optically thin emission probes the whole ejecta. For stripped envelope SNe (SESNe; stars which have lost their hydrogen envelope, leading to Type Ib SNe, and sometimes also their helium envelope, leading to Type Ic SNe), studies of nebular line profiles began in earnest with the linked gamma-ray burst (GRB)/Type Ic-broadline (Ic-BL) SN 1998bw \citep{mazzali2001nebular,Maeda2002,Mazzali2005,Maeda2006}. For regular SESNe, initial sample studies of line profiles were carried out by \citet{maeda2008asphericity}, \citet{modjaz2008double} and \citet{taubenberger2009nebular}. These studies mostly focused on line profiles of the [O I] $\lambda\lambda6300,\,6364$ doublet, uncovering large variations. \citet{taubenberger2009nebular} also showed that at (very) late times the profile of the Mg I] $\lambda4571$ feature generally resembles that of the [O I] doublet, once a correction for the doublet nature is applied. More recently, \citet{fang2022core} looked at correlations between [O I] $\lambda\lambda6300,\,6364$ doublet and the [Ca II] $\lambda\lambda7291,\,7323$ doublet to investigate links between progenitor carbon-oxygen core masses with the ejecta geometry. \citet{fang2023inferring} followed up on this, searching for further constraints on the relations between the progenitor mass and explosion energy using nebular phase spectroscopy. \citet{fang2023aspherical} investigated the deviations from symmetry using the line profiles of the [O I] and [Ca II] doublets, finding the degree of asphericity to be larger for higher-mass carbon-oxygen cores.

Spectral modelling in the nebular phase involves many complex processes. The densities at these late times are low enough that Non-Local Thermodynamic Equilibrium (NLTE) must be accounted for, yet the radiative transfer effects like photoionization can still be relevant. 
The radioactive decay of species like $^{56}$Ni injects particles at MeV energies while the cooling energies in the optical and infrared are at $\lesssim 1\,$eV, covering over 6 orders of magnitude. The modelling of the nebular phase therefore requires large computational efforts typically involving the iterative solution of the non-thermal electron degradation spectrum, NLTE rate equations for $\gtrsim 10^5$ levels, and radiative transfer in $\gtrsim 10^6$ lines and continua. Advanced 1D NLTE radiative transfer codes like \texttt{SUMO} \citep{jerkstrand201144Ti,jerkstrand2012progenitor} and \texttt{CMFGEN} \citep{Hillier2012} have been extensively used to compute such models -- but in these the complex 3D morphologies of CCSN ejecta can only be approximately accounted for by parameterized mixing.

In \citet{vanbaal2023modelling} we introduced \texttt{ExTraSS} (EXplosive TRAnsient Spectral Simulator), a new 3D NLTE spectral synthesis code. Building upon the 3D gamma-ray deposition and photon packet transport platform in spherical coordinates developed in  \citet{jerkstrand2020properties}, \texttt{ExTraSS} solves the non-thermal degradation equations and the NLTE rate equations in each cell, and then generates spectra, currently in the optically thin limit. In \citet{vanbaal2023modelling}, the code was applied to generate viewing-angle dependent nebular spectra of a 3D explosion model of a low-mass He-core. Such 3D CCSN ejecta models, evolved to the homologous phase, are now becoming available in increasing number \citep{wongwathanarat2015three,wongwathanarat2017production,muller2018multidimensional,stockinger2020three,gabler2021infancy} and offer opportunities to directly compare synthetic light curves and spectra of state-of-the-art explosion models against observations.

NLTE modelling of stripped-envelope SNe at nebular times has been carried out for Type Ib/IIb SNe \citep{Fransson1989,Houck1996,maurer2012hydrogen,jerkstrand2015late,dessart2021nebular,dessart2023modeling,ergon2022spectral,barmentloo2024nebular}, normal Type Ic SNe \citep{mazzali2010type} and Ic-BL SNe \citep{Sollerman2000,mazzali2001nebular,Maeda2006,Mazzali2004,jerkstrand2017long, dessart2019simulations}. With the exception of the 2D modelling of SN 1998bw by \citet{Maeda2006}, all of these studies were done with 1D models. The recent work by \citet{vanbaal2023modelling} takes the step to bring stripped-envelope SN nebular modelling into full 3D, investigated the 3D explosion of a $3.3\,M_\odot$ He-star progenitor from \citet{woosley2019evolution} and \citet{ertl2020explosion}. In 3D, one gets around the main issue with 1D modelling, namely the need to parameterize the mixing -- and instead have a direct, realistic morphology representation. Using 3D ab-initio simulations of the explosions and ensuing hydrodynamics instead allows full focus on the microphysics, and ensures that when this becomes accurately enough treated, direct tests of modern explosion models against observations become feasible.

In this paper we expand upon \citet{vanbaal2023modelling} by using \texttt{ExTraSS} to study a grid of nine Type Ib SN models of neutrino-driven explosions, hydrodynamically simulated with the \texttt{Prometheus-HotB} code in 3D, varying the initial mass of the He-core ($3.3-11\,M_\odot$) and the explosion energy ($0.5-3\,$B). The grid is computed with some improvements to the physics in \texttt{ExTraSS}, in particular the addition of photoionization effects, important for some of the neutral lines, by an on-the-spot treatment. We carry out comparisons of the predicted nebular line profile properties against observations of over 20 Type Ib SNe, demonstrating how new inferences on which combinations of He-core masses and explosion energies are realized in nature can be made by the use of line profile properties that can only be probed by 3D modelling.

The paper is organized as follows; in Section \ref{sec:models} we describe the nine supernova explosion models, upgrades to the \texttt{ExTraSS} code, and the synthetic spectra computations and line metrics. In Section \ref{sec:results} we analyze the model spectra, starting with variations of line profile properties with viewing angle and by comparing the model line profile properties to those measured from a large set of archival observed spectra. We also discuss the NIR He-lines and their predicted visibility. In Section \ref{sec:discussion} we evaluate and discuss our findings, and in Section \ref{sec:conclusion} we present a summary.

\section{Modelling} \label{sec:models}  
\begin{table*}
	\centering
	\caption{Overview of all the models. $M_\text{preSN}$ is the stellar mass at the onset of core-collapse, and $M_\text{CO}$ is the mass of the carbon-oxygen core at this stage. $N_r$ is the number of radial zones at the end of the \texttt{P-HotB} simulation, which is given by $t_\text{end}$. $E_\text{kin}$ is the explosion energy. $M_\text{ej}$ is the total ejecta mass and $M_\text{Ni}$ is the ejected $^{56}$Ni mass. $M_\text{NS,5}$ is the mass of the neutron star 5 seconds into the explosion, $M_\text{fallback}$ is the mass accreted onto the neutron star after 5 seconds, and $M_\text{CR,final}$ is the baryonic mass of the compact remnant (either a neutron star [NS] or black hole [BH], denoted in the CR column). $v_\text{CR,final}$ is the kick velocity of the compact remnant. The masses for $M_\text{NS,5}$ and $M_\text{CR,final}$ are the \textit{baryonic} masses.}
	\label{tab:model_overview}
    \setlength\tabcolsep{0pt}
     \begin{tabular*}{.98\linewidth}[H]{@{\extracolsep{\fill}}l ccc ccc ccc ccc c}
        \hline
        \multicolumn{1}{c}{Model} & $M_\text{preSN}$ & $M_\text{CO}$ & $N_r$ & $t_\text{end}$ & $E_\text{kin}$ & $M_\text{ej}$ & $M_\text{Ni}$ & $M_\text{NS,5}$ & $M_\text{fallback}$ & $M_\text{CR,final}$ & CR & $v_\text{CR,final}$   \\ 
          & ($M_\odot$) & ($M_\odot$) &  & (s) & ($10^{51}\,$erg) & ($M_\odot$) & ($M_\odot$) & ($M_\odot$) & ($M_\odot$) & ($M_\odot$) &   & (km$\,$s$^{-1}$)  \\
        \hline
        HEC-33L  & 2.65 & 1.75 & 2597 & 1001 & 0.47 & 1.204 & 0.045 & 1.41 & 0.04 & 1.45 & NS & 213 \\
        HEC-33   & 2.65 & 1.75 & 2512 &  999 & 1.05 & 1.308 & 0.084 & 1.34 & --    & 1.34 & NS & 135 \\
        HEC-33H  & 2.65 & 1.75 & 2535 &46840 & 2.98 & 1.470 & 0.170 & 1.18 & --    & 1.18 & NS &  27 \\
        HEC-60L  & 4.42 & 3.15 & 2000 & 3411 & 0.51 & 2.051 & 0.048 & 1.60 & 0.77 & 2.37 & NS & 225 \\
        HEC-60   & 4.42 & 3.15 & 2022 & 3214 & 0.99 & 2.621 & 0.094 & 1.49 & 0.31 & 1.80 & NS & 394 \\
        HEC-60H  & 4.42 & 3.15 & 3007 & 3388 & 3.27 & 3.142 & 0.258 & 1.28 & --   & 1.28 & NS & 97 \\
        HEC-110L & 6.95 & 5.45 & 2592 & 1414 & 0.52 & 2.096 & 0.008 & 1.97 & 2.89 & 4.86 & BH & 37 \\
        HEC-110  & 6.95 & 5.45 & 1917 & 1369 & 1.08 & 3.303 & 0.095 & 1.78 & 1.87 & 3.65 & BH & 102 \\
        HEC-110H & 6.95 & 5.45 & 1920 & 1339 & 3.15 & 5.069 & 0.280 & 1.54 & 0.34 & 1.88 & NS & 200 \\
		\hline
	\end{tabular*}
\end{table*}
\subsection{Explosion modelling}
In this work we use nine different explosion simulations, with three different progenitor stars at three different explosion energies. The progenitor stars are helium stars based on \citet{ertl2020explosion}'s models for initial He-core masses of $3.3\,M_\odot$, $6.0\,M_\odot$ and $11.0\,M_\odot$ at the end of core-hydrogen burning. After stripping their hydrogen envelope these progenitors were further evolved to the presupernova stage with standard mass loss by \citet{woosley2019evolution}. Some more details of the progenitor stars at the onset of core collapse is given in Table \ref{tab:model_overview} and in Appendix \ref{app:progdata}.

The 3D explosion modelling of these progenitors with the \texttt{Prometheus-HotB} (\texttt{P-HotB}) CCSN code\footnote{The \texttt{Prometheus} hydrodynamics module of the code was developed by \citet{fryxell1991instabilities} and \citet{muller1991high,muller1991instability} and is supplemented for CCSN simulations by microphysics added by \citet{janka1996neutrino} and \citet{kifonidis2003non,kifonidis2006non}, including an approximate neutrino transport implemented by \citet{scheck2006multidimensional} and \citet{arcones2007nucleosynthesis}.} followed the methodology applied in a series of previous works by \citet{wongwathanarat2010hydrodynamical,muller2012parametrized,wongwathanarat2013three,wongwathanarat2015three,wongwathanarat2017production}. By adapting the values of time-dependent neutrino luminosities at the inner grid boundary, the explosion energy was tuned to the chosen values of approximately $0.5\,$B, $1\,$B and $3\,$B. The symmetry of the original 1D progenitor models was broken about $10\,$ms after core bounce by applying small-amplitude random (cell-by-cell) perturbations to seed the growth of post-shock instabilities in the neutrino-heating layer \citep[see][]{wongwathanarat2010hydrodynamical,wongwathanarat2013three}. A more detailed look at \texttt{P-HotB} is given in Appendix \ref{app:BonusRenderings}.

An overview of basic properties of the explosion models is provided by Table \ref{tab:model_overview}. The CCSN simulations were carried out with an angular resolution of $2^{\circ}$ and the radial grid had a zone number at the end of the simulations as given in the table. This ensured a relative radial resolution $\Delta r / r$ of better than $1\%$ at all radii and times. Model HEC-33 was previously used by \citet{vanbaal2023modelling}. The model names contain information on the initial He-core mass of the progenitor. A suffix H indicates the high-energy models, a suffix L the low-energy models while no suffix indicates those with intermediate energy.

The variation in the explosion energy leads to differences in the amount of $^{56}$Ni+tracer element X\footnote{Element X is an extra species added to the small $\alpha-$nuclei network used in the \texttt{P-HotB} simulations (see Appendix \ref{app:BonusRenderings} for details), to track the production of neutron-rich nuclear species at low electron fractions $Y_e < 0.49$, to save time on otherwise expensive detailed nucleosynthesis calculations. In the rest of the paper, when we mention $^{56}$Ni, it refers to $^{56}$Ni+element X unless explicitly mentioned otherwise. We stress that the mass of $^{56}$Ni+element X must be considered as upper limit of the true $^{56}$Ni mass, because not all of the tracer material is likely to end up as $^{56}$Ni.} produced. For the $3.3\,M_\odot$ explosions, this ranges from $0.045-0.17\,M_\odot$; for the $6.0\,M_\odot$ explosions this ranges from $0.048-0.258\,M_\odot$ and for the $11.0\,M_\odot$ explosions from $0.008-0.28\,M_\odot$. This results in variations in the energy available at nebular times, which will affect e.g. the ionization levels and temperature across the nebula. 

\begin{figure*}
    \centering
    \includegraphics[width=.32\linewidth]{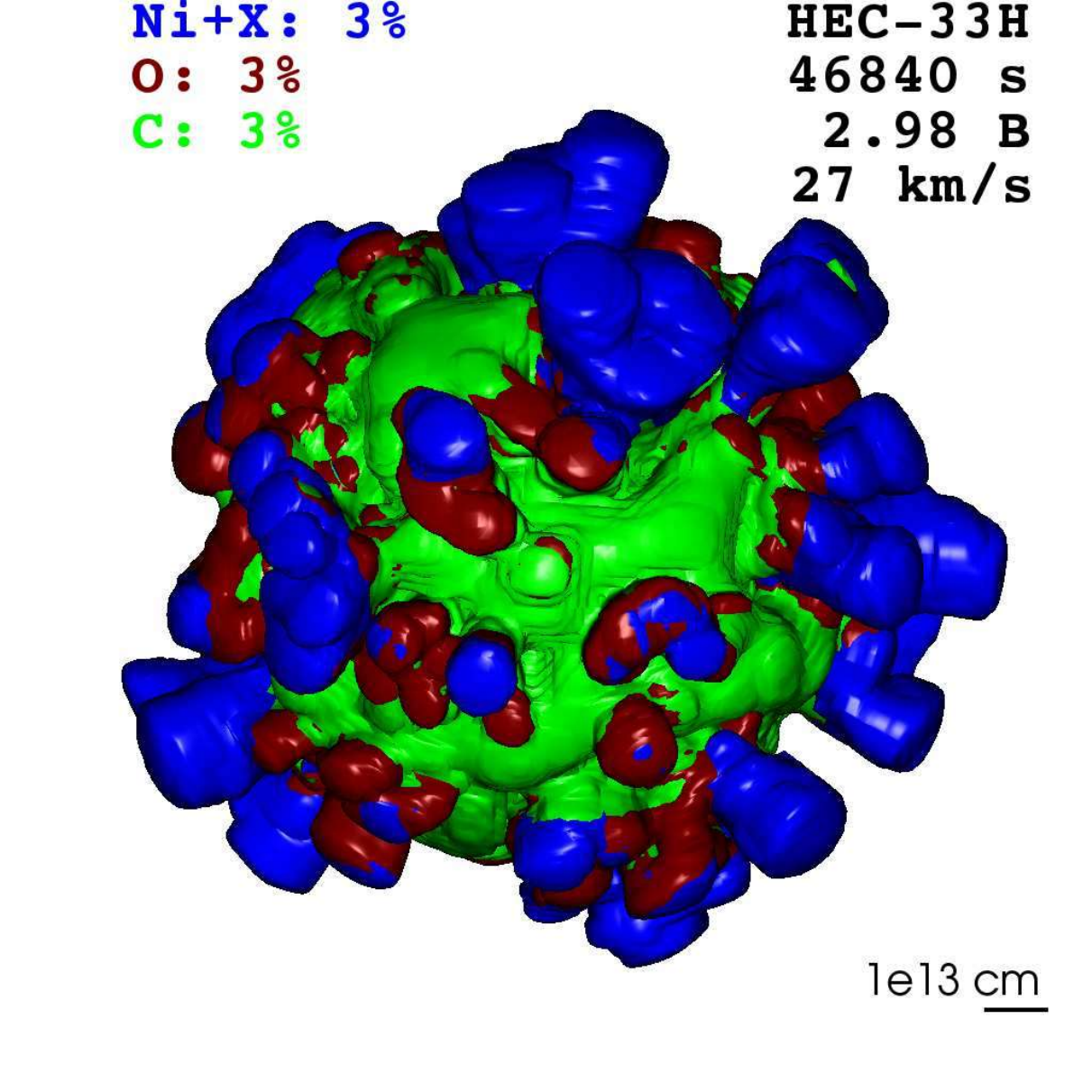}   
    \includegraphics[width=.32\linewidth]{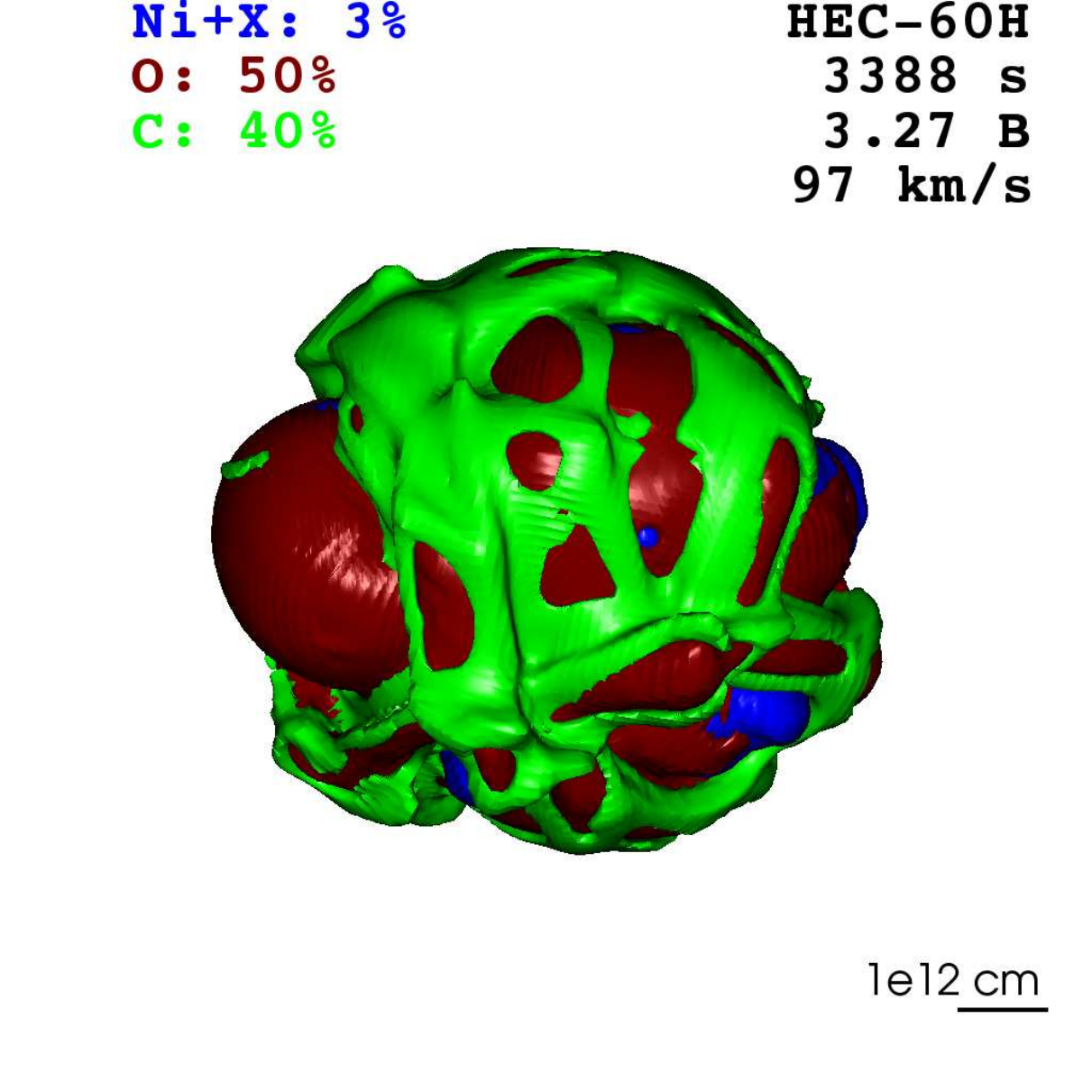}   
    \includegraphics[width=.32\linewidth]{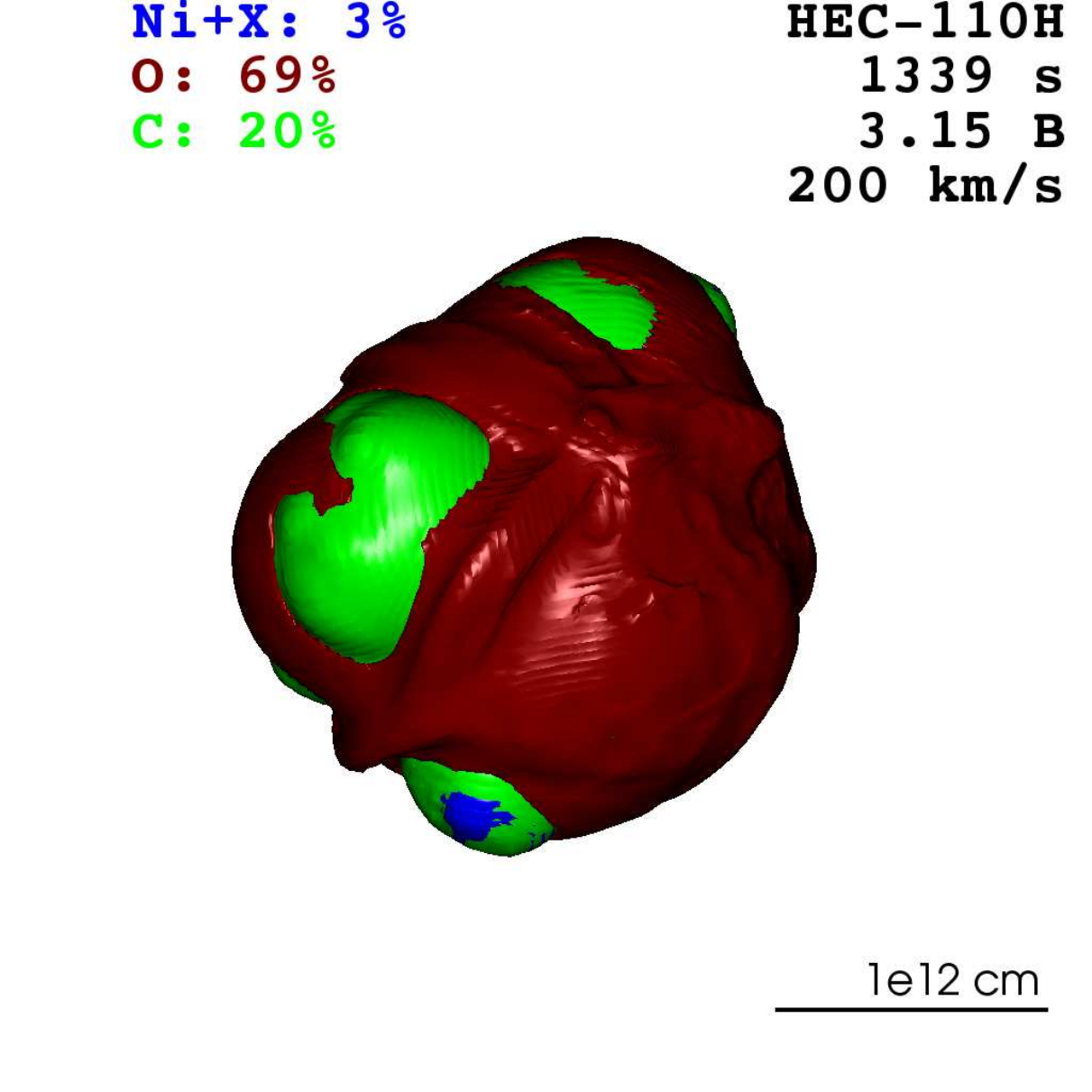}   
    \includegraphics[width=.32\linewidth]{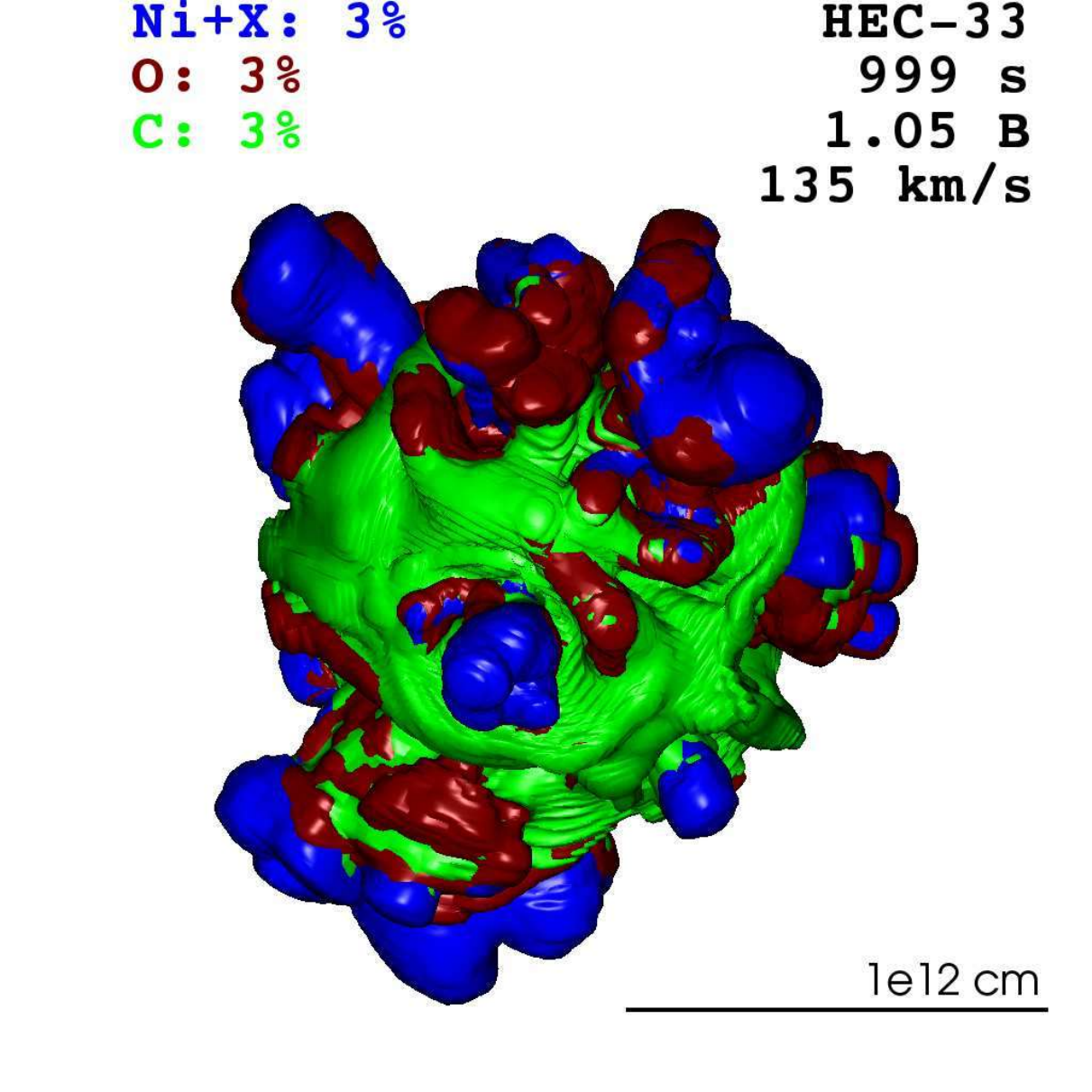}   
    \includegraphics[width=.32\linewidth]{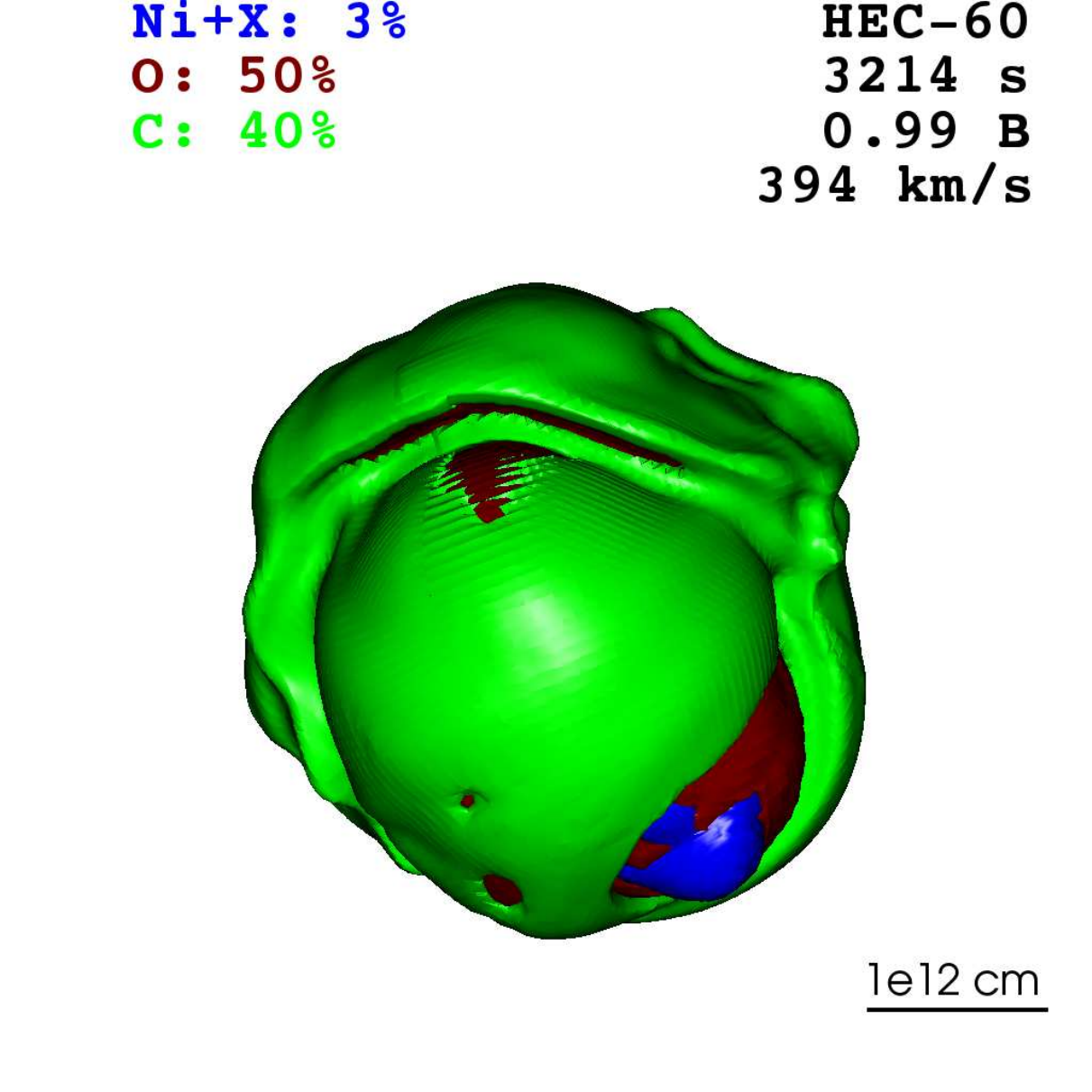}   
    \includegraphics[width=.32\linewidth]{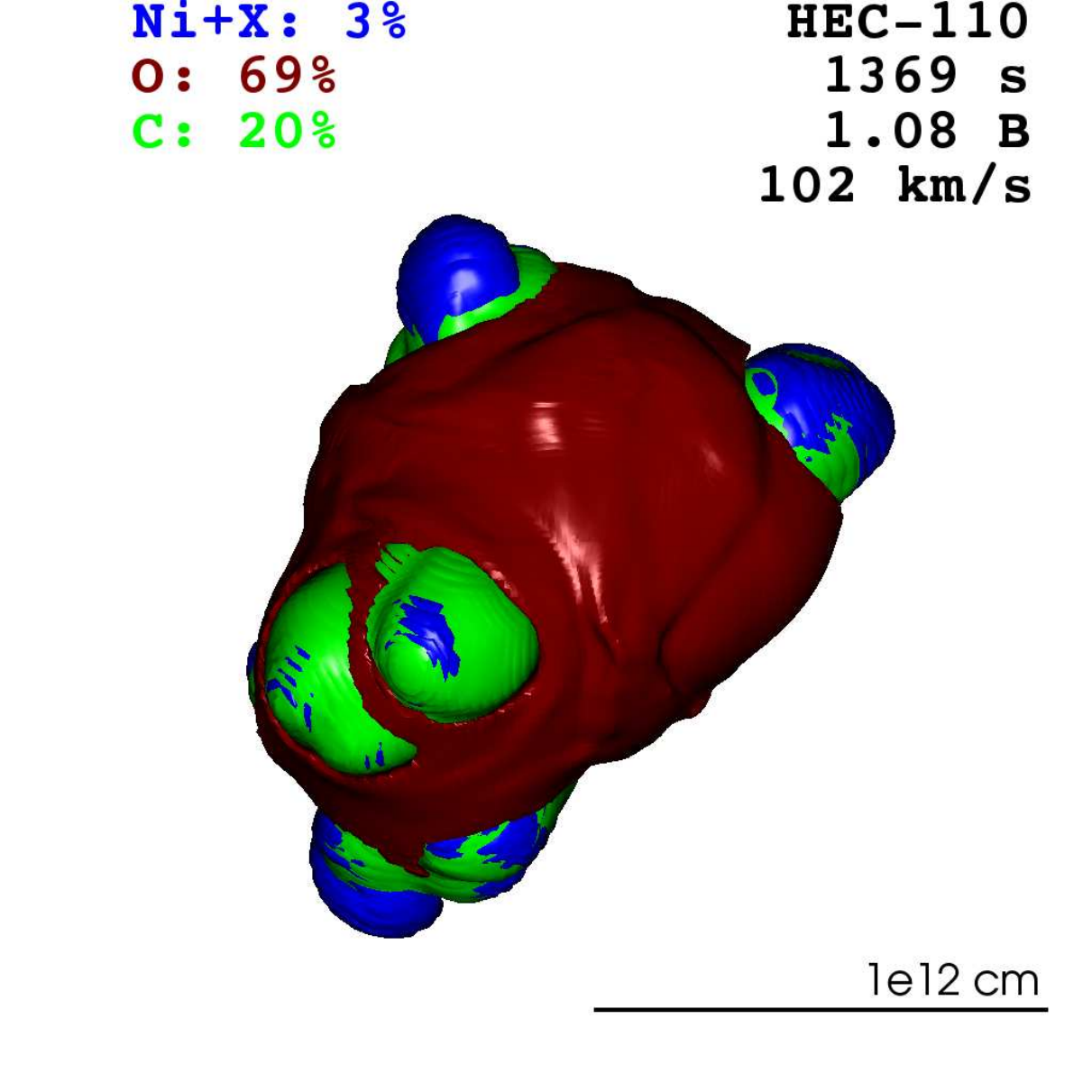}   
    \includegraphics[width=.32\linewidth]{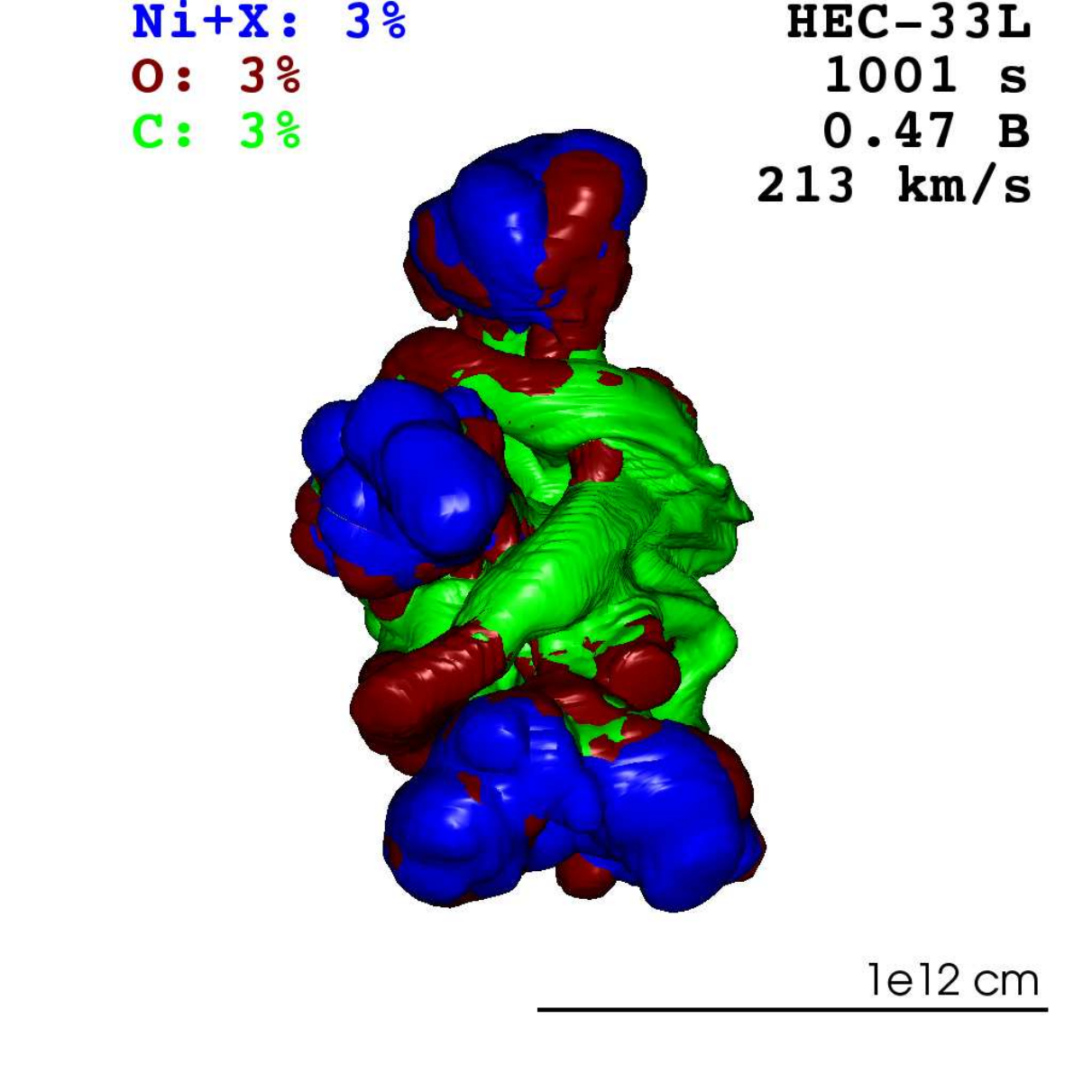}   
    \includegraphics[width=.32\linewidth]{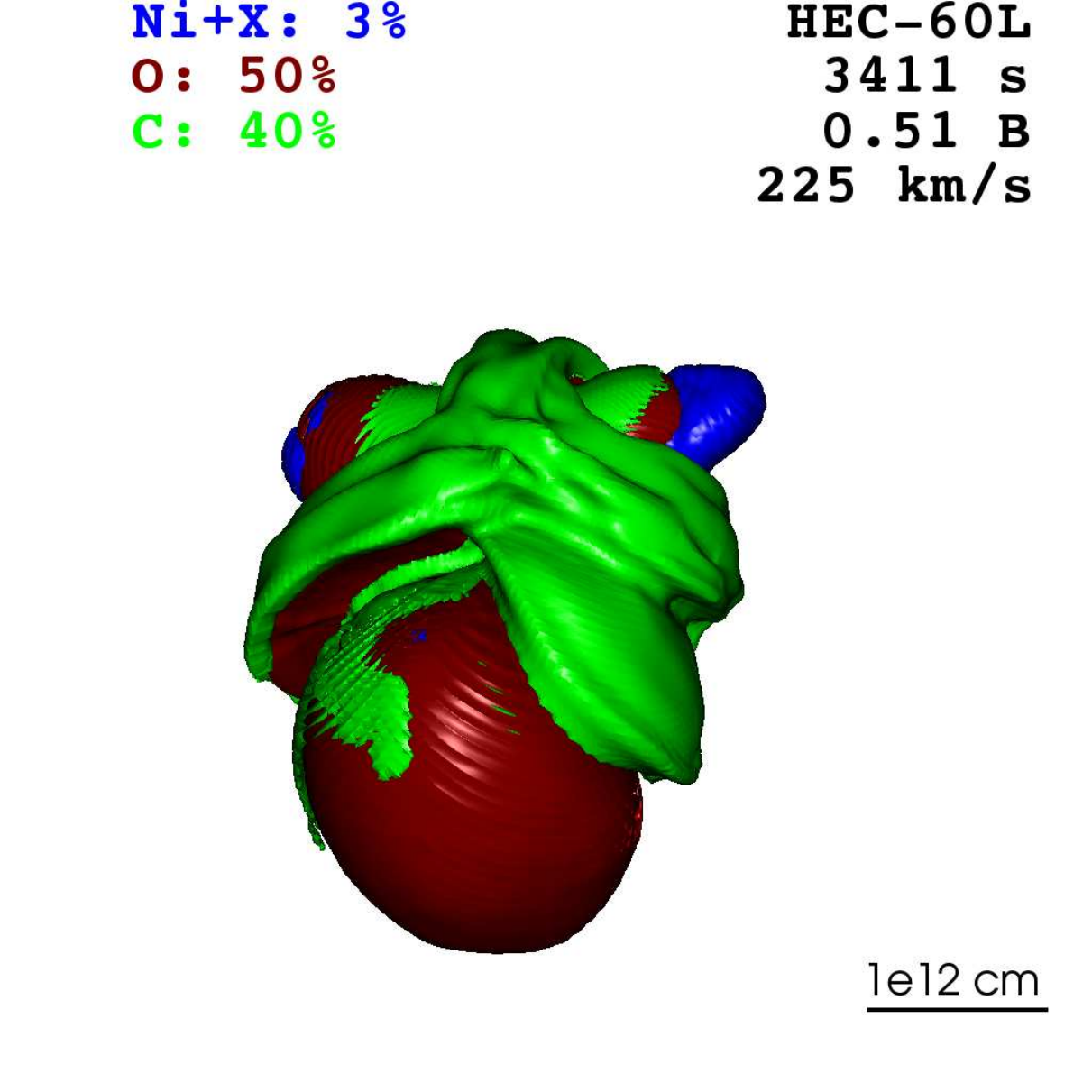}   
    \includegraphics[width=.32\linewidth]{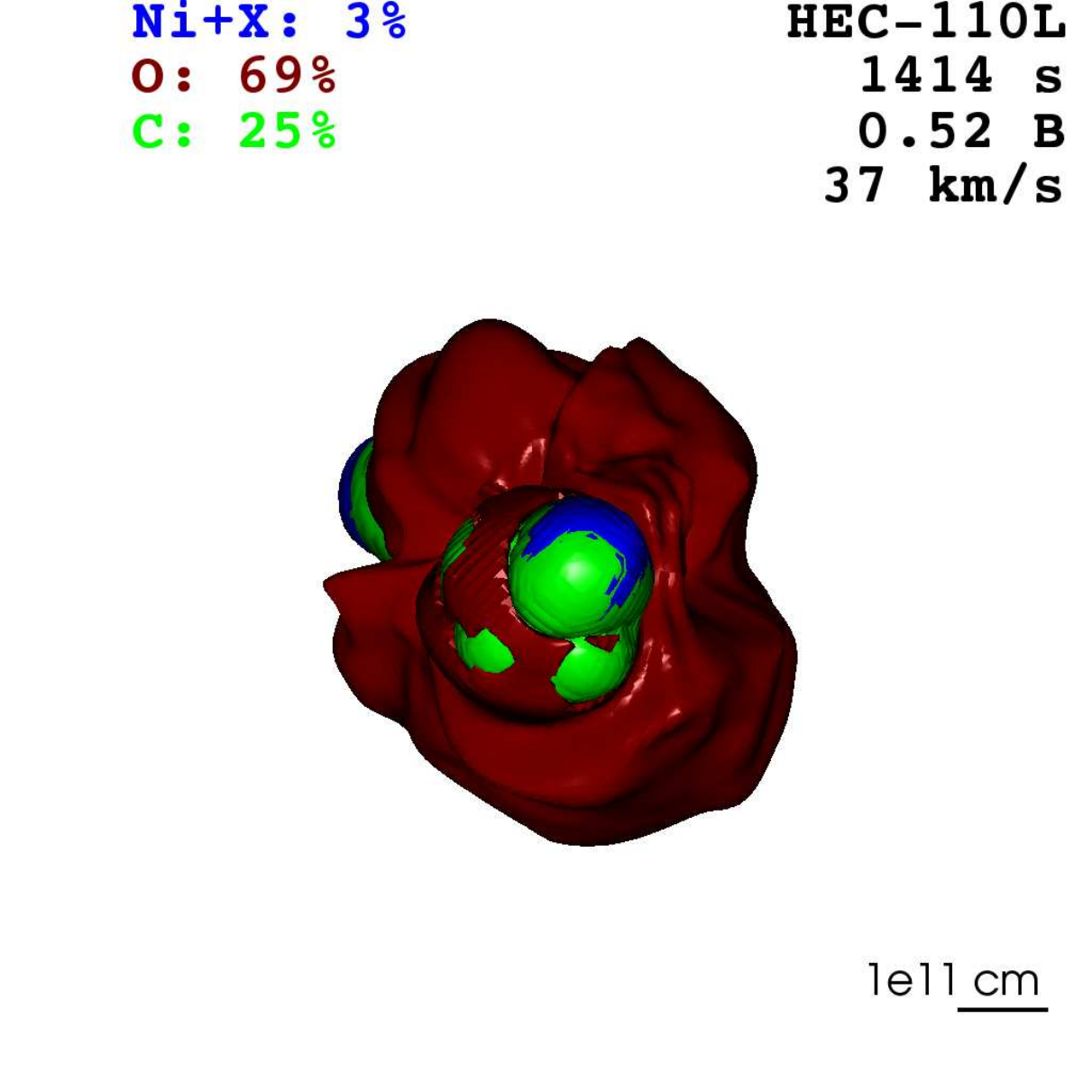}   
    \caption{The 3D renderings of all explosion models, showing iso-surfaces of C (green), oxygen (red) and Ni+X (blue), for the element mass fractions specified in the Figure legends, at the end of our 3D SN simulations with the \texttt{P-HotB} code. Each row has similar explosion energies (from top to bottom, 3, 1 and 0.5 Bethe), while each column has the same progenitor star (from left to right, 3.3, 6.0 and 11 $M_\odot$ He-cores). The final \texttt{P-HotB} time ($t_\text{end}$ in Table \ref{tab:model_overview}), explosion energy (B$\equiv10^{51}\,$erg) and compact object kick velocity are given in the top-right of each panel; each panel is orientated such that the neutron star motion is vertically upwards. The mass fractions of the different elements on the isosurfaces are given on the top-left of each panel. Note that the isosurfaces enclose (roughly) the same fractions of the total masses of an element in all cases, for which reason the local mass fractions on the isosurfaces differ. Note also that the scales between the different renderings vary, due to the different expansion velocities at time $t_\text{end}$ of the simulations. Angle-averaged chemical compositions of each explosion model are plotted in Figures \ref{fig:composition_mixing_hec33}-\ref{fig:composition_mixing_hec11}.}
    \label{fig:3D-Rendering}
\end{figure*}

In Figure \ref{fig:3D-Rendering} 3D renderings of the explosion models are shown. In each of the renderings, the isosurfaces for constant mass fractions of $^{56}$Ni are shown in blue, of oxygen in red and carbon in green. For example, a $3\%$ oxygen isosurface outlines the surface around the computational grid cells where the composition is at least $3\%$ oxygen. The values of the isosurfaces change between the different models such that in each model (roughly) the same fraction of that element is enclosed. This allows us to highlight both the global asymmetries in the model as well as differences between chemical elements. In each panel the ejecta is orientated such that the neutron star's movement is directly upwards. Additional renderings of all models are shown in Appendix \ref{app:BonusRenderings}.

For the three HEC-33 models, the higher energy model shows the most symmetric ejecta on a global scale, although clear $^{56}$Ni fingers are present at various places. The medium energy model shows more deviation from sphericity with slightly broader $^{56}$Ni fingers spaced out more than in the high energy model. For the low energy model the ejecta are predominantly found along the neutron star's kick vector, and even fewer but broader $^{56}$Ni fingers appear.

For the HEC-60 models, the oxygen and carbon isosurfaces are shown at $50\%$ and $40\%$ respectively as this is roughly the region where these contours overlap with the $3\%$ nickel one. For the all energies but especially the medium model it can be seen that the $^{56}$Ni fingers struggle to penetrate past this isosurface out into the envelope of the ejecta, with only a few hints of the fingers piercing through the carbon/oxygen dominant regions. For the high energy model a few more $^{56}$Ni fingers succeed at this, but they are still unable to reach the outer envelope. Hidden underneath these carbon/oxygen layers is still an asymmetric distribution of $^{56}$Ni, although on the global scale the low energy model displays strong asymmetries even for the carbon/oxygen layers.  

For the HEC-110 models, a similar trend exists: only a few $^{56}$Ni fingers make an appearance into the carbon/oxygen envelope for all three models, and on a global scale the low energy model is the most asymmetric of the three. The distribution of $^{56}$Ni in this HEC-110L model is also quite bipolar and only a small fraction of the overall ejecta (see also Table \ref{tab:model_overview}), and there are large portions of the ejecta without any $^{56}$Ni.

\begin{figure*}
    \centering
    \includegraphics[width=\linewidth]{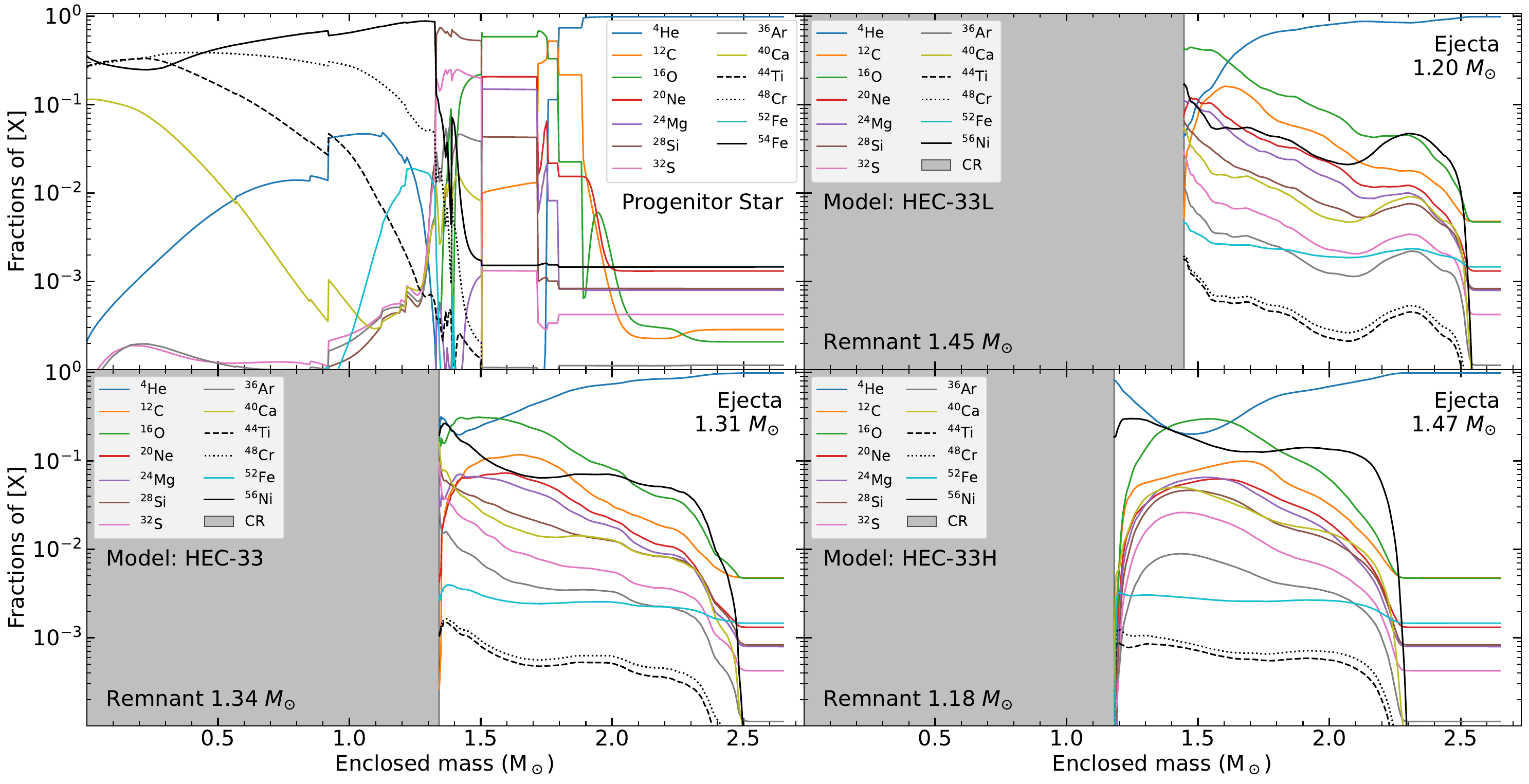}
    \caption{Chemical compositions of the $3.3\,M_\odot$ progenitor star (top left) and of the angle-averaged ejecta for the three different explosion energies of $0.5\,$B (HEC-33L, top right), $1\,$B (HEC-33, bottom left) and $3\,$B (HEC-33H, bottom right). For each panel the 12 $\alpha-$nuclei from $^{4}$He to $^{52}$Fe are shown, as well as $^{54}$Fe for the progenitor (the dominant component of the Fe-core) and $^{56}$Ni for the SN ejecta (which is the source of $\gamma$-rays through radioactive decay). The shaded area in the ejecta panels indicates the mass which has formed the compact object, with the remnant mass given.}
    \label{fig:composition_mixing_hec33}
\end{figure*}

\begin{figure*}
    \centering
    \includegraphics[width=\linewidth]{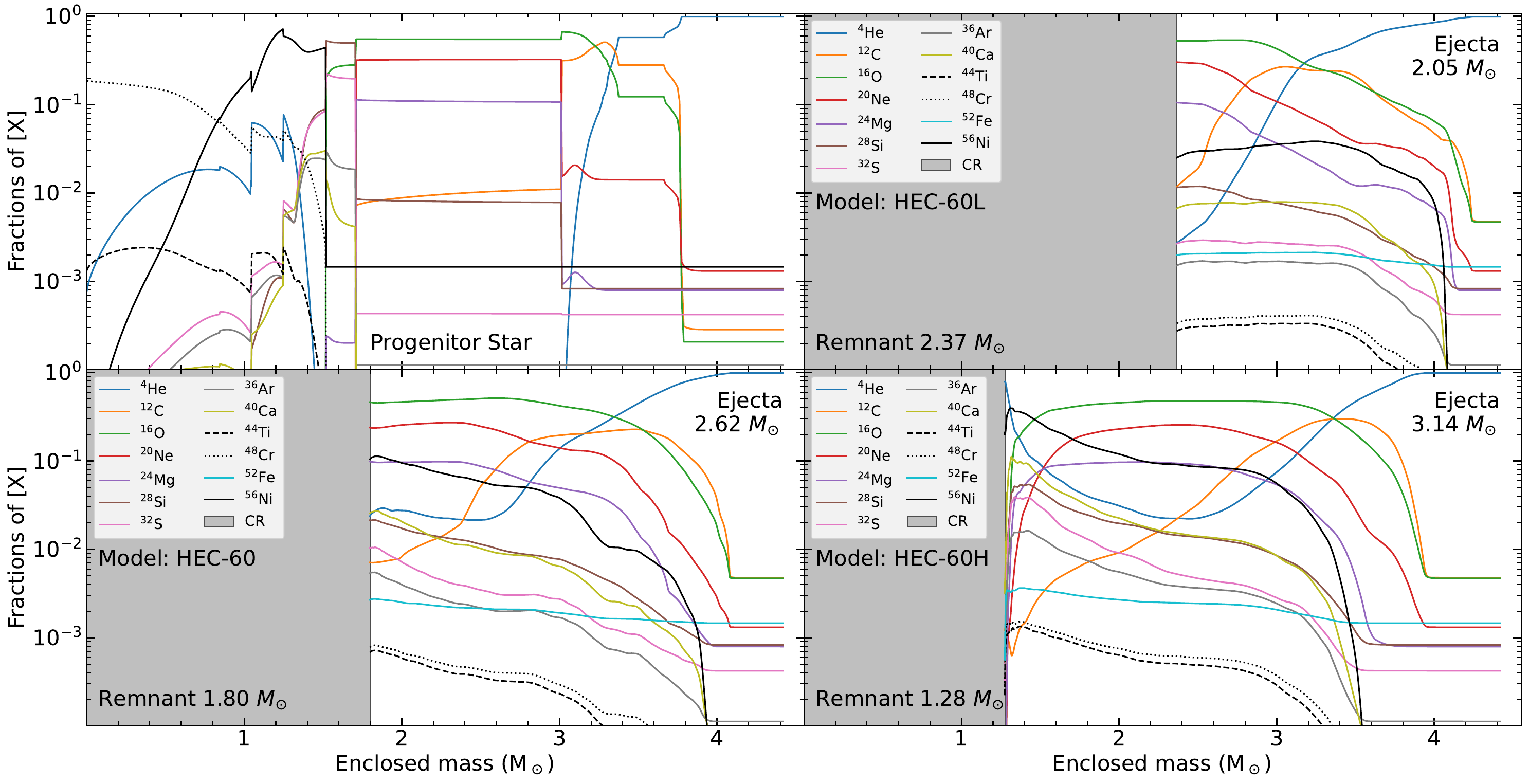}
    \caption{Same as Figure \ref{fig:composition_mixing_hec33}, but for the progenitor and SN ejecta models of the $6.0\,M_\odot$ He-core star. The legend for the progenitor star is the same as in Figure \ref{fig:composition_mixing_hec33}.}
    \label{fig:composition_mixing_hec60}
\end{figure*}

\begin{figure*}
    \centering
    \includegraphics[width=\linewidth]{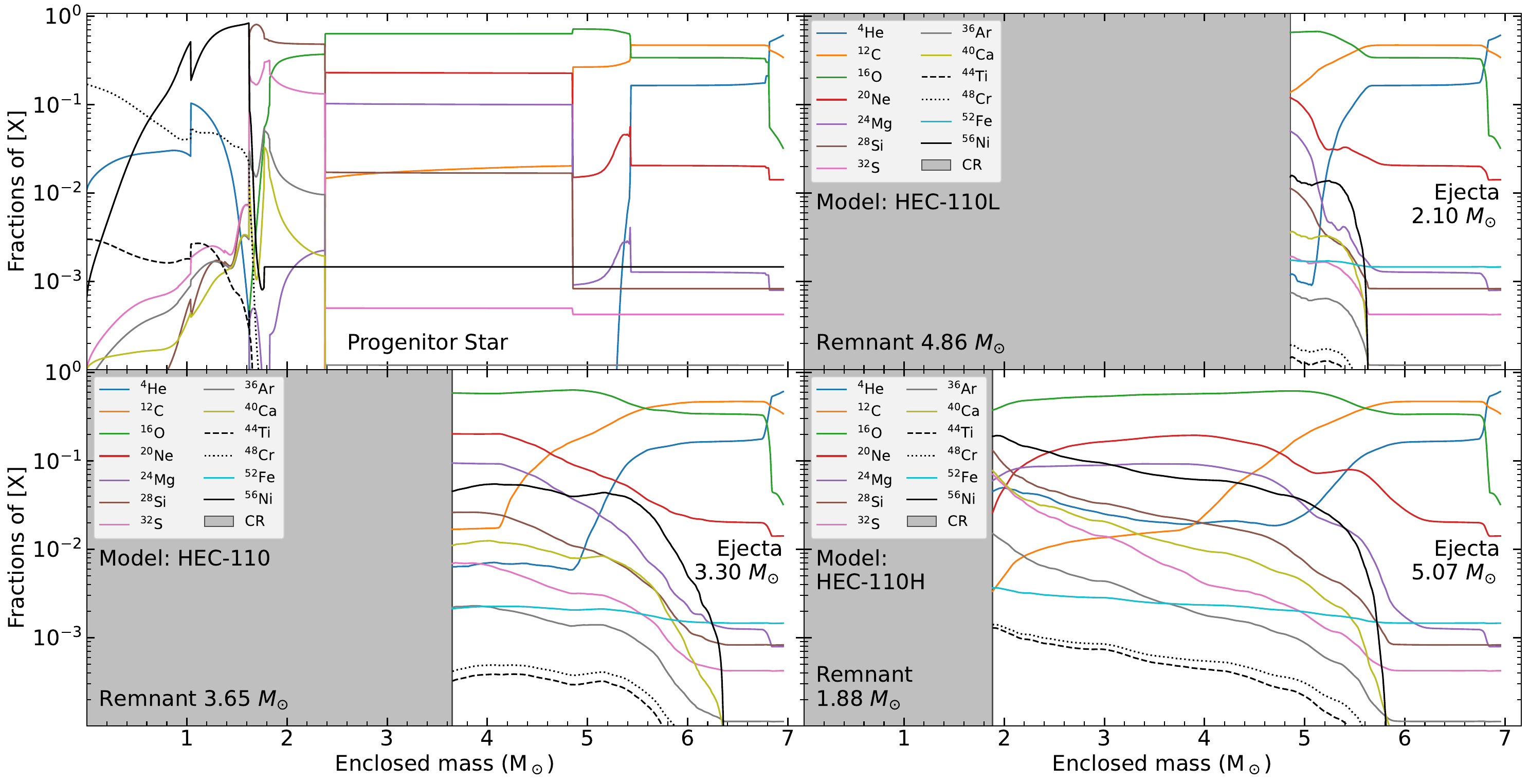}
    \caption{Same as Figure \ref{fig:composition_mixing_hec33}, but for the progenitor and SN ejecta models of the $11.0\,M_\odot$ He-core star. The legend for the progenitor star is shown in Figure \ref{fig:composition_mixing_hec33}.}
    \label{fig:composition_mixing_hec11}
\end{figure*}

In Figure \ref{fig:composition_mixing_hec33} the chemical composition for the $3.3\,M_\odot$ progenitor is shown alongside the composition for each of the three explosion energies; in Figures \ref{fig:composition_mixing_hec60} and \ref{fig:composition_mixing_hec11} these compositions are shown for the $6.0\,M_\odot$ and $11.0\,M_\odot$ models respectively. In each of these figures, the top left panel shows the progenitor structure while the other three panels show the angle-averaged element distributions for the three different explosion energies. The shaded area indicates the mass forming the central compact remnant; the model names and remnant masses are denoted in each panel. The HEC-110L and HEC-110 models have remnant masses well over the maximum possible NS star mass ($\sim\,2\,M_\odot$, \citealt{cromartie2020relativistic,romani2021PSR}) and will form black holes -- such fallback black hole formation accompanied by a SN explosion was recently achieved also with 3D explosion models with more sophisticated neutrino treatment than in our parametric neutrino-driven explosion models \citep{chan2020impact,burrows2023black}. Most of the mass over $2\,M_\odot$ is added by fallback accretion over comparatively long time-scales however (see Table \ref{tab:model_overview}), and thus does not directly affect the explosion itself. 


For the models in Figures \ref{fig:composition_mixing_hec33} and \ref{fig:composition_mixing_hec60} it can be seen that the outer He-rich envelope remains relatively unmixed, with the more energetic explosions having a larger region unaffected by mixing. This is because the envelope gets expelled at lower velocities for the less energetic explosions, which allows for the core ejecta to be mixed into a larger fraction of the envelope. This can be seen as the 'flatlining' of the outermost ejecta in Figures \ref{fig:composition_mixing_hec33} and \ref{fig:composition_mixing_hec60}. For the HEC-110 models, this flatlining occurs in the C/O core rather than the He envelope as the C/O core is much more massive. 

In Figure \ref{fig:composition_mixing_hec11}, we see that for HEC-110 the systematic energy dependence of the mixing is not present, as the outer He-rich envelope is already much smaller (only the outer $\sim0.1\,M_\odot$ of the progenitor is He-dominated) and remains unmixed. Instead, underneath this thin He envelope the $11\,M_\odot$ models show an additional C/O rich layer that is also lacking any mixing with radioactive $^{56}$Ni.

The outward mixing of $^{56}$Ni is non-monotonic in the $11.0\,M_\odot$ models, as only a small fraction of the ejecta in the low explosion energy model contain (significant) $^{56}$Ni, while in the $3.3$ and $6.0\,M_\odot$ models the low explosion models have the largest outward mixing of $^{56}$Ni. Across the three progenitor models, it can be seen that towards higher explosion energies, relatively more He is expelled in the innermost ejecta, yet only a smaller part of this can be attributed to inward mixing of the envelope. Instead, at higher explosion energies the inner regions achieve higher entropy, which causes $\alpha-$rich freeze-out in neutrino-heated ejecta to produce more He (in addition to producing more $^{56}$Ni), and this is the main contributor to the high He-fractions in the innermost ejecta. The nucleosynthesis network in \texttt{P-HotB} is discussed in more detail in Appendix \ref{app:BonusRenderings}.

\subsection{Spectral synthesis}
Each of the models is taken from the time at the end ($t_\text{end}$, see Table \ref{tab:model_overview}) of the \texttt{P-HotB} simulation to $200\,$days after explosion with \texttt{ExTraSS} \citep[see][for an introduction of the code]{vanbaal2023modelling}. This is done by assuming homologous expansion (i.e. the only velocity component is in the radial direction) from $t_\text{end}$, so the densities are simply decreased by $\left(200\,\text{days}/t_\text{end}\right)^{-3}$. The $^{56}$Ni bubble expansion effect \citep[see][]{gabler2021infancy} is not accounted for. As discussed in \citet{vanbaal2023modelling}, for SESNe this bubble expansion effect is expected to be smaller than for the H-rich SN models, as the expansion velocities of the ejecta are much lower in H-rich SN.

The spectral synthesis code \texttt{ExTraSS} is described in \citet{vanbaal2023modelling}. In brief, the code calculates the deposition of gamma-rays and leptons from radioactive decay of $^{56}$Ni and $^{56}$Co across the nebula. It then solves for the distribution of non-thermal electrons by using the Spencer-Fano subroutine developed by \citet{kozma1992gamma}, also used and further developed in the \texttt{SUMO} code \citep{jerkstrand201144Ti,jerkstrand2012progenitor}. These solutions are iterated with solutions to thermal equilibrium and Non-Local Thermodynamic Equilibrium (NLTE) level populations by accounting for the non-thermal and thermal collisional processes, spontaneous radiative decay (under the Sobolev approximation), recombination, and photoionization induced by local (but no non-local) UV emissivity (details below). From these level populations the wavelength-dependent emission in each cell is generated. The final spectra are then generated by considering the Doppler shifts of all cells for different viewing directions \citep[see][for details]{jerkstrand2020properties}. For this work, we computed spectra in the wavelength range of $3000\,$--$25000\,\angstrom$ at a resolution of $R=1000$, which gives us observables for all lines observable from the ground. Tests were done to ensure that higher resolution did not significantly change the computed line properties. For each of our models, we consider a set of 400 viewing angles ($20\times20$ in the azimuthal$\times$polar coordinates). We calculate the variations of the two most key line properties -- line shift and line width\footnote{See Eqs. 2 and 3 in \citet{vanbaal2023modelling} for the definitions of $v_\text{shift}$ and $v_\text{width}$. $v_\text{shift}$ corresponds to the line centroid shift, while $v_\text{width}$ corresponds to the width of the feature. These equations are also defined in \citet{jerkstrand2020properties} as Eqs. 7 and 8.} for each of these angles. 

This work also includes some upgrades to \texttt{ExTraSS} $-$ in particular, a new 'on-the-spot' treatment for photoionization caused by local recombination UV emissivity. Most of the free-bound emissivity is at UV/blue wavelengths, where optical depths are typically significant and the mean-free-path can be on the order of a cell length or less. We create energy bins for the recombination emissivity and use the hydrogenic approximation \citep{rybicki1979radiative} for all cross sections. We consider photoionization from the ground multiplet plus the next eight excited states, for computational reasons. Allowing for more levels does not significantly impact the total photoionization rate but does strongly influence the runtimes. For each wavelength, the fraction of the emissivity to be absorbed in any given cell is calculated from the total optical depth of the cell, using $R_\text{cell}=V^{-1/3}_\text{cell}$. The excess energy put towards the 'on-the-spot' photoionization is accounted for as a heating source in the temperature calculations. With this treatment, we can get an approximate effect of the radiation field in place without doing a full radiative transfer calculation, which we leave for future work. 

In addition, the radioactive decay is now also taken into account for the chemical composition of each cell, with the Ni, Co and Fe abundances calculated correctly from the radioactive decay chain \citep[this erroneously did not occur in][]{vanbaal2023modelling} . A minor error in the ionization structure of Ti was also corrected. As a result of these code updates and fixes, the results for the HEC-33 model that was previously shown in \citet{vanbaal2023modelling} are also somewhat altered but the main conclusions therein are unchanged as the study was focused on Ca, O, C and Mg lines.

\section{Results} \label{sec:results}
In \citet{vanbaal2023modelling}, an in-depth analysis of the physical conditions arising in the HEC-33 3D model, and variation of temperature and ionization with radius and position angle, was carried out. For the model grid here, the corresponding distributions and associated analysis can be found in Appendix \ref{app:PhysicalCon}\footnote{The updates to \texttt{ExTraSS} have slightly altered the conditions in the HEC-33 model but no significant changes have occurred.}. 

Here, we go straight to investigating the variations of the $v_\text{shift}$ and $v_\text{width}$ properties with viewing angles for two of the models before we turn to comparing the entire set against observationally inferred values for those line properties. For this we will focus on the four (generally) strongest features; the singlet features Mg I] $\lambda4571$ and [C I] $\lambda8727$, and the doublet features [O I] $\lambda\lambda6300,\,6364$ and [Ca II] $\lambda\lambda7291,\,7323$. In Appendix \ref{app:observationdata} a full overview of the used observational data is given. We will also compare the NIR spectra of our models to investigate the behaviour of the He lines in particular, to see how these might change for the different SESNe models we have, and whether our models can inform the long-standing debate on presence or absence of helium in Type Ic SNe.

\subsection{Model variability}
\begin{figure*} 
    \includegraphics[width=.495\textwidth]{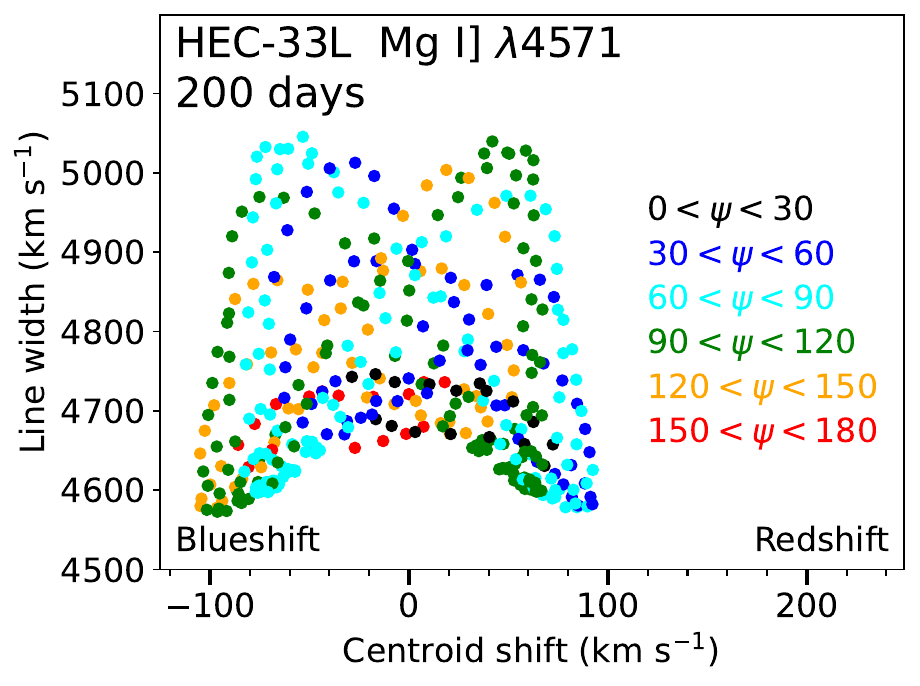}
    \includegraphics[width=.495\textwidth]{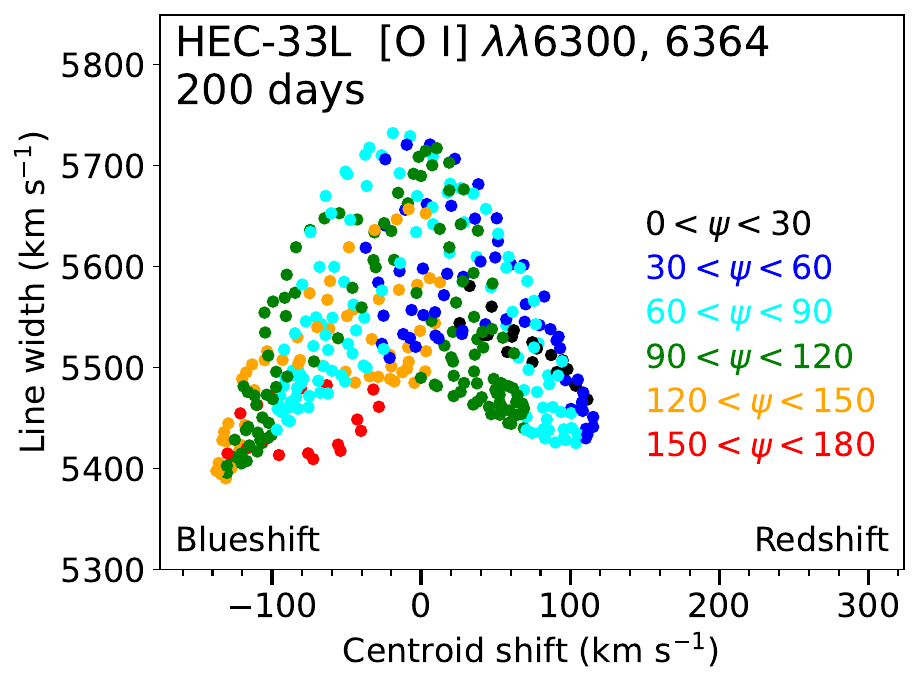}
    \includegraphics[width=.495\textwidth]{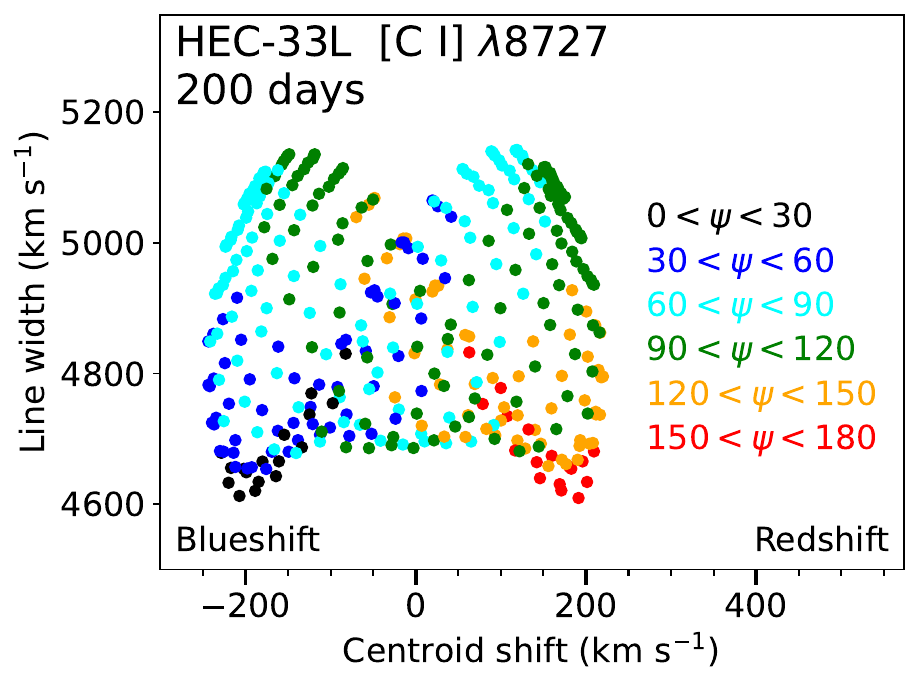}
    \includegraphics[width=.495\textwidth]{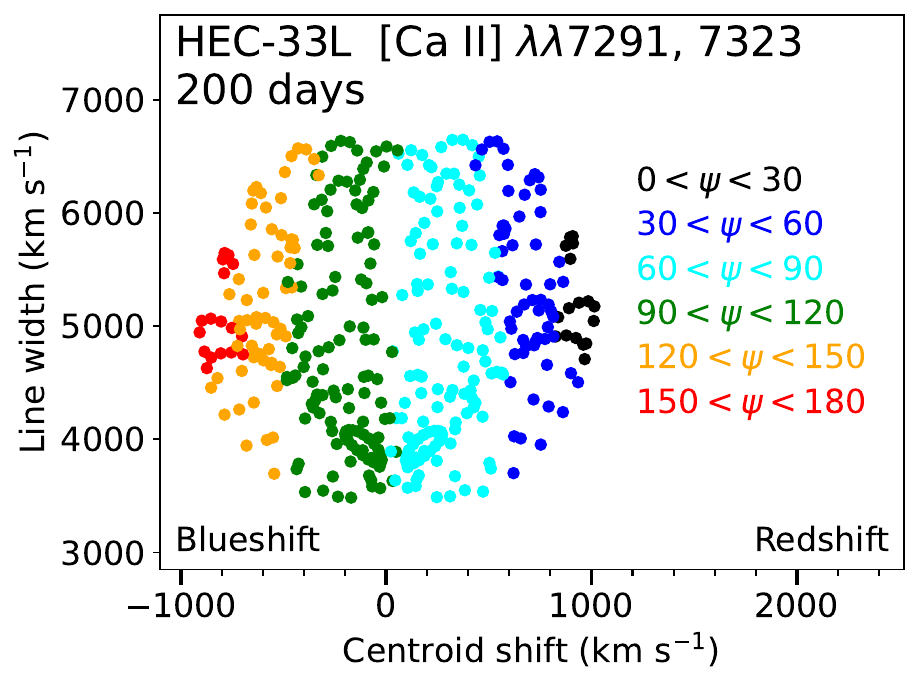}
    \includegraphics[width=\textwidth]{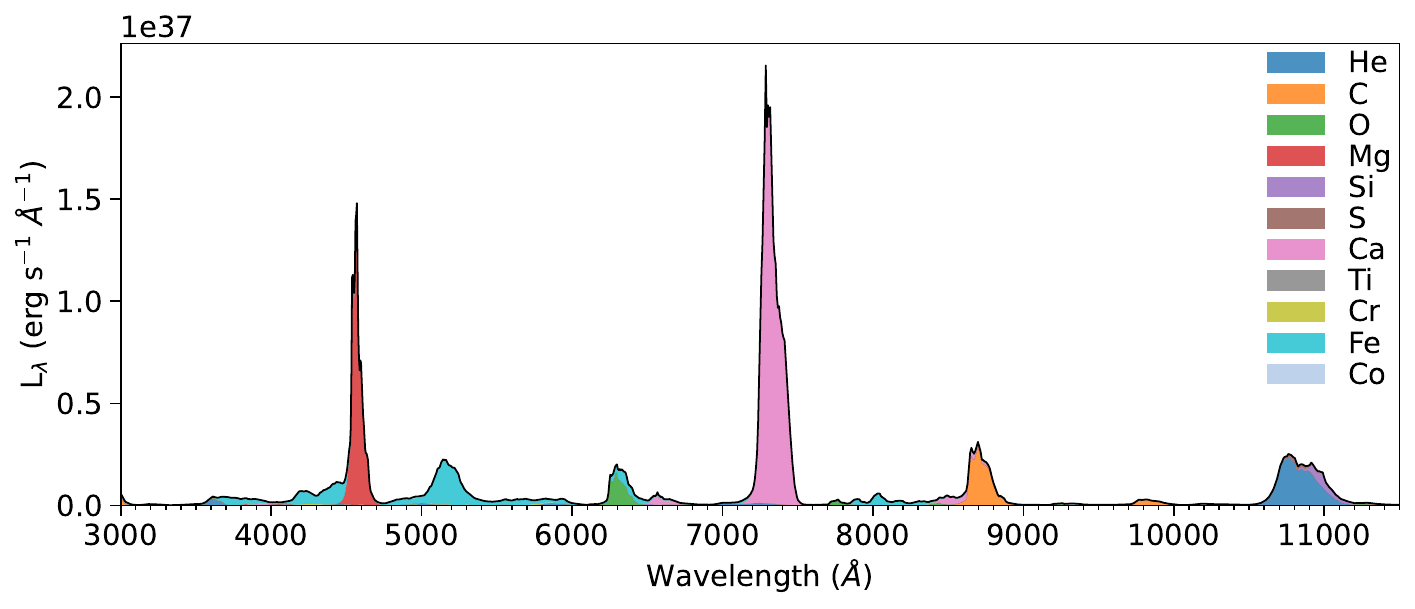}
    \caption{The line properties of the HEC-33L model for the four spectral features Mg I] (top left), [C I] (mid left), [O I] (top right) and [Ca II] (mid right) are shown for all of our viewing angles, color coded for the angle $\Psi$ which is the angle between the direction vector to the viewer and the neutron star motion vector, i.e. the black points (small $\Psi$) correspond to viewing angles where the neutron star is moving almost directly towards the observer. Centroid shifts are given relative to the rest wavelength of the singlet features and relative to the transition-strength weighted rest wavelength for the doublet features (i.e. 6316$\,\angstrom$ for [O I] and 7304$\,\angstrom$ for [Ca II]). Positive (negative) centroid shifts correspond to a redshifted (blueshifted) centroid. The bottom panel shows the full spectrum for this model in the wavelength range $3000-11500\,\angstrom$, for the observer that is most directly approached by the neutron star.}
    \label{fig:33L-shiftwidth}
\end{figure*}
In \citet{vanbaal2023modelling} we investigated the line shift and line width properties for the HEC-33 model in detail and found that [Ca II] had a completely different trend from the other three main lines, which themselves were surprisingly displaying centroid shifts aligned with the NS kick vector rather than anti-aligned shifts. Here we will study the HEC-33L and HEC-60 models at some depth, to compare the effects of different explosion energy for the same star (HEC-33 in Paper I against HEC-33L here) as well as the effect of a more massive star exploding with the same energy (HEC-33 against HEC-60). 

In Figure \ref{fig:33L-shiftwidth} we show the shifts and widths for the four features together with the overall spectral output for the HEC-33L model $-$ the chosen viewing angle is the viewing angle which has the smallest $\Psi$, i.e. the observer for whom the neutron star is most directly approaching. In the spectrum, the four features can clearly be seen, although the [O I] doublet is quite weak.

The line widths of the Mg I] and [C I] lines vary between $\sim4600-5100\,\text{km}\,\text{s}^{-1}$ and are about $2000\,\text{km}\,\text{s}^{-1}$ lower than in the HEC-33 model \citep[exploding with twice the energy and studied in][]{ vanbaal2023modelling}. The [O I] doublet has a similar decrease in widths and varies between $\sim5400-5700\,\text{km}\,\text{s}^{-1}$. The [Ca II] behaviour is qualitatively different to the other three lines, displaying a much larger variation in line widths, $\sim 3500-6600\,\text{km}\,\text{s}^{-1}$; the widest of these profiles are actually as wide as some of the [Ca II] profiles in HEC-33.

For the centroid shifts, Mg I] and [O I] display a similar range of values of up to roughly $\pm100\,\text{km}\,\text{s}^{-1}$, and neither show a clear correlation with the viewing angle relative to the NS kick (i.e. the colours are not cleanly separated into bands). [C I] has shifts that are about twice as large, reaching $\pm200\,\text{km}\,\text{s}^{-1}$, and show some degree of correlation with viewing angle, as the viewing positions where the observer is approached by the neutron star (black dots) consistently give blueshifted profiles, while the viewing positions where the neutron star moves away from the observer (red dots) give redshifted profiles. This behaviour was also apparent in the HEC-33 model in \citet{vanbaal2023modelling}, in which also Mg I] and [O I] showed such correlation. The centroid shifts for [Ca II] are much bigger, reaching up to $\pm1000\,\text{km}\,\text{s}^{-1}$ -- twice the maximum shift values in HEC-33 ($\pm 500\,\text{km}\,\text{s}^{-1}$) despite the explosion energy being twice as small. The values are well separated by the angle $\Psi$ between the neutron star kick velocity vector and the vector direction from the explosion center to the observer position, albeit in the opposite way from [C I], as for [Ca II] the low $\Psi$ observers see redshifted profiles, whereas viewing locations from where the neutron star moves away observe [Ca II] blueshifts. While [Ca II] has unique cooling properties -- being able to take over the cell emission even at low abundances -- it is likely the unique explosive formation site of Ca which causes it to behave differently from the non-explosively synthesized elements.

In both HEC-33 and HEC-33L, the variations with viewing angle for the [Ca II] line properties are very different than for the other three features. While in HEC-33 [O I], [C I] and Mg I] all behaved similarly, in HEC-33L the [C I] profiles are quite different from Mg I] and [O I]. This gives us a first idea of how quantities vary, and do not vary, depending on model. Overall the line widths for HEC-33L are lower, which is expected for the lower explosion energy. The self-consistent 3D modelling now gives us quantified ranges for the line widths depending on He-core mass and explosion energy -- a new analysis tool we will use in Section \ref{sec:obscomp} when comparing to observations.

\begin{figure*} 
    \includegraphics[width=.495\textwidth]{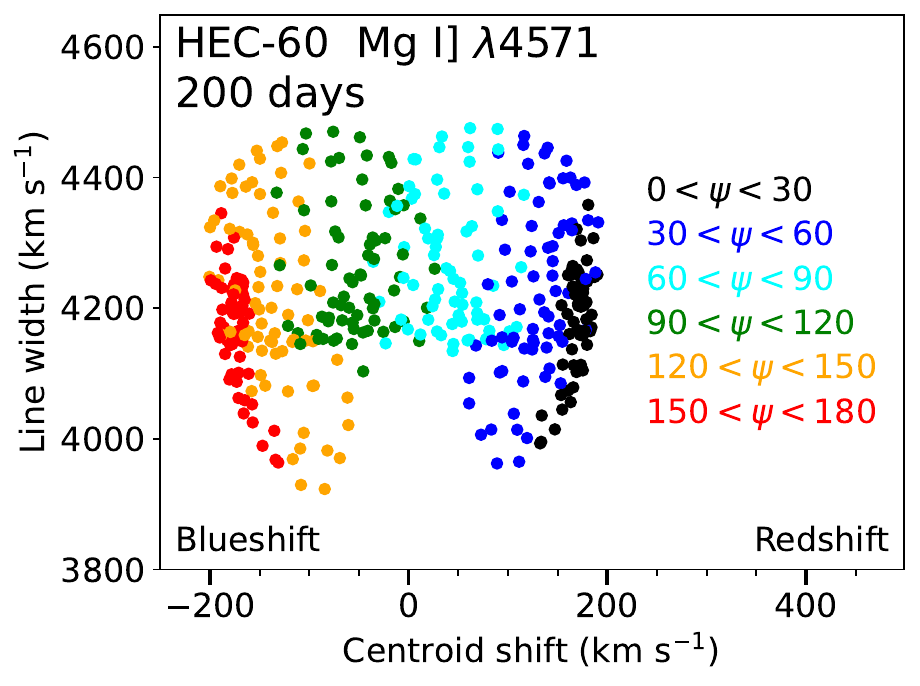}
    \includegraphics[width=.495\textwidth]{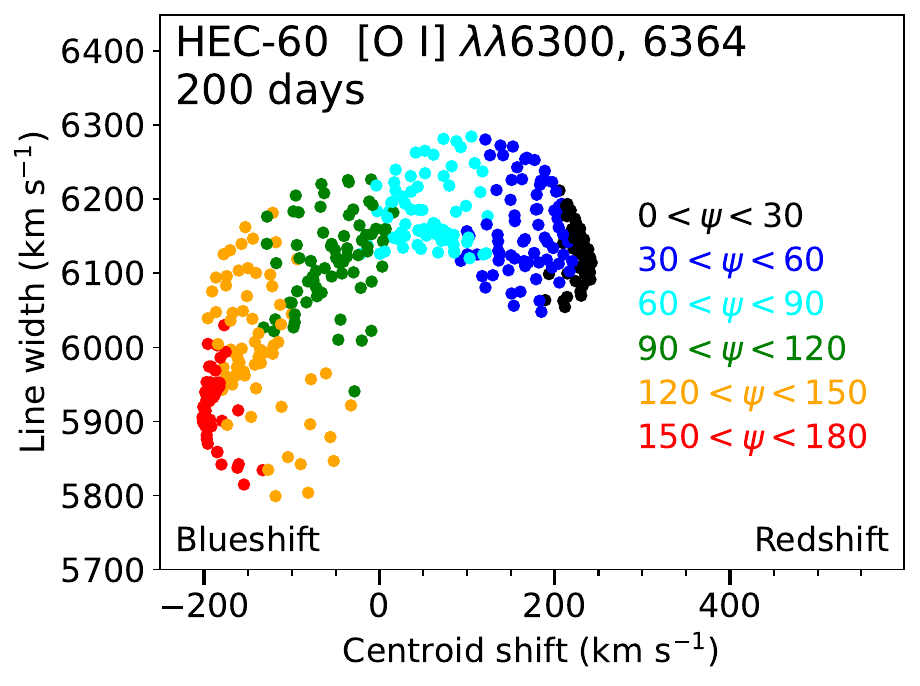}
    \includegraphics[width=.495\textwidth]{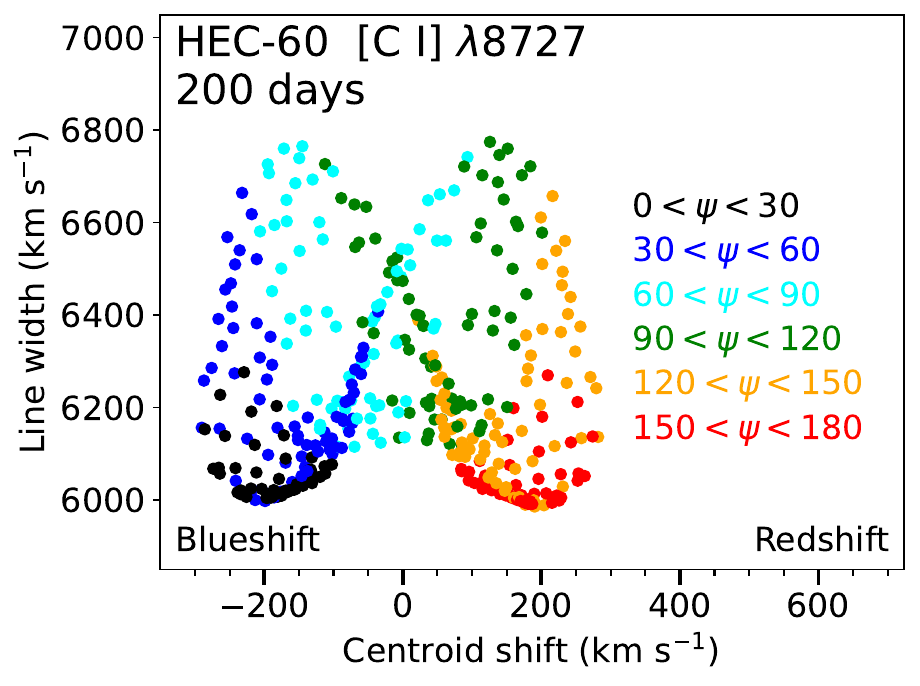}
    \includegraphics[width=.495\textwidth]{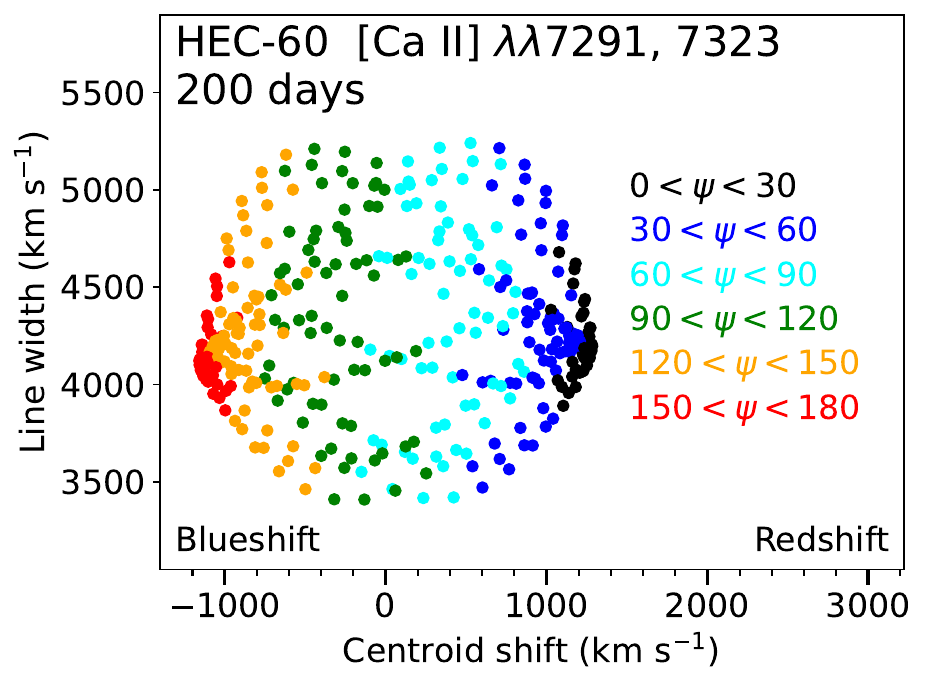}
    \includegraphics[width=\textwidth]{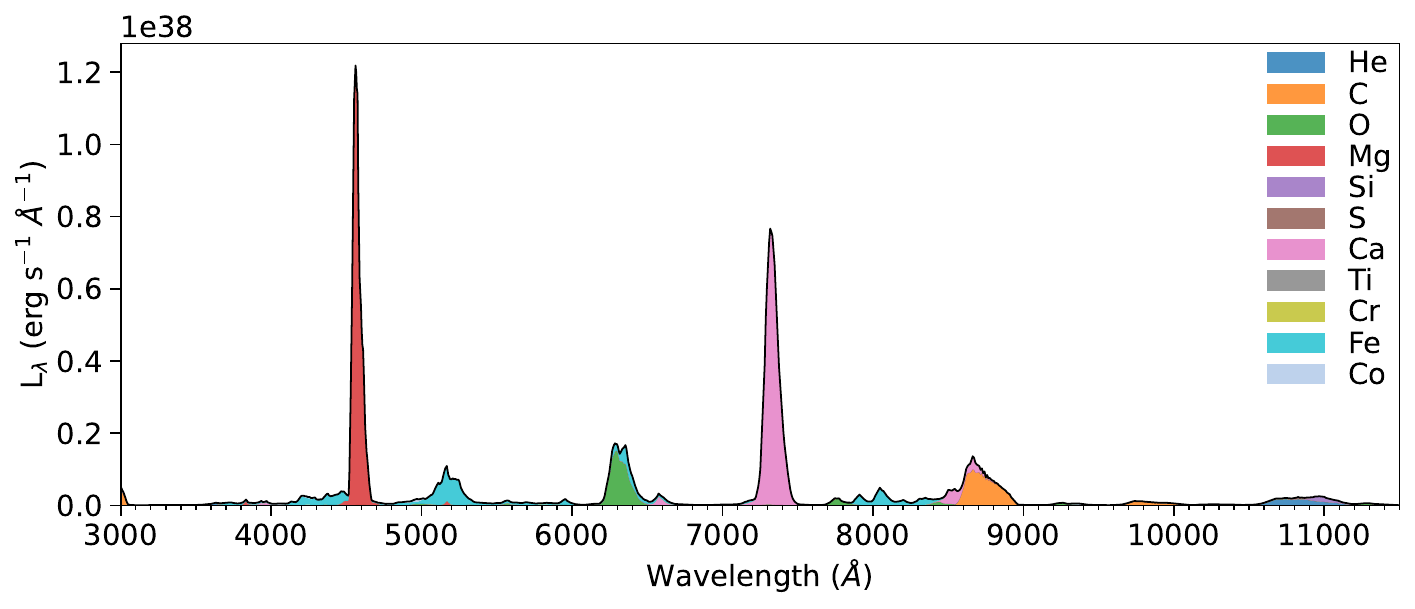}
    \caption{The line properties of the HEC-60 model for the four spectral features Mg I] (top left), [C I] (mid left), [O I] (top right) and [Ca II] (mid right), in the same pattern as in Figure \ref{fig:33L-shiftwidth}, together with the $3000-11500\,\angstrom$ spectrum for the viewing angle most closely aligned to the NS kick vector.}
    \label{fig:60-shiftwidth}
\end{figure*}

In Figure \ref{fig:60-shiftwidth} we display these viewing angle effects for the line properties for the HEC-60 model, which has a similar explosion energy ($\sim1\,$B) as the HEC-33 model used in \citet{vanbaal2023modelling} but originates in a He-core originally almost twice as massive at the onset of He burning ($6.0$ vs $3.3\,M_\odot$) and 60\% more massive at explosion ($4.42$ vs $2.65\,M_\odot$, Table \ref{tab:model_overview}). In the spectrum for this model (shown for the observer that is most directly approached by the neutron star), the four features Mg I], [Ca II], [O I] and [C I] are the four strongest features. 

What can immediately be seen from the distributions of centroid shifts, is that in this model each of the four lines have shifts quite well correlated with viewing angle, and not just [Ca II] as for HEC-33L. Mg I] and [O I] now also show redshifted centroids for for observer locations in the hemisphere that the neutron star moves to (signalling that the bulk of the emitting elements moves away from the observer), and blueshifted centroids when the neutron star moves away from the observer, although they only reach values of $\pm200\,\text{km}\,\text{s}^{-1}$ rather than the $\pm1000\,\text{km}\,\text{s}^{-1}$ values which are found for [Ca II] (similar to the shifts in HEC-33L). [C I] can reach slightly higher centroid shift values than Mg I] and [O I]. Additionally, although [C I] is separated by viewing angle, again observers approached by the neutron star find blueshifted profiles and observers on the other side see redshifted profiles, as with HEC-33L. 

Regarding line widths, for HEC-60 we find that Mg I] is much narrower than [C I] with a range of values of $3900-4500\,\text{km}\,\text{s}^{-1}$, while the carbon line has widths of $6000-6800\,\text{km}\,\text{s}^{-1}$. This makes the [C I] profile the broadest profile in this model, even exceeding [O I] ($5800-6300\,\text{km}\,\text{s}^{-1}$) despite the latter being a doublet feature. [Ca II] again shows the largest range of widths for the different viewing angles, ranging from $3500-5200\,\text{km}\,\text{s}^{-1}$, with the narrowest profiles being a bit narrower than the Mg I] cases and the widest cases being a bit broader. These line widths can be compared to those in HEC-33 which are $6000-7300\,\text{km}\,\text{s}^{-1}$ (Mg I] and [C I]), $7000-8200\,\text{km}\,\text{s}^{-1}$ ([O I]) and $6200-7200\,\text{km}\,\text{s}^{-1}$ ([Ca II]). The lower $E/M_{\rm ejecta}$ ratio in HEC-60 (by a factor 2) is reflected in lower line widths for all the lines.

Between the two models displayed here as well as the model from \citet{vanbaal2023modelling}, it appears that the line profile properties can be quite unpredictable, both in terms of which lines behave similar and different for a given model, and in particular how variation with viewing angle behaves. This can be linked partially to the neutron star kick velocities -- when the kick velocity is low the hemispheric asymmetries of ejected chemical elements will be small and the line properties will be more stochastic. When the kick velocity becomes large, systematics should start appearing. However, these systematics can still be opposite for different elements, as it depends on the role that element plays in the explosive burning. Elements produced by explosive burning should preferentially be ejected in the opposite direction of the neutron star kick vector \citep{wongwathanarat2013three}, and thus observers that see the neutron star recede will detect blueshifted lines for these elements. Conversely, the lighter elements (e.g. $^{12}$C) which are not (significantly) produced in the explosive nucleosynthesis are found further out in the ejecta which is more symmetric. In these regions, secondary effects like mixing and NLTE physics instead become a more dominant component in setting the line centroid shifts, which can cause observers which see the neutron star receding to also see redshifted lines for such elements. 

One thing that stands out though is that [Ca II] tends to have more shifted lines than C, O, and Mg. As Ca is the only one of these elements explosively produced, this raises interesting possibilities to use [Ca II] line profile asymmetries for diagnosis of the explosion engine. Developing such an analysis tool is challenging though -- the [Ca II] emission behaviour arises from some combination of explosive nucleosynthesis asymmetries, hydrodynamic mixing effects, and the fact that calcium has unique NLTE cooling capabilities compared to O, Mg, and C, and can thus become active also in cells in which it has a low abundance. This introduces the potential for a strong "non-linear" mapping between the intrinsic Ca distribution and its emissivity -- we will return to this more in the next section.

Our spectral outputs in the optical range are still quite different from the usual Type Ib observations, as (most) models find very strong Mg I] and [Ca II] emission, with relatively weak [O I] emission. Furthermore, the Ca II NIR triplet emission in our models is quite weak as well compared to the [C I] feature, while observationally generally the triplet feature is the dominant component in this mixed feature. We have however improved \texttt{ExTraSS} with an 'on-the-spot' photoionization effect, which has largely removed the [Ca I] $\lambda6573$ feature and somewhat weakened the Mg I] feature, as both of these elements are sensitive to photoionization.

In our model predictions, we find that the [C I] $\lambda8727$ feature typically dominates over the Ca II NIR triplet, yet in most observations this effect is much smaller. Instead most -- if not all -- observations have the Ca II NIR triplet as the dominant component in this blend.

\subsection{Observational comparison} \label{sec:obscomp}
With quantified predictions for line shifts and widths shown for a total of nine hydrodynamic models \citep[two of which are analyzed in some depth above, and one in][]{vanbaal2023modelling} it now becomes of interest to compare to a set of observed Type Ib SNe. This defines the first spectroscopic test of modern 3D CCSN explosion models against observations, and holds potential for putting interesting constraints on the link between SESNe and their progenitor stars and explosion energies.  The full list of observed spectra used in this comparison is given in Table \ref{tab:obsSN_data}. For each spectrum, we carry out the integrations defining line shifts and line widths in the same manner as on the model spectra. Although we correct spectra for extinction and redshift, we are not sensitive to extinction as we do not compare line strengths but only line profile shapes. We are instead sensitive to any background contamination (both line width and shift) and redshift uncertainty (line shift). A more detailed description of our fitting technique for the observational calculations is given in Appendix \ref{app:observationdata}.

For each of our models, we turn the set of 400 viewing angles into a contour outline containing all of them. We define two model contours for each line, one using the contributions by the specific element only (solid) and one using the full spectrum (including also any contamination by nearby lines, dashed). While the latter may seem most suitable for comparison to observations which only tell us the total spectrum, it becomes problematic in cases where our model has poor estimates of the relative line strengths in the spectral region. For example, the lack of global radiative transfer in the model makes predictions for the Ca II triplet, which blends with [C I] 8727, unreliable as it is sensitive both to fluorescence following HK-line absorption as well as line-to-line scattering within the triplet itself. 

We split the analysis into three groups, separated by progenitor star, for better clarity between the models, but for each group we consider the full observational dataset.

\subsubsection{HEC-33 models}
\begin{figure*}
    \includegraphics[width=.495\textwidth]{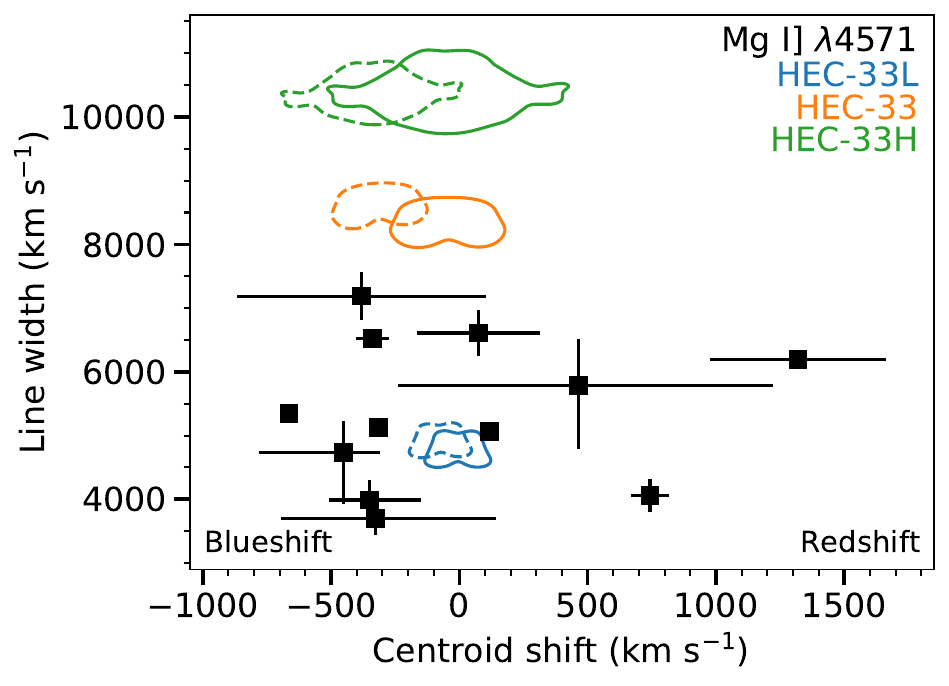}
    \includegraphics[width=.495\textwidth]{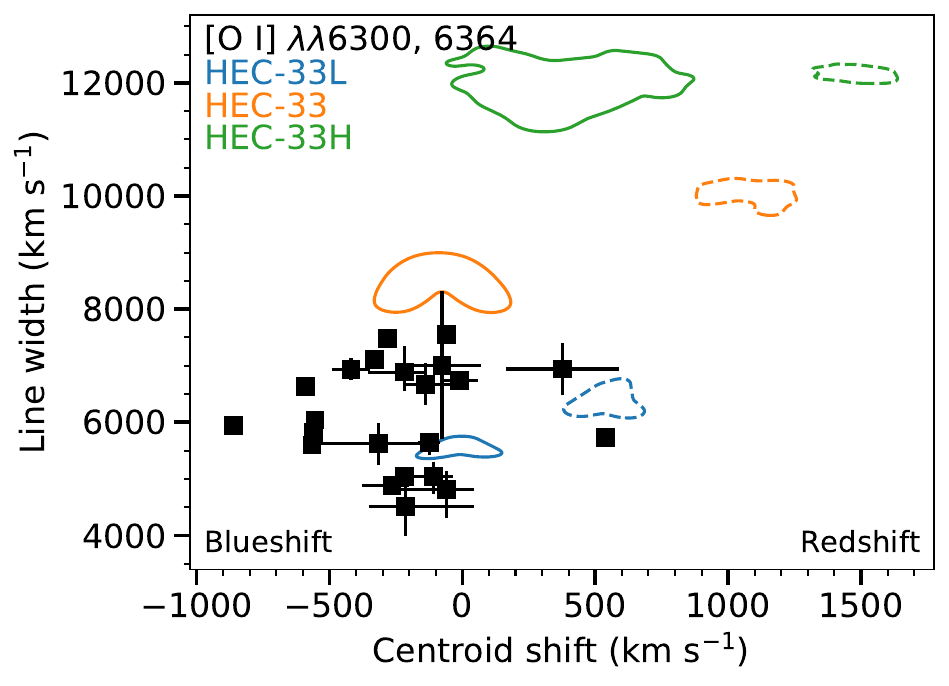}
    \includegraphics[width=.495\textwidth]{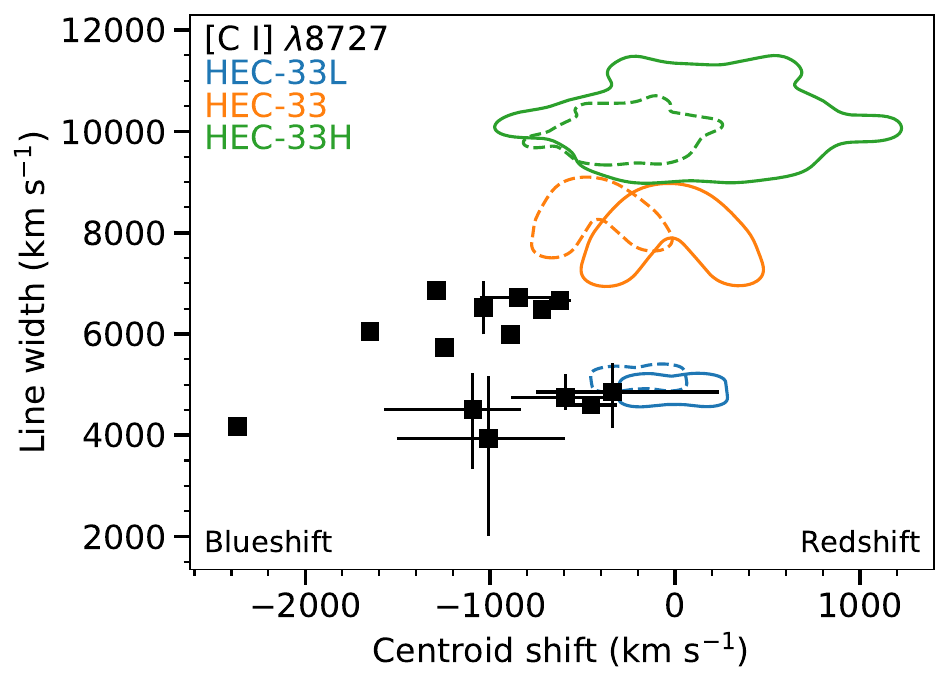}
    \includegraphics[width=.495\textwidth]{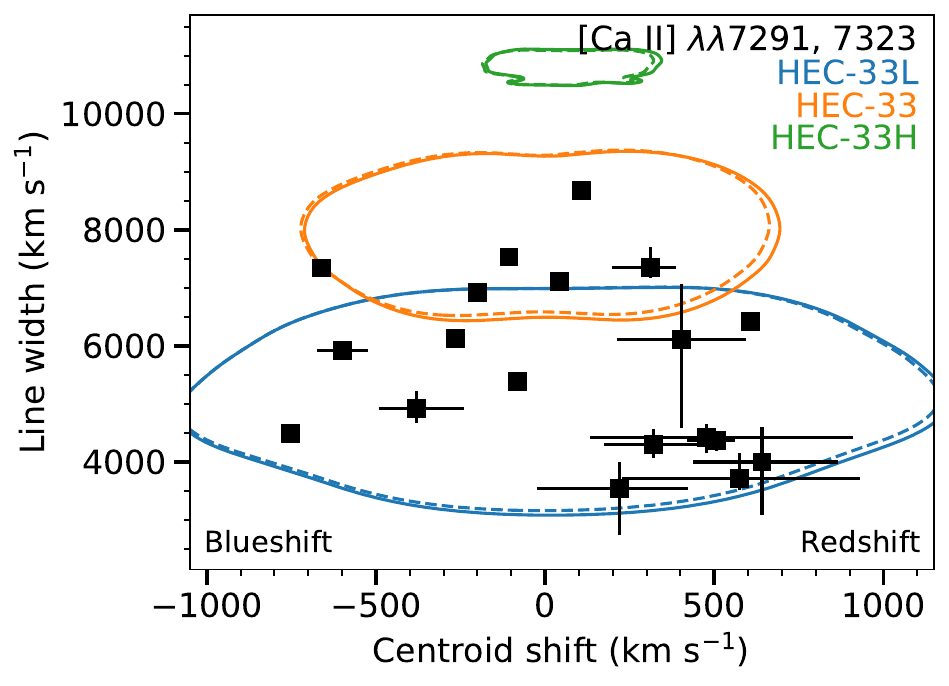}
    \caption{The line centroid shifts and line widths for the HEC-33 models, shown as contours, compared to the observational SNe (see Appendix \ref{app:observationdata} for a list of all included SNe). On the left the singlet features Mg I] $\lambda4571$ (top) and [C I] $\lambda8727$ (bottom) are shown, and on the right the doublet features [O I] $\lambda\lambda6300,\,6364$ (top) and [Ca II] $\lambda\lambda7291,\,7323$ (bottom). Observations (black squares) from the same SN are grouped together; the error bars indicate variation between different observations from the same SN. Different colours denote the three explosion energies. The solid contours show the line properties calculated from just that element, while the dashed contours are calculated using all emission.}
    \label{fig:HEC-33_contour}
\end{figure*}
In Figure \ref{fig:HEC-33_contour} the line shifts and widths for the HEC-33 models are displayed as contour outlines, split for the four strongest features (Mg I], [C I] for the singlets and [O I], [Ca II] for the doublets).  In each panel a comparison to observed SN is shown (see Appendix \ref{app:observationdata} for the full list as well as more details on the observational calculations) -- if one SN has multiple epochs, the bars denote the range between these with the square denoting the mean values. As denoted inside each panel, a positive (negative) centroid shift corresponds to a redshifted (blueshifted) centroid.

For the Mg I] line it can be seen that our models predict quite small centroid shifts of up to $\sim500\,\text{km}\,\text{s}^{-1}$, and when considering the full spectrum (dashed lines) our models tend to give a small blueshift due to contamination from Fe. Most of the observed SNe do also have blueshifted centroids, with a similar centroid shift, although there are a few which do yield a (large) redshifted Mg I] feature. For the observed SNe, the line widths display quite a range of values, from $\sim3500\,\text{km}\,\text{s}^{-1}$ to $\sim7800\,\text{km}\,\text{s}^{-1}$; only the low energy model of HEC-33L finds line widths in this range, with HEC-33 being somewhat over ($\sim8000\,\text{km}\,\text{s}^{-1}$) and HEC-33H being much higher still. 

For the [C I] panel it can be seen that for all observations we find large centroid shifts towards the blue side, which can largely be explained by the presence of the Ca II NIR triplet which is observationally generally the stronger component in this blend. Our models do show shifts towards the blue side as well when considering the full emission component rather than just the [C I] feature, but the Ca II feature is not very strong in these three models and thus we find lower centroid shifts than in most observations. For the line widths, the observations are roughly the same range of values as for Mg I] ($\sim3500\,\text{km}\,\text{s}^{-1}$ to $\sim7800\,\text{km}\,\text{s}^{-1}$), which can be matched by the HEC-33L model while the HEC-33 model is slightly wider and the HEC-33H model predicts significantly wider profiles.

\begin{figure*}
    \includegraphics[width=.495\textwidth]{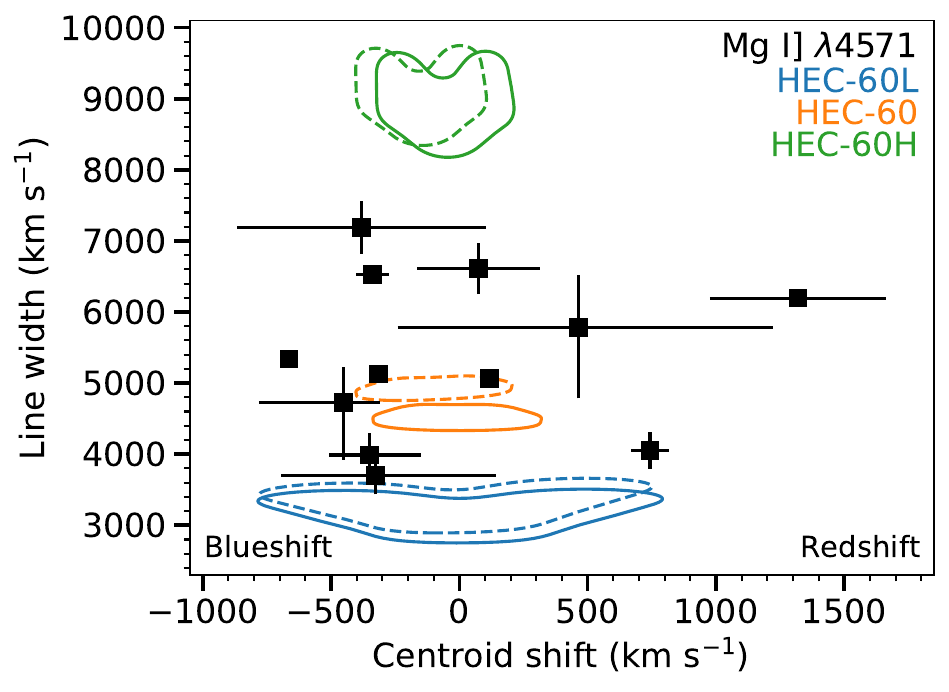}
    \includegraphics[width=.495\textwidth]{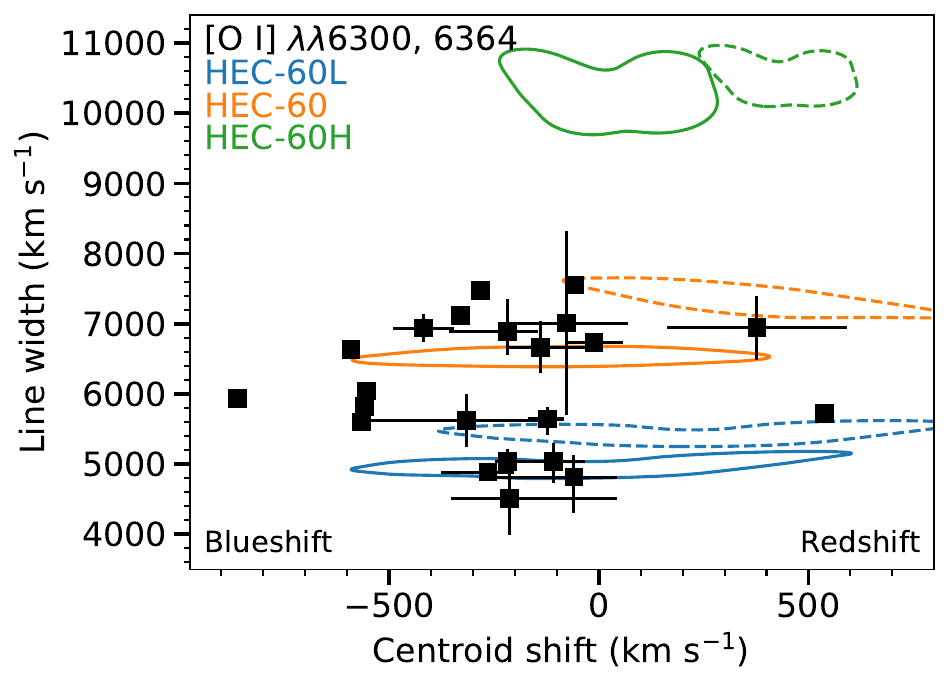}
    \includegraphics[width=.495\textwidth]{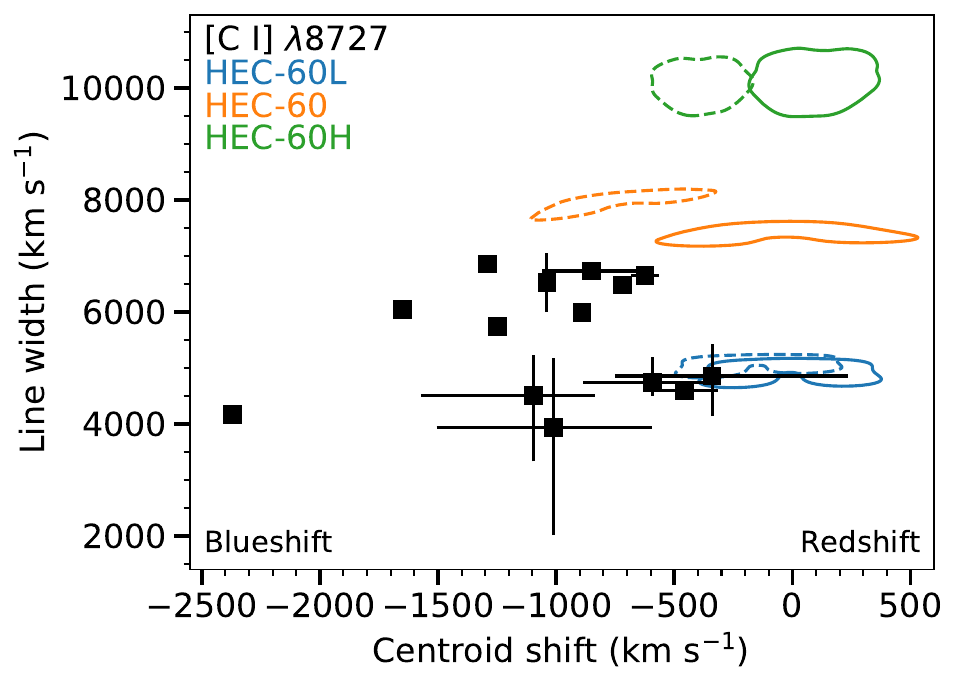}
    \includegraphics[width=.495\textwidth]{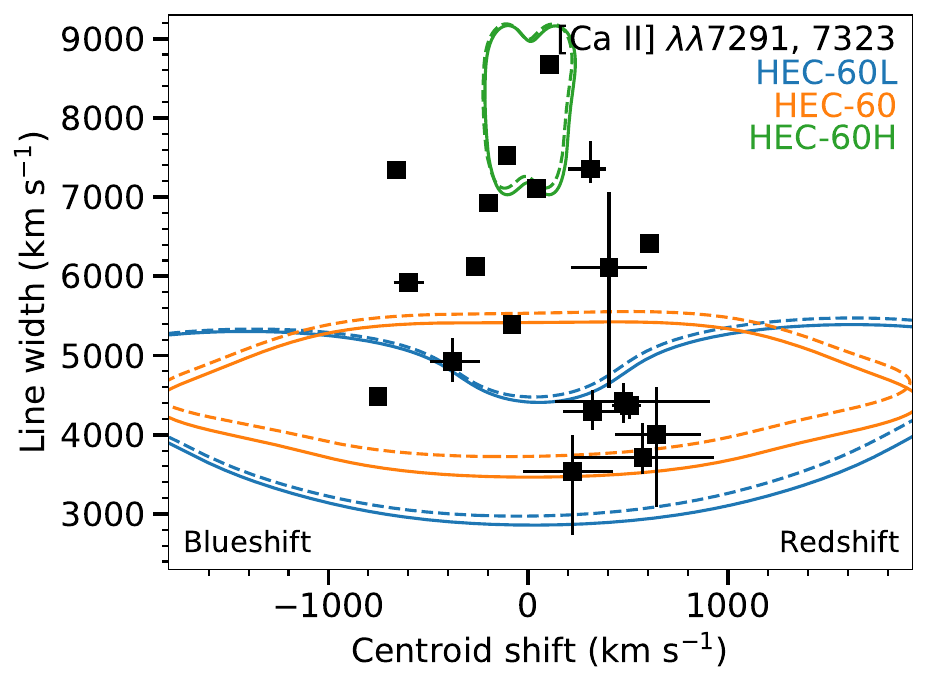}
    \caption{The line centroid shifts and line widths for the HEC-60 models, shown as contours, compared to the observational SNe (see Appendix \ref{app:observationdata} for a list of all included SNe). On the left the singlet features Mg I] $\lambda4571$ (top) and [C I] $\lambda8727$ (bottom) are shown, and on the right the doublet features [O I] $\lambda\lambda6300,\,6364$ (top) and [Ca II] $\lambda\lambda7291,\,7323$ (bottom).  Observations (black squares) from the same SN are grouped together; the error bars indicate variation between different observations from the same SN. Different colours denote the three explosion energies. The solid contours show the line properties calculated from just that element, while the dashed contours are calculated using all emission.}
    \label{fig:HEC-60_contour}
\end{figure*}

Similar trends are present for [O I] and [Ca II], where the HEC-33L model can match the observational widths and HEC-33 is slightly too high (for [O I]) or can also match some of the objects (for [Ca II]). For the centroid shifts in the models, the two doublets display two different behaviours; the [Ca II] feature is barely impacted by contaminating features due to its strength, while in our HEC-33 models the [O I] feature is quite weak and even small contaminations from Fe II and Ca I drag the calculated model shifts far to the red side. In observations, the [O I] and [Ca II] features are both (usually) strong and thus such contaminating effects do not occur -- this means comparison to our solid contours are most informative. The data shows centroid shifts of up to $\sim1000\,\text{km}\,\text{s}^{-1}$, with the [O I] distribution somewhat weighted towards preferential blueshifts, while the [Ca II] distribution is largely symmetric around zero. 

Lastly, between the different panels it can be noted that for Mg I], [C I] and [O I], the centroid shift and line width distribution spreads increase at higher explosion energies, while for [Ca II] the opposite trend is seen. This effect does get reduced a bit when considering the full emission and not just the element-specific profiles. Interestingly, from the observational data it appears that all features are capable of achieving quite high centroid shifts and there is no element which is always seen at very minor shifts, although for [C I] the contamination from the Ca II NIR triplet makes it hard to compare as that feature is quite weak in our HEC-33 models.

\subsubsection{HEC-60 models}
In Figure \ref{fig:HEC-60_contour} the line shifts and widths for Mg I], [C I] (left column), [O I] and [Ca II] (right column) are shown for the HEC-60 models, together with the observational data. As before, a positive (negative) centroid shift corresponds to a redshifted (blueshifted) centroid. 

For Mg I] we find that the medium (orange) energy model has quite a narrow range of widths compared to the low (blue) and especially high (green) energy models. The widths predicted by the high energy model are somewhat too high for what the observations show, while both the low and medium energy models fall in the low-to-intermediate end of the observed width distribution. For the centroid shifts, the low energy model can match almost all of the observed values, while the medium and high energy models show a more narrow range of centroid shifts and can only reach up to $\sim400\,\text{km}\,\text{s}^{-1}$. When accounting for the full spectrum (dashed lines) it can be seen that the line widths for the low and medium energy models go up by $\sim500\,\text{km}\,\text{s}^{-1}$, bringing them in closer agreement with the observed values.

For [C I], as with the HEC-33 models we again see that when considering the carbon emission only (solid lines) predictions do not match the large centroid shifts observed. When accounting for the total emission, this picture shifts somewhat, with the models trending a bit towards the observed centroid shifts, but they are only capable of matching the least blueshifted observations. The spread of centroid shifts values predicted due to viewing angle variation is $\sim1200\,\text{km}\,\text{s}^{-1}$, which roughly matches with the range found in the observations (excluding SN 2006gi at $-2400\,\text{km}\,\text{s}^{-1}$ at the blueshifted end). 
In terms of widths, the high energy model results in larger widths than the observations, the medium energy model is on the high side, while the low energy model matches well within the observed distribution.

For [O I] it can be seen that the difference between the full spectrum and the pure [O I] contours is smaller than it was for the HEC-33 models, due to the larger strength of the [O I] feature in the HEC-60 models. For the line widths, it can be seen that the high energy model gives much higher values than observed, while the other two models are broadly compatible with the observations. A similar picture appears for the centroid shifts, as the high energy model has the smallest centroid shifts while the low energy model has the largest shifts and can match almost all of the observed values. 

The [Ca II] centroid shifts in the models follow a similar picture with explosion energy as the other features for the HEC-60 models, with the high energy model predicting small shifts and the low energy model the largest shifts, even upwards of $\sim1500\,\text{km}\,\text{s}^{-1}$. The low and medium energy models predict larger centroid shifts than the observations show. Interestingly, for the line widths, the high energy model is not completely out of the range found in observations like it is for the other three features, while the low and medium energy models are again compatible with the lower end of the observed values.

Between the different $6.0\,M_\odot$ models it can be noticed that higher explosion energies lead to much higher line widths but also narrower centroid shift distributions for every feature (unlike the $3.3\,M_\odot$ models, where only [Ca II] displayed this trend). However, unlike the $3.3\,M_\odot$ models, the [Ca II] feature is here the most narrow feature, or shared most narrow together with Mg I]. Especially the low energy model also finds large differences in the centroid shift for all features with viewing angles and is capable of broadly matching the observed values, or even exceeding them.

\subsubsection{HEC-110 models}
\begin{figure*}
    \includegraphics[width=.495\textwidth]{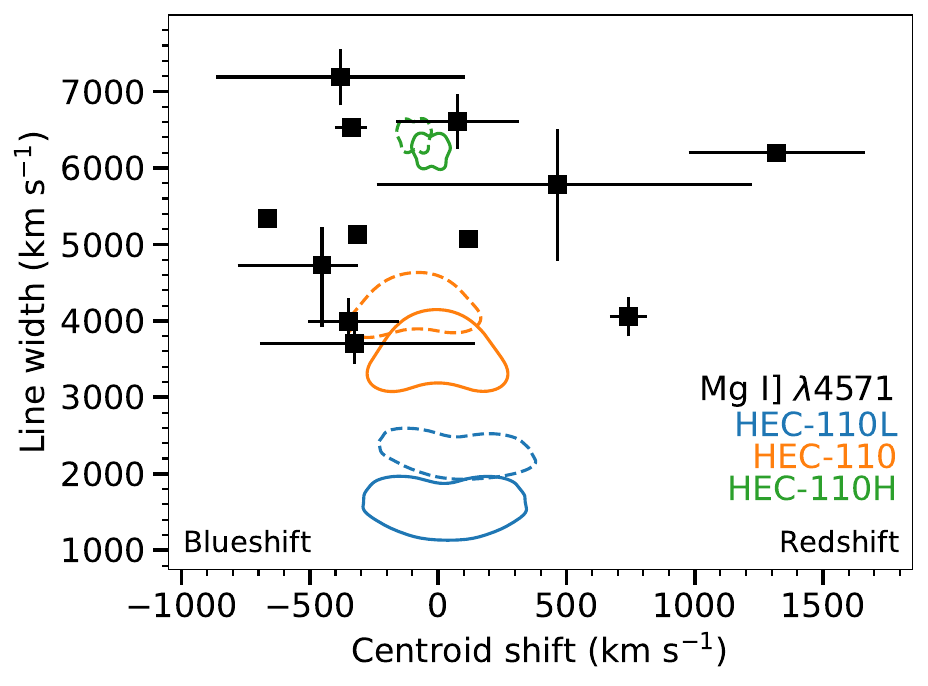}
    \includegraphics[width=.495\textwidth]{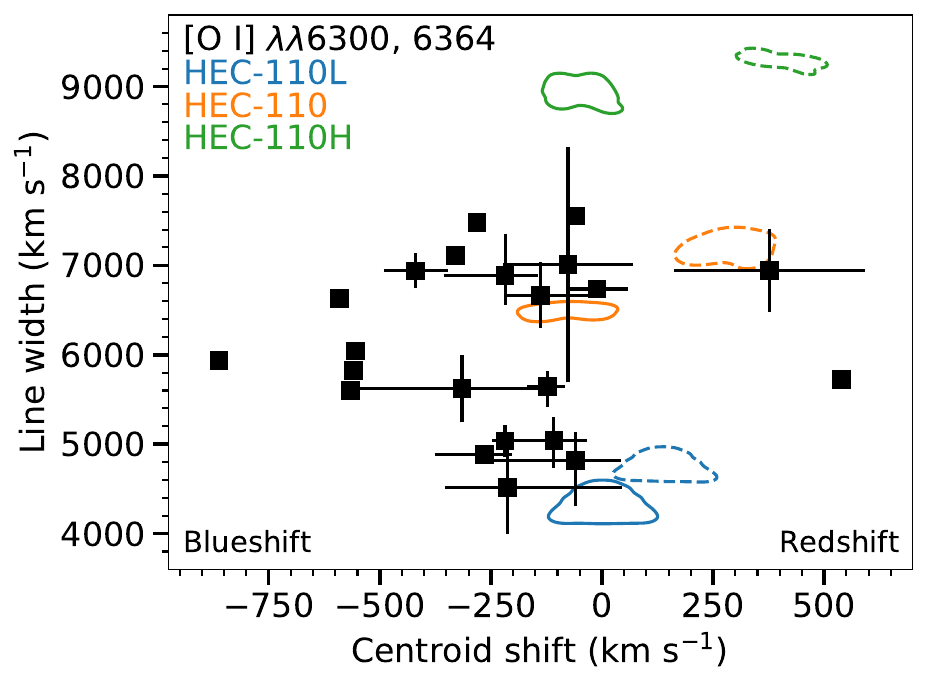}
    \includegraphics[width=.495\textwidth]{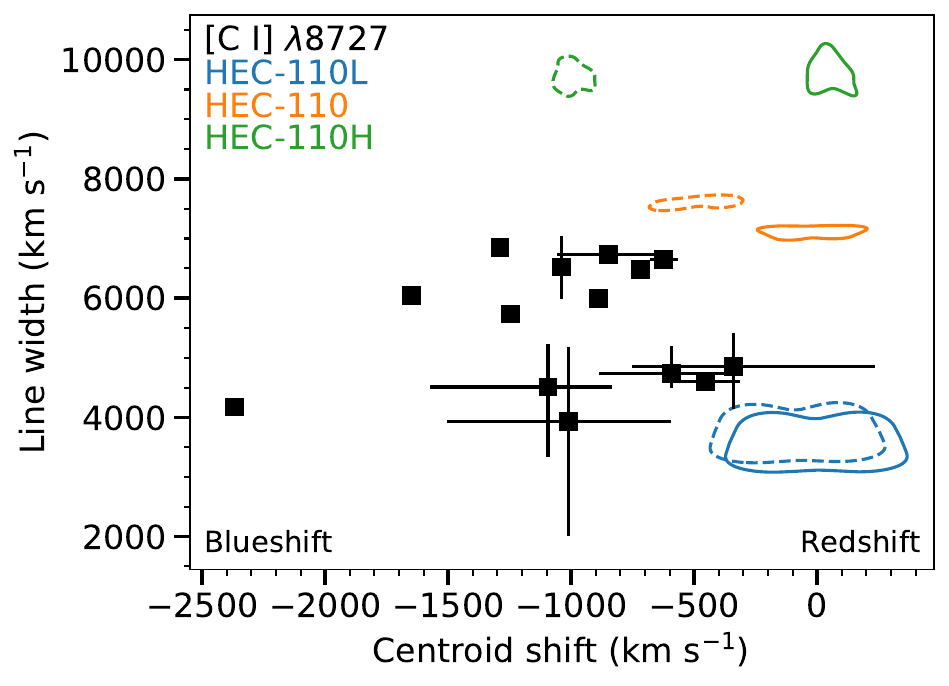}
    \includegraphics[width=.495\textwidth]{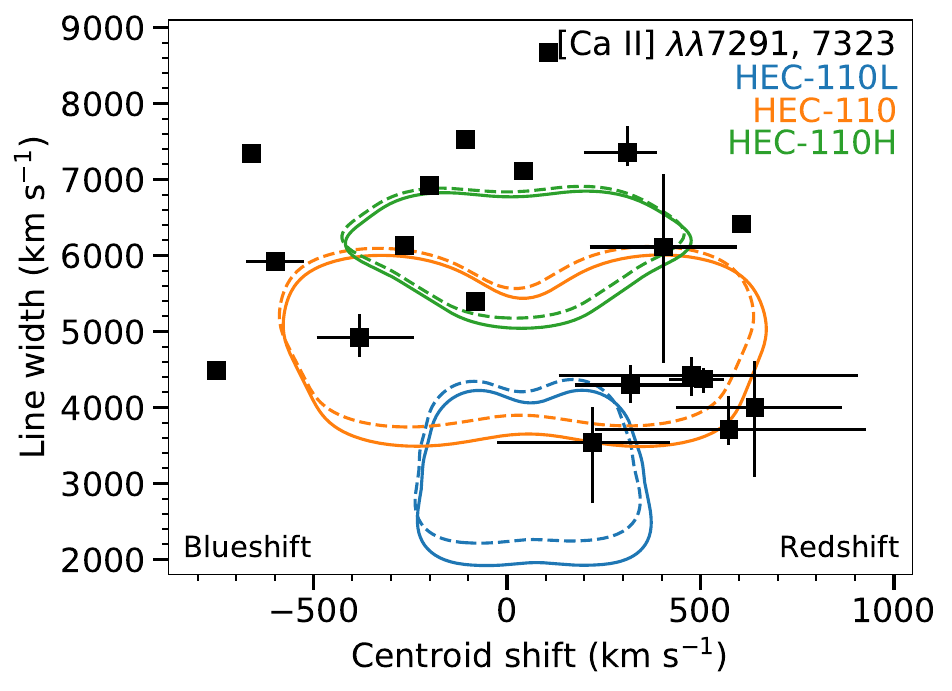}
    \caption{The line centroid shifts and line widths for the $11.0\,M_\odot$ models, shown as contours, compared to the observational SNe (see Appendix \ref{app:observationdata} for a list of all included SNe). On the left the singlet features Mg I] $\lambda4571$ (top) and [C I] $\lambda8727$ (bottom) are shown, and on the right the doublet features [O I] $\lambda\lambda6300,\,6364$ (top) and [Ca II] $\lambda\lambda7291,\,7323$ (bottom).  Observations (black squares) from the same SN are grouped together; the error bars indicate variation between different observations from the same SN. Different colours denote the three explosion energies. The solid contours show the line properties calculated from just that element, while the dashed contours are calculated using all emission.}
    \label{fig:HEC-11_contour}
\end{figure*}
The line shifts and widths for Mg I], [C I] (left column), [O I] and [Ca II] (right column) are shown in Figure \ref{fig:HEC-11_contour} for the $11.0\,M_\odot$ models, together with the observational data. It can be noted that compared to Figures \ref{fig:HEC-33_contour} and \ref{fig:HEC-60_contour}, the model values for $v_\text{shift}$ are here smaller across the board, as are the maxima of $v_\text{width}$.

For Mg I] in particular, the low energy model cannot match the observed widths when considering only the Mg I] emission, although when considering the full spectrum this model does end up among the observed values. The pure Mg I] feature is, however, exceptionally narrow, reaching not even half of the narrowest observed line widths. The medium energy model barely makes it into the observational range, while the high energy model has a very small contour but ends up towards the higher end of the observed values. Neither of these two models are impacted much when considering the full spectrum instead. The centroid shifts for these models are generally quite low, up to maybe $\sim300\,\text{km}\,\text{s}^{-1}$, putting some of the observed shifts out of their reach.

For the [C I] feature, considering only the [C I] emission causes the HEC-110 models to be unable to match the observational shifts, due to the presence of the Ca II NIR triplet. When considering the full spectrum, the low energy model again ends up among some of the observed values, while this time the medium energy model gets the centroid shifts in the right ballpark but the widths are slightly higher than observed, but not by much. For the high energy model, the width of this feature is much too high, although the impact of the Ca II NIR feature does put the centroid shift into a good regime.

In the [O I] panel, the contamination from Fe and Ca causes more of a shift for the full spectrum for the medium and high energies than for the low energy model. For the line widths, it can be seen that the low energy model just about matches the narrowest observed profiles, while the medium energy model sits quite nicely in the middle of the observed range. The high energy model still results in too wide profiles, as it also did for the HEC-33 and HEC-60 models. None of our three models display significant centroid shifts (excluding the shifts due to contaminating lines), leaving the HEC-110 models unable to reach the range of observed widths. 

Lastly, for [Ca II] we see very little difference between the element-only profiles and the full spectrum ones. For this feature, it is the low energy model which struggles to reach the observed line widths, as some of the viewing angles for this model result in narrower lines than observed ($v_\text{width}\lesssim 3500\,\text{km}\,\text{s}\,^{-1}$), while the medium energy model is capable of matching the lower range of observations. The high energy model also falls within the observed range, although it doesn't reach the higher end of the observations, which none of the HEC-110 models do. The high and in particular medium energy model also can reach centroid shifts close to the observed range, while HEC-110L can only achieve about half of the observed range.

For the different HEC-110 profiles, it should be noted that apart from [Ca II], and to a lesser extent Mg I], the impact of the viewing angle on the line properties is quite small, as there is considerably less variation of $v_\text{shift}$ and $v_\text{width}$ than for the HEC-33 and HEC-60 model sets. Even for [Ca II], the range of values within the HEC-110 models is not as large as for their lighter counterparts.

\subsubsection{Summary of models} \label{sssec:modelsummary} 
Between all the different models,  only a few of them are capable of matching some parts of the observed properties of Type Ib SNe. For example, the widest observed [Ca II] profiles are only (partially) matched by the low and medium energy $3.3\,M_\odot$ models, as well as the high energy $6.0\,M_\odot$ one. The Mg I] feature, which is quite strong in our models but can be tricky to identify in observations, can be partially matched by the $3.3\,M_\odot$ low energy model, the $6.0\,M_\odot$ medium and maybe low energy models, as well as the $11.0\,M_\odot$ high and medium energy models.

Taking the mean and median values of the (absolute) centroid shifts in the observations allows for another probe into the models. For Mg I], the observations have a mean value of $460\,\text{km}\,\text{s}^{-1}$ (median $370\,\text{km}\,\text{s}^{-1}$), which only the HEC-60L model can comfortably reach. For [O I], we have a mean centroid shift of $310\,\text{km}\,\text{s}^{-1}$ (median $270\,\text{km}\,\text{s}^{-1}$) which our low and intermediate mass models can reach. For [Ca II], the observations have a mean centroid shift of $380\,\text{km}\,\text{s}^{-1}$ (median $380\,\text{km}\,\text{s}^{-1}$), which can be achieved by six models: HEC-33L, HEC-33, HEC-60L, HEC-60, HEC-110 and HEC-110H. For [C I] we run into the issue of the blending with the Ca NIR triplet; in the observations this feature has a mean centroid shift of $1010\,\text{km}\,\text{s}^{-1}$ (median $950\,\text{km}\,\text{s}^{-1}$), but none of our models have a strong enough impact from the Ca NIR feature to achieve this kind of centroid shift.

One interesting pattern which appears for the HEC-33 and HEC-60 models is that our [Ca II] centroid shifts display much larger variations with viewing angle for the lowest explosion energies, while the high explosion energies display very little centroid shift variation. In Table \ref{tab:Ca-velocity-onecolumn} we list the velocity vectors of all Ca, just the emitting [Ca II] component, the angle $\beta$ between those two, the total ejecta vector and the $\gamma$-ray deposition vector. The vector for the emitting [Ca II] is generally somewhat smaller than the overall Ca vector, but they are both significantly bigger than the overall ejecta vectors. This indicates that Ca is ejected more asymmetrically than the overall ejecta, and also that this holds true also for the emitting component. The models where the [Ca II] (and Ca) vectors are biggest are also the models which show the most variation in their centroid shifts, showing that our centroid shift calculations picks up on these asymmetries. The origin of these asymmetries largely lies with the explosion-related hydrodynamics (see Table \ref{tab:Ca-velocity-onecolumn}), rather than with NLTE-related effects \citep{wongwathanarat2013three}.

\begin{table}
    \centering
    \caption{An overview of the magnitudes of the velocity vectors $\vec{v}$ for the overall calcium distribution, the emitting [Ca II] $\lambda\lambda7291,\,7323$, the angle $\beta$ between these two, and the magnitude of the vectors $\vec{v}$ for the total ejecta and for the $\gamma$-ray deposition from our models. All velocity values are given in km$\,$s$^{-1}$.} 
    \setlength\tabcolsep{8pt}
    \begin{tabular}{l ccc cc}
    \hline
    Model  & Ca & [Ca II] & $\beta$ & $\Sigma M_\text{ejecta}$ & $\gamma-\text{ray}$ \\
    \hline
    HEC-33L  & 1574 & 1075 & $3.5^{\circ}$ & 251 & 604  \\
    HEC-33   &  769 &  630 & $24^{\circ}$ & 136 & 415  \\
    HEC-33H  &   36 &   83 & $41^{\circ}$ &  11 &  69  \\
    HEC-60L  & 1744 & 1597 & $6.0^{\circ}$ & 255 & 923  \\
    HEC-60   & 1528 & 1218 & $0.24^{\circ}$ & 265 & 694  \\
    HEC-60H  &  244 &  166 & $6.6^{\circ}$ &  39 & 108  \\
    HEC-110L &  214 &  231 & $44^{\circ}$ &  95 & 152  \\
    HEC-110  &  646 &  431 & $6.9^{\circ}$ & 106 & 228  \\
    HEC-110H &  278 &  333 & $18^{\circ}$ &  71 & 199  \\
    \hline
    \end{tabular}
    \label{tab:Ca-velocity-onecolumn}
\end{table}

This could potentially be an interesting probe by its connection to the explosive burning region, and how for our models it seems that the models which produce less $^{56}$Ni are more asymmetric (except for HEC-110L). With our parametrized neutrino-driven CCSN models, there is a tendency for low explosion energies to develop larger asymmetries -- HEC-110L loses a large part of the asymmetric inner ejecta due to fallback and thus breaks this trend. As shown by \citet{wongwathanarat2013three}, in particular the explosively produced elements (e.g. $^{40}$Ca) get ejected much more asymmetrically in these cases. Self-consistent models \citep[see e.g.][]{burrows2020overarching,bollig2021selfconsistent} tend to find more asymmetric explosions as well if the explosions take longer to be launched.

The angle $\beta$ between the total Ca vector and the [Ca II] vector is quite variable between the models. For HEC-33 we had previously shown in \citet[Fig 14]{vanbaal2023modelling} that there is an offset between these two and speculated that this was due to the relatively low NS kick vector combined with NLTE effects. Both HEC-33H and HEC-110L have an even lower NS kick vector (see Table \ref{tab:model_overview}) and both find a larger $\beta$, which would indeed indicate that this is -- at least partially -- caused by the weak NS kick. For HEC-110H it is less clear where the relatively large $\beta$ originates from, although it could be because this model is more spherical (low $\vec{v}\ \Sigma\,M_\text{ejecta}$, see Table \ref{tab:Ca-velocity-onecolumn}) yet has a lot of $^{56}$Ni which leads to a strong NLTE effect -- it is also one of the few models where $\vec{v}$ is larger for [Ca II] than for pure Ca. 

There are some patterns between the different models which emerge, such as the consistently (much) too wide [O I] and [C I] + Ca II NIR features for the high energy models. The [O I] in particular might be because the feature tends to be quite weak in our models as instead [Ca II] is the dominant cooler. Having a weaker central component will cause the wings to appear as significant contributions to the overall profile and thus strongly influence the width of the features. For the blended [C I] + Ca II NIR feature, it could be that usually in observations the Ca triplet is a stronger emission component, which should be present more in the inner regions, while in our models we are dominated by carbon emission which should be found in the outer regions, and thus have a wider profile. We performed the [C I] calculations by pretending that the line centroid should be at $8727\,\angstrom$, and we clearly see that our models cannot reach the blueshifts which the observations find, which indicates that our models are either overestimating [C I] or underestimating the Ca II triplet, or both.  

Reviewing all nine models against the Type Ib SNe considered here, it becomes clear that several models are unable to match line properties for the four main features in the nebular spectra -- thus appearing to be ruled out. The most obvious discrepancies are for HEC-33H and HEC-60H which consistently predict too broad lines. From this one would infer that He-cores of mass $\lesssim 6\,M_\odot$ do not explode with energies up towards 3 B but something significantly lower, $\lesssim1\,$B. It is by these first multi-D models where the $^{56}$Ni mixing is self-consistently computed and represented that allows this conclusion to be drawn. HEC-110H gives a bit more complex picture, with two features being much broader than observed, one being at the high-end and one being in the mid-range -- but also here the indication is that an energy of 3 B seems too high. 

The HEC-33 model is generally somewhat too broad, which would indicate that low-mass He-cores are not able to explode with 1 B, but generally something lower. This is in line with the self-consistent explosion simulations of low-mass cores by \citet{muller2018multidimensional,muller2019three} and \citet{ertl2020explosion}. The discrepancy to observations is here, however, lower, so energies just a bit below 1 B cannot be ruled out. On the other hand, HEC-110L is generally somewhat too narrow, indicating that massive He-cores (HEC-110 has a pre-SN mass of $7.0\,M_\odot$) are not generally exploding with such low energies, $\lesssim\,0.5\,$B.

The remaining four models (HEC-33L, HEC-60L, HEC-60 and HEC-110) give line widths broadly consistent with the observations, lending support to the picture that He-cores on the low-mass side ($3.3-6.0\,M_\odot$) explode with $0.5-1\,$B, and more massive ones ($6.0-11.0$) with $\sim 1\,$B. At least some of them also give centroid shifts in agreement with data $-$ although in particular for [Ca II] they tend to predict somewhat too large centroid shifts. 

Between the nine models, quite clearly there is not one model which is capable of matching the observed set, and some combination of He-stars and explosion energies around the models we have used here are needed to explain the full diagram of observations. However, several of our models here appear incapable of matching the Type Ib SNe properties, indicating that some of these models are not truly realistic representations of such SNe. 

\begin{figure*}
    \includegraphics[width=.99\textwidth]{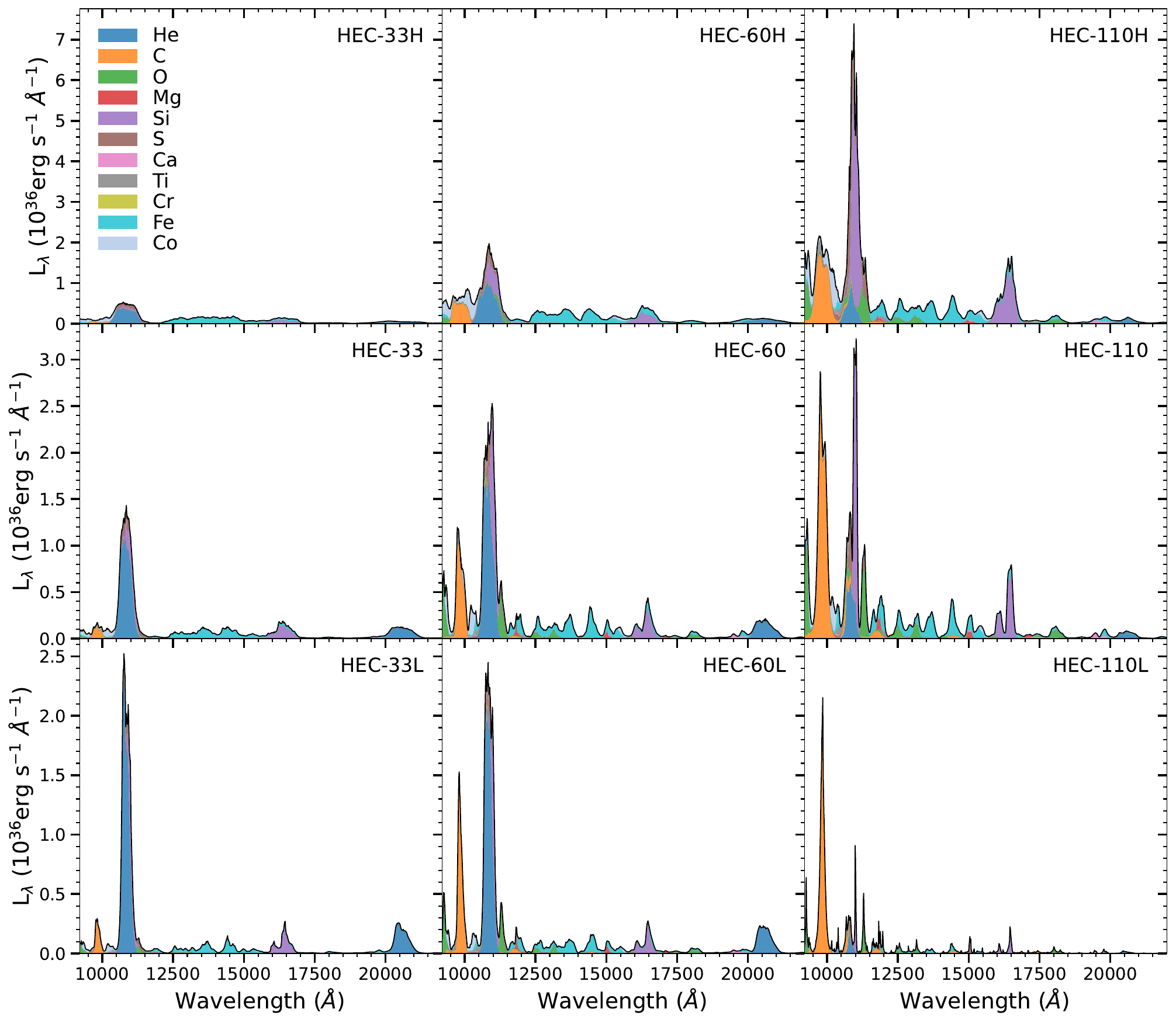}
    \caption{The NIR spectra ($9200-22500\,\angstrom$) for each of our models, arranged in the same order as in Figure \ref{fig:3D-Rendering} with each panel also labelled in the top right corner. The top-left panel denotes the colour scheme for the different elements. Note that every row has its own y-axis scale. The relevant He lines are the He I 1.08$\,\mu$ and 2.06$\,\mu$ lines.}
    \label{fig:HeNIR3x3}
\end{figure*}
\subsection{He lines in NIR}
Sometimes the classification of SNe into either Type Ib or Ic can be ambiguous \citep[e.g.][]{piro2014transparent,folatelli2014supernova,fremling2018oxygen}, in particular in the photospheric phase as the conditions for the appearance of He lines can be tricky to predict \citep[see also e.g.][]{sauer2006properties,dessart2012nature,williamson2021modeling}. One potential solution to this classification problem could perhaps be solved by moving past the photospheric phase into the nebular phase, and consider the He I lines formed at $10833\,\angstrom$ and $20587\,\angstrom$. In the photospheric phase, the study of \citet{shahbandeh2022carnegie} showed that a feature around $1\,\mu$m shows up for all SESNe although its origins are unclear, as several elements can blend into the He I feature. Instead the $2.06\,\mu$m feature is the ideal candidate to determine the presence of He.

With the nine different models at our disposal we can use \texttt{ExTraSS} to make predictions for the spectral output in the NIR range to investigate if there is a possibility of using nebular phase spectra to confirm or reject the presence of He in the nebula. The critical physics that govern the formation of He lines are the 3D morphology and the non-thermal ionization and excitation of He -- which we here treat for the first time. At the same time, the impact of radiative transfer is expected to be quite small in the NIR, so \texttt{ExTraSS} in its current capacity is a suitable tool to explore this question.

In Figure \ref{fig:HeNIR3x3} we show the NIR spectra for each model (in the same configuration as in Figure \ref{fig:3D-Rendering}, with each model denoted in the top right corner) for the observer for whom the neutron star is most directly approaching (i.e. $\Psi=0$). It should be noted that each row has its own y-axis. The [C I] $\lambda\lambda9824,\,9850$ feature, which is particularly noticeable for the more massive and lower energy models, is sometimes also seen in late-time observations of SESNe \citep[e.g.][]{jerkstrand2015late}, although not usually this strong.

Regarding the He lines, it can be noted that in none of the high energy models (top row) the $2.06\,\mu$m line is detectable, and it also appears to be absent in the HEC-110L model (bottom right). For the HEC-110 model (middle right) this feature is very weak, while in the remaining four models it is perhaps an order of magnitude weaker than the $1.08\,\mu$m line. This is somewhat unfortunate, as the $1.08\,\mu$m line is contaminated by Si and S emission at similar wavelengths. For the HEC-33 and HEC-60 model suites (left and middle columns in Figure \ref{fig:HeNIR3x3}) the He line is the dominant component in this emission feature at $\sim11000\,\angstrom$, but for the HEC-110 models (right column) this is not the case and the feature at this wavelength is instead dominated by Si (and to a lesser extent, S).

For the HEC-33 models, the $1.08\,\mu$m line has some variation in its peak strength with explosion energy, with the lower energy resulting in a line peaking at $\sim2.5\,\times 10^{36}\,\text{erg}\,\text{s}^{-1}\,\angstrom^{-1}$, while the HEC-33H model reaches a peak luminosity that is roughly five times lower. Interestingly, for the HEC-60 models such an effect does not appear; each of the three HEC-60 models peaks at roughly the same luminosity ($\sim2-2.5\,\times 10^{36}\,\text{erg}\,\text{s}^{-1}\,\angstrom^{-1}$). For the HEC-60 models, however, we see a bit more Fe emission between $12500-16000\,\angstrom$, and the [C I] doublet mentioned above appears much stronger as well.

Except for the HEC-33H and HEC-110L models, some Si emission from the [Si I] $\lambda\lambda16073,\,16459$ doublet can also perhaps be detected, although the emission from this appears to be a factor several lower than the emission component at $\sim11000\,\angstrom$ in all the models.

The HEC-110 models have a quite small amount of He in the ejecta ($0.31\,M_\odot$, see Figure \ref{fig:composition_mixing_hec11} and Appendix \ref{app:PhysicalCon}), and in both the high and low energy models there appears to be very little He emission, while in the medium energy model this emission would be very hard to detect as it gets overwhelmed (mostly) by the [Si I] $\lambda10994$ feature. So despite these models containing He, the nebular phase spectra would not conclusively show this. Conversely, the lower mass models do predict that He I would be the dominant component in this region. The presence of He in any SN would be more conclusive if it can be detected in the uncontaminated He I $2.06\,\mu$m line, which is unfortunately much weaker. 

For SESNe, several NIR spectra are available; e.g. SN 2000ew \citep{gerardy2002carbon}, SN 2007gr \citep{hunter2009extensive}, SN 2013ge \citep{drout2016double}, SN 2020oi \citep{rho2021NIR} and SN 2022xxf \citep{Kuncarayakti2023broadlined}. From these, SN 2000ew is lacking observations for the He $1.08\,\mu$m line and detects only a very weak $2.06\,\mu$m feature; for SN 2013ge the He lines appear absent although the spectra are very early; for SN 2020oi the focus is even earlier. SN 2007gr is a Type Ic SN which has NIR spectra at both 176 and 377 days. In the former, a feature around $1.1\,\mu$m is detected but its origin is ambiguous while there is no feature at $2.06\,\mu$m. In the 377 day spectrum there is still a feature around $1.1\,\mu$m, but aside from the [Si I] feature around $1.6\,\mu$m the spectrum is too noisy to identify any lines. SN 2022xxf is another Type Ic which has NIR coverage up to $16500\,\angstrom$ at $114\,$d, and only detects a blend of features at $1.1\,\mu$m which may or may not contain He I $1.08\,\mu$m \citep{shahbandeh2022carnegie,Kuncarayakti2023broadlined}.

Between the nine models in our study, there are several which we can discard as problematic -- HEC-33H, HEC-60H and HEC-110L (see Section \ref{sssec:modelsummary}). The remaining six all clearly show a feature around $1.1\,\mu$m, although the dominant element in this feature does differ as it goes from He dominated towards Si dominated for more massive progenitors. The light progenitors in particular all predict a weak but noticeable $2.06\,\mu$m feature which is clearly absent in the 177 day spectrum in SN 2007gr \citep{hunter2009extensive}. Although this is but one object, it does indicate that the NIR is indeed a possible probe to identify between Type Ib and Ic cases. 

The NIR is a powerful tool for this since He lines do not show up in the optical \citep[see e.g.][]{mazzali1998feature} by nebular times, but the He NIR lines can be non-thermally excited. However, NIR observations at nebular times are rare for SESNe and we thus cannot perform a good comparison between observations and our models.

For the NIR spectra, the radiative transfer effect is expected to be small, especially for the He I lines as these are allowed transitions in a region which should be optically thin at our $200\,$day epoch. This allows us to make some predictions on the line strengths in this regime to investigate if there is a potential window here which could be used to help classify Type Ib/c spectra more firmly into their separate groups. It seems that the $2.06\,\mu$ line, which is isolated yet weak, would be the favoured line to look for due to the potential presence of [Si I] $\lambda10994$ blending into He I $\lambda10833$, although that line itself is significantly stronger. Tentatively, our spectra seem to indicate that the transition from 'has He lines' to 'no He lines' will be quite gradual, as the presence of Si and weakness of the $2.06\,\mu$ line can still lead to ambiguity in these spectra. However, if the $2.06\,\mu$ line were detected it is a clear indication that (some) He is present in the ejecta. As discussed above, in SN 2007gr there is only a feature at $1.1\,\mu$m while there is no feature around $2\,\mu$m -- as this object is classified as a Type Ic SN this is to be expected. It's possible that the progenitor of this SN was an even more massive He-core than considered for our work here, which evolved into a carbon-rich star as was suggested by \citet{valenti2008carbon}.

To some degree, the line luminosities predicted by Figure \ref{fig:HeNIR3x3} will vary under viewing angle, yet the relative contributions between the different elements will not -- so a feature dominated by He in one viewing angle will be dominated by He in all viewing angles. While the peak luminosity might vary somewhat across the viewing angles, the relative strength between the two He I features will not really change; the further away from $1:6$ this ratio is between the two features, the more contaminated the $1.08\,\mu$ feature is (by e.g. Si/S). What is potentially more harmful for our predictions is the lack of dust and molecules in our models, which could come into play at $200\,$days in the NIR \citep[see e.g.][]{sarangi2018dust,rho2021NIR,ravi2023NIR}.

\section{Discussion} \label{sec:discussion}
\subsection*{Models versus observations}
In Section \ref{sec:obscomp} for the [O I] doublet in particular, there appears to be a bias in the line centroids from the observations towards the blueshifted side by a few hundred km$\,$s$^{-1}$. \citet{taubenberger2009nebular} found that there is a systematic blueshift for the [O I] doublet which decreases with time, although even for spectra past 200$\,$days a minor blueshift persist. It is only a few $\angstrom$, which matches this slight bias we find, and which can likely be attributed to the inner ejecta not being fully transparent as was shown by \citet{jerkstrand2015late}. However, such a systematic blueshift does not appear for Mg I], where it should have been even stronger according to \citet{jerkstrand2015late} -- although we have few early spectra with Mg I].

Our models are compared to a compilation of observations, with all our models calculated at $200\,$days post-explosion, while our observations span a range of $131-416\,$days. Some of the variation in the $v_\text{shift}$ and $v_\text{width}$ for the same object could be due to the evolving state of the nebula. Despite this large discrepancy in timing epochs, making these comparisons still allows for a ballpark estimation if the input models are capable of matching observations. As running one model for one epoch is already quite expensive\footnote{Running one model to full completion in \texttt{ExTraSS} takes roughly $30\,000\,$cpu$\,$h.}, we did not do a time-evolution comparison with \texttt{ExTraSS}.

The work by \citet{fang2023aspherical} found that larger carbon-oxygen cores leads to a larger degree of asphericity, which we only partially recover. For [Ca II] in particular this is most clear, as we instead find that our $6.0\,M_\odot$ models have the largest centroid shifts, and not the more massive $11.0\,M_\odot$ He-cores. This difference can be attributed to the 3D nature of our models. Our work indicates that the current 3D neutrino-driven CCSN models do not linearly lead to an increase of asphericity with a larger He-core or larger explosion energy.  

In our models, it appears that in particular [Ca II] $\lambda\lambda7291,\,7323$ displays a correlation between the maximum centroid shift and ejecta asymmetry. Some of our models are predicting larger centroid shifts than what is observed, which is worrisome. It could be that our ejecta structure for the explosive nucleosynthesis is too asymmetric compared to what happens and that this causes such a discrepancy; alternatively the Ca is simply being mixed too much with lighter elements and taking over the emission at higher velocities which creates this larger asymmetry. It might also be that these large centroid shifts will disappear when radiative transfer is accounted for through the improved physical modelling, in which case these large centroid shifts are simply an artefact of the optically thin approximation.

Another alternative explanation is that for the more energetic explosions -- which synthesize more Ca -- the Ca ends up being distributed more symmetrically; this is generally true for the overall ejecta structure and it could explain why in particular the low energy models predict such huge centroid shifts for some viewing angles. Another unlikely option is that SNe which would show such drastic centroid shifts have simply never been observed, as these typically have low $^{56}$Ni ejecta and would be quite faint. It is also possible that some regime in between our tested models (either by progenitor mass of explosion energy) would show better agreement to the observations than our nine models here do.

Recent work by \citet{drout2023observed} and \citet{gotberg2023stellar} led to the first identification of several intermediate-mass helium stars in the LMC and SMC, with masses of $\sim2-8\,M_\odot$, which could be progenitors for the models considered in our work. The identification of these stars is important, as they will allow for fine-tuning of our understanding of He stars and their evolution towards the pre-supernova stage. Although their work shows a surprisingly low mass-loss rate for these helium stars (with rates much lower than what is assumed for our progenitor models), this is likely influenced by their lower metallicity.

\subsection*{Model variability}
When comparing the different viewing angle effects in the different models for different elements, it becomes clear that every model has its own quirks and it is unpredictable which elements $-$ if any $-$ are going to be nicely separated by viewing angle for their $v_\text{shift}$ and $v_\text{width}$ values. It is unpredictable which elements are going to behave similarly, as well as which elements (anti-)align with the NS kick vector for their centroid shifts. This indicates that the local conditions inside the nebula are more important than we estimated in \citet{vanbaal2023modelling}, where we thought that some of this could (largely) be attributed to the relatively small NS kick. Here instead we find that for every model (also the 6 not shown in detail) this unpredictability is found, indicating that this is likely a phenomenon which occurs during the nebular epoch rather than a model-specific outcome.

However, for elements which are strongly influenced by the explosive nucleosynthesis and the delay before explosive runaway occurs (i.e. heavy elements past $^{24}$Mg in particular, \citealt{wongwathanarat2013three}), the separation by viewing angle will line up as expected (i.e. approaching neutron star observes a receding line) more often than for elements which are (predominantly) found in the outer ejecta.

This is partially shown by the angle $\beta$ between the vectors $\vec{v}$ for Ca and [Ca II] (Table \ref{tab:Ca-velocity-onecolumn}), where several models show very minor differences between these two while other models show large discrepancies. Models with larger $\beta$ angles are models where the NLTE condition effects inside the nebula are relatively strong compared to the asymmetric explosive ejection, either because the high explosion energy shortens the time during which asymmetries form or because the inner, most asymmetric ejecta are lost. In particular for such events the viewing angle effects become much more difficult to predict. 

In several of our models, the centroid shift of the [Ca II] feature displays large variety with the viewing angles. These large asymmetries for Ca could perhaps be related to the small $\alpha-$network used in \texttt{P-HotB}, which could lead to Ca being produced in the wrong locations or at too high rates. It is known that such a network overproduces $^{44}$Ti \citep{sieverding2023production}, and the study by \citet{sandoval2021three} showed that a much larger network influences the production of key species like $^{28}$Si and $^{56}$Ni as well as the influence of having different isotopes for the same species. For Ca specifically however, it is unknown what the impact is of the small $\alpha-$network.

In our models, the other elements (C, O, Mg) do not show the high centroid shifts that Ca does. While Ca tends to predict centroid shifts that can exceed the observed values, the models predict centroid shifts which are on a similar scale as the observations for the lighter elements. With the growing amount of models available this opens the pathway to start investigating the radial mixing which takes place, or potentially links to the progenitor structure, for these elements. What is clear however already from our set is that having a much more asymmetric ejecta for these lighter elements would be problematic to conform to the observations.

\subsection*{\texttt{P-HotB} into \texttt{ExTraSS}}
One of the first choices we have to make with our models in the translation between \texttt{P-HotB} and \texttt{ExTraSS} is how to deal with Element X, the neutron-rich species tracer. For our models here we turned all of Element X into $^{56}$Ni, as this leads to more realistic $^{56}$Ni masses compared to observationally estimated values -- although the HEC-60H and HEC-110H models might be a bit too high. To test the impact of this choice, we also ran the HEC-60 model once while turning Element X into $^{56}$Fe directly, reducing the $^{56}$Ni mass from $0.094\,M_\odot$ to $0.032\,M_\odot$. This only has a marginal effect on the $v_\text{shift}$ and $v_\text{width}$ values for this model as well as the spectral shape which is remarkably similar to the spectrum shown in Figure \ref{fig:60-shiftwidth}. This shows that how exactly we choose to deal with Element X does not have a large impact on our findings, and that these results are quite independent from the $^{56}$Ni mass in the ejecta, even when this drops by a factor of three.

Work by \citet{prentice2019investigating} shows that bolometric light curve models give similar ejecta masses as our models ($1.2-4.8\,M_\odot$ in their SESNe sample) as well as similar $^{56}$Ni masses, if less extreme ($0.026-0.19\,M_\odot$). This shows that our models are in the right ballpark for ejecta masses and synthesized $^{56}$Ni, as the total $^{56}$Ni mass has only a minor effect on the spectral profiles. This allows for the investigation of our combinations of ejecta masses and explosion energies, and compare the spectral properties obtained with observations, to try and further our understanding of SESNe.

When looking at the nine different models, it can be noted that in several cases the compact object left behind after the explosion is too massive to be a NS. The HEC-110L and HEC-110 models have remnants of $4.86\,M_\odot$ and $3.65\,M_\odot$ respectively, which would form black holes, while the HEC-60L model has a remnant mass of $2.37\,M_\odot$. This is the \textit{baryonic} mass of the compact object however, rather than the gravitational mass which is $\sim2.1\,M_\odot$ and thus within the range of observed NS masses \citep{cromartie2020relativistic,romani2021PSR}. The formation of the black hole from bound-ejecta accretion onto the proto-NS takes place on a longer timescale than the unbinding of the star (see Table \ref{tab:model_overview}; most accretion occurs more than 5 seconds after the explosion for the HEC-110L and HEC-110 models), which means that our parametrized neutrinosphere does not suddenly disappear due to this collapse. This is critical since it is the energy from this neutrinosphere which will unbind the star to create our ejecta, so if a black hole would form before our target explosion energy was reached we would not get an explosion at all.

\section{Conclusions} \label{sec:conclusion}
We have presented an analysis of the nebular-phase spectra of nine 3D He-core explosion simulations, covering a He$_\text{ZAMS}$ mass range of $3.3-11\ M_\odot$ (pre-SN masses between $2.7-7.0\ M_\odot$, ejecta masses between $1.2-5.1\ M_\odot$) and an explosion energy range of $0.5-3$ B. The models are exploded by the neutrino-heating mechanism and evolved to the homologous phase with the \texttt{Prometheus-HotB} code and then have physical conditions and emergent spectra computed with the \texttt{ExTraSS} code. We have added an 'on-the-spot' photoionization treatment to \texttt{ExTraSS}, to approximately account for this effect, which is important for some of the neutral lines.  

We investigate the variations of the line widths ($v_\text{width}$) and shifts ($v_\text{shift}$) for different viewing angles, focusing on the four features Mg I] $\lambda4571$, [C I] $\lambda8727$, [O I] $\lambda\lambda6300,\,6364$ and [Ca II] $\lambda\lambda7291,\,7323$. With the self-consistent 3D modelling, both these metrics give us a powerful new way to determine how the progenitor mass and explosion energy affect observables, and what can be learnt from comparison to  observations. For the observational comparisons we compile a data set with 59 nebular spectra of 24 Type Ib SNe from the literature.

We find that combinations of low progenitor mass and high explosion energy give nebular line widths significantly too broad compared to observations. If He-cores with pre-SN mass $\lesssim\,3\,M_\odot$ contribute to the Type Ib SN population, the explosion energy needs to be low, $\sim0.5\,$B to match observed line widths. For $3\,M_\odot \lesssim M_\text{pre-SN} \lesssim\,4.5\,M_\odot$, the corresponding upper limit is a little bit above $\sim1\,$B. At the same time, if pre-SN cores as massive as $7\,M_\odot$ contribute to the Type Ib population, they must explode with $\sim 1-3\,$B, as less energetic explosions create too narrow lines. We can thus outline a domain of allowed combinations of mass and energy for He-core explosions.

Regarding line asymmetries, we find, for a given model, relatively similar maximum shift values for Mg I] 4571, [O I] 6300, 6364 and [C I] 8727. For the $3.3\,M_\odot$ He-core, these are relatively small, $\lesssim\,300\,\text{km}\,\text{s}^{-1}$, for the $6.0\,M_\odot$ He-core $400-800\,\text{km}\,\text{s}^{-1}$, and for the $11.0\,M_\odot$ He-core again small, $\lesssim\,300\,\text{km}\,\text{s}^{-1}$ -- with the values stated for models with energies in reasonable agreement with observations. Observational data show a mean shift value of about $460\,\text{km}\,\text{s}^{-1}$ for Mg I] 4571 (median $370\,\text{km}\,\text{s}^{-1}$), and good symmetry between blueshifts and redshifts in the distribution, which reduces the risk of contaminations or radiative transfer effects. This is too large (the mean observed value is expected to be about half the maximum model one) for the 3.3 and 11$\,M_\odot$ models, but in line with predictions from the 6.0$\,M_\odot$ model. For [O I] 6300, 6364 the data show a somewhat lower mean shift (mean $310\,\text{km}\,\text{s}^{-1}$, median $270\,\text{km}\,\text{s}^{-1}$), although a blueshift preference suggests non-clean emission. The [C I] 8727 line is observationally strongly blended with the Ca II NIR triplet, and our insufficient modelling of the triplet prevents a meaningful comparison.

The [Ca II] 7291, 7323 feature behaves qualitatively differently in the models. This is likely due to two effects; its different nucleosynthesis (explosive rather than hydrostatic as carbon, oxygen, and magnesium) and its unique cooling capabilities which allow it to emit strongly also in regions of low abundance. The vectors shown in Table \ref{tab:Ca-velocity-onecolumn} indicate that the explosive nucleosynthesis is largely responsible for this large asymmetry for Ca relative to the overall ejecta.  

The maximum shift values of the [Ca II] feature are $500-1000\,\text{km}\,\text{s}^{-1}$ in the $3.3\,M_\odot$ model, $\sim1500\,\text{km}\,\text{s}^{-1}$ in the $6.0\,M_\odot$ model, and $\sim500\,\text{km}\,\text{s}^{-1}$ in the $11.0\,M_\odot$ model, significantly larger than for the C, O, and Mg lines. Observational data show good blueshift-redshift symmetry (enabling a direct comparison) with a mean value of about $380\,\text{km}\,\text{s}^{-1}$ (median $380\,\text{km}\,\text{s}^{-1}$), reasonably in line with the $3.3\,M_\odot$ and $11.0\,M_\odot$ model predictions, but too small by a factor $\sim$2 for the $6.0\,M_\odot$ model. These are qualitatively new tests that for the first time allow direct comparison of predictions from current CCSN neutrino explosion theory with respect to the degree of asymmetry of SN ejecta as inferred from spectra. We see that current models predict asymmetries roughly in the right ballpark compared to observational data -- although detailed comparisons, object by object, and using yet more models (progenitors and energies) and accurate spectral modelling, are needed to draw more specific conclusions. 

Although we find different patterns for HEC-33L and HEC-60 than we did previously for HEC-33 \citep{vanbaal2023modelling}, it remains true that these four features behave differently from each other with viewing angle variation. The centroid shifts for the two models highlighted here still show the [C I] feature to be aligned with the NS motion rather than anti-aligned, indicating that this surprising alignment from \citet{vanbaal2023modelling} was not isolated to that model.

What has become clear from our models is that the more asymmetric the explosion, in particular for Ca, the better the separation by viewing angle relative to the neutron star kick vector for the centroid shifts. Our models suggest that in particular low-energy explosions from low and intermediate mass progenitors will create highly asymmetric (Ca) ejecta, leading to the large centroid shifts. In contrast, the Ca line shifts of the most massive models are strongly reduced by the fact that the low and intermediate energy cases produce black holes by fallback. In  these cases the fallback accretion carries a large fraction of the asymmetrically distributed Ca into the black hole. This opens up the possibility to probe the neutron star kick vector through the properties of the [Ca II] feature for lower mass progenitors.

Lastly, the formation of the nebular NIR lines of He I $1.08\,\mu$m and $2.06\,\mu$m was studied in the models. The formation of these depend both on 3D mixing and non-thermal effects -- which the models here are the first ones to have a self-consistent treatment for. We find that for our most massive model ($11.0\,M_\odot$), the He I $1.08\,\mu$m line is overwhelmed by the [Si I] $\lambda10994$ feature while the He I $2.06\,\mu$m feature is either very weak or absent. Thus, if the He is also not strongly visible during the photospheric phase in such a model, it would remain hidden also in the nebular phase. For the light and intermediate mass models, however, both He features are strong and clearly observable. The $2.06\,\mu$m feature is clean, and while the $1.08\,\mu$m feature still blends with silicon, it is dominant in these models. The search for these nebular NIR lines in Type Ic SNe would add importation information to the long-standing debate of hidden helium in these events. Additionally, a study of the He lines in Type Ib SNe would give another important diagnostic for the SN morphology.

\section*{Acknowledgements}

The authors thank Stan Woosley for his work on the pre-supernova modelling, and Stan Barmentloo for his collaboration on the Type Ib observational database. BvB would like to express gratitude to Quentin Pognan, Beatrice Guidici, Daniel Kresse and Michael Gabler for their discussions and feedback on this work. The authors acknowledge support from the European Research Council (ERC) under the European Union’s Horizon 2020 Research and Innovation Programme (ERC Starting Grant No. [803189], PI: A. Jerkstrand). The computations were enabled by resources provided by the National Academic Infrastructure for Supercomputing in Sweden (NAISS) at the PDC Center for High Performance Computing, KTH Royal Institute of Technology, partially funded by the Swedish Research Council through grant agreement no. 2018-05973. This work was supported by the German Research Foundation (DFG) through the Collaborative Research Centre ``Neutrinos and Dark Matter in Astro- and Particle Physics (NDM),'' Grant SFB-1258$\,$--$\,$283604770, and under Germany's Excellence Strategy through the Cluster of Excellence ORIGINS EXC-2094$\,$--$\,$390783311.

\section*{Data Availability}

The Data underlying this article will be shared on reasonable request to the corresponding author.

\noindent The nebular phase spectra included in this work were all obtained from \href{https://www.wiserep.org}{WiSeREP} -- \citealt{wiserep2012}. 



\bibliographystyle{mnras}
\bibliography{paper} 

\begin{thebibliography}{}
\makeatletter
\relax
\def\mn@urlcharsother{\let\do\@makeother \do\$\do\&\do\#\do\^\do\_\do\%\do\~}
\def\mn@doi{\begingroup\mn@urlcharsother \@ifnextchar [ {\mn@doi@}
  {\mn@doi@[]}}
\def\mn@doi@[#1]#2{\def\@tempa{#1}\ifx\@tempa\@empty \href
  {http://dx.doi.org/#2} {doi:#2}\else \href {http://dx.doi.org/#2} {#1}\fi
  \endgroup}
\def\mn@eprint#1#2{\mn@eprint@#1:#2::\@nil}
\def\mn@eprint@arXiv#1{\href {http://arxiv.org/abs/#1} {{\tt arXiv:#1}}}
\def\mn@eprint@dblp#1{\href {http://dblp.uni-trier.de/rec/bibtex/#1.xml}
  {dblp:#1}}
\def\mn@eprint@#1:#2:#3:#4\@nil{\def\@tempa {#1}\def\@tempb {#2}\def\@tempc
  {#3}\ifx \@tempc \@empty \let \@tempc \@tempb \let \@tempb \@tempa \fi \ifx
  \@tempb \@empty \def\@tempb {arXiv}\fi \@ifundefined
  {mn@eprint@\@tempb}{\@tempb:\@tempc}{\expandafter \expandafter \csname
  mn@eprint@\@tempb\endcsname \expandafter{\@tempc}}}

\bibitem[\protect\citeauthoryear{{Arcones}, {Janka}  \& {Scheck}}{{Arcones}
  et~al.}{2007}]{arcones2007nucleosynthesis}
{Arcones} A.,  {Janka} H.~T.,   {Scheck} L.,  2007, \mn@doi [\aap]
  {10.1051/0004-6361:20066983}, \href
  {https://ui.adsabs.harvard.edu/abs/2007A&A...467.1227A} {467, 1227}

\bibitem[\protect\citeauthoryear{{Arnett}}{{Arnett}}{1996}]{arnett1996supernovae}
{Arnett} D.,  1996, {Supernovae and Nucleosynthesis: An Investigation of the
  History of Matter from the Big Bang to the Present}

\bibitem[\protect\citeauthoryear{{Arzoumanian}, {Chernoff}  \&
  {Cordes}}{{Arzoumanian} et~al.}{2002}]{arzoumanian2002velocity}
{Arzoumanian} Z.,  {Chernoff} D.~F.,   {Cordes} J.~M.,  2002, \mn@doi [\apj]
  {10.1086/338805}, \href
  {https://ui.adsabs.harvard.edu/abs/2002ApJ...568..289A} {568, 289}

\bibitem[\protect\citeauthoryear{{\VAN{Baal}{Van}{van} Baal}, {Jerkstrand},
  {Wongwathanarat}  \& {Janka}}{{\VAN{Baal}{Van}{van} Baal}
  et~al.}{2023}]{vanbaal2023modelling}
{\VAN{Baal}{Van}{van} Baal} B. F.~A.,  {Jerkstrand} A.,  {Wongwathanarat} A.,
  {Janka} H.-T.,  2023, \mn@doi [\mnras] {10.1093/mnras/stad1488}, \href
  {https://ui.adsabs.harvard.edu/abs/2023MNRAS.523..954V} {523, 954}

\bibitem[\protect\citeauthoryear{{Barmentloo}, {Jerkstrand}, {Iwamoto},
  {Hachisu}, {Nomoto}, {Sollerman}  \& {Woosley}}{{Barmentloo}
  et~al.}{2024}]{barmentloo2024nebular}
{Barmentloo} S.,  {Jerkstrand} A.,  {Iwamoto} K.,  {Hachisu} I.,  {Nomoto} K.,
  {Sollerman} J.,   {Woosley} S.,  2024, \mn@doi [arXiv e-prints]
  {10.48550/arXiv.2403.08911}, \href
  {https://ui.adsabs.harvard.edu/abs/2024arXiv240308911B} {p. arXiv:2403.08911}

\bibitem[\protect\citeauthoryear{{Boggs} et~al.,}{{Boggs}
  et~al.}{2015}]{boggs201544ti}
{Boggs} S.~E.,  et~al., 2015, \mn@doi [Science] {10.1126/science.aaa2259},
  \href {https://ui.adsabs.harvard.edu/abs/2015Sci...348..670B} {348, 670}

\bibitem[\protect\citeauthoryear{{Bollig}, {Yadav}, {Kresse}, {Janka},
  {M{\"u}ller}  \& {Heger}}{{Bollig} et~al.}{2021}]{bollig2021selfconsistent}
{Bollig} R.,  {Yadav} N.,  {Kresse} D.,  {Janka} H.-T.,  {M{\"u}ller} B.,
  {Heger} A.,  2021, \mn@doi [\apj] {10.3847/1538-4357/abf82e}, \href
  {https://ui.adsabs.harvard.edu/abs/2021ApJ...915...28B} {915, 28}

\bibitem[\protect\citeauthoryear{{Burrows}, {Radice}, {Vartanyan}, {Nagakura},
  {Skinner}  \& {Dolence}}{{Burrows} et~al.}{2020}]{burrows2020overarching}
{Burrows} A.,  {Radice} D.,  {Vartanyan} D.,  {Nagakura} H.,  {Skinner} M.~A.,
   {Dolence} J.~C.,  2020, \mn@doi [\mnras] {10.1093/mnras/stz3223}, \href
  {https://ui.adsabs.harvard.edu/abs/2020MNRAS.491.2715B} {491, 2715}

\bibitem[\protect\citeauthoryear{{Burrows}, {Vartanyan}  \& {Wang}}{{Burrows}
  et~al.}{2023}]{burrows2023black}
{Burrows} A.,  {Vartanyan} D.,   {Wang} T.,  2023, \mn@doi [\apj]
  {10.3847/1538-4357/acfc1c}, \href
  {https://ui.adsabs.harvard.edu/abs/2023ApJ...957...68B} {957, 68}

\bibitem[\protect\citeauthoryear{{Ceverino} \& {Klypin}}{{Ceverino} \&
  {Klypin}}{2009}]{ceverino2009role}
{Ceverino} D.,  {Klypin} A.,  2009, \mn@doi [\apj]
  {10.1088/0004-637X/695/1/292}, \href
  {https://ui.adsabs.harvard.edu/abs/2009ApJ...695..292C} {695, 292}

\bibitem[\protect\citeauthoryear{{Chan}, {M{\"u}ller}  \& {Heger}}{{Chan}
  et~al.}{2020}]{chan2020impact}
{Chan} C.,  {M{\"u}ller} B.,   {Heger} A.,  2020, \mn@doi [\mnras]
  {10.1093/mnras/staa1431}, \href
  {https://ui.adsabs.harvard.edu/abs/2020MNRAS.495.3751C} {495, 3751}

\bibitem[\protect\citeauthoryear{{Cromartie} et~al.,}{{Cromartie}
  et~al.}{2020}]{cromartie2020relativistic}
{Cromartie} H.~T.,  et~al., 2020, \mn@doi [Nature Astronomy]
  {10.1038/s41550-019-0880-2}, \href
  {https://ui.adsabs.harvard.edu/abs/2020NatAs...4...72C} {4, 72}

\bibitem[\protect\citeauthoryear{{Dessart}}{{Dessart}}{2019}]{dessart2019simulations}
{Dessart} L.,  2019, \mn@doi [\aap] {10.1051/0004-6361/201834535}, \href
  {https://ui.adsabs.harvard.edu/abs/2019A&A...621A.141D} {621, A141}

\bibitem[\protect\citeauthoryear{{Dessart}, {Hillier}, {Li}  \&
  {Woosley}}{{Dessart} et~al.}{2012}]{dessart2012nature}
{Dessart} L.,  {Hillier} D.~J.,  {Li} C.,   {Woosley} S.,  2012, \mn@doi
  [\mnras] {10.1111/j.1365-2966.2012.21374.x}, \href
  {https://ui.adsabs.harvard.edu/abs/2012MNRAS.424.2139D} {424, 2139}

\bibitem[\protect\citeauthoryear{{Dessart}, {Hillier}, {Sukhbold}, {Woosley}
  \& {Janka}}{{Dessart} et~al.}{2021}]{dessart2021nebular}
{Dessart} L.,  {Hillier} D.~J.,  {Sukhbold} T.,  {Woosley} S.~E.,   {Janka}
  H.~T.,  2021, \mn@doi [\aap] {10.1051/0004-6361/202141927}, \href
  {https://ui.adsabs.harvard.edu/abs/2021A&A...656A..61D} {656, A61}

\bibitem[\protect\citeauthoryear{{Dessart}, {Hillier}, {Woosley}  \&
  {Kuncarayakti}}{{Dessart} et~al.}{2023}]{dessart2023modeling}
{Dessart} L.,  {Hillier} D.~J.,  {Woosley} S.~E.,   {Kuncarayakti} H.,  2023,
  \mn@doi [\aap] {10.1051/0004-6361/202346626}, \href
  {https://ui.adsabs.harvard.edu/abs/2023A&A...677A...7D} {677, A7}

\bibitem[\protect\citeauthoryear{{Drout} et~al.,}{{Drout}
  et~al.}{2016}]{drout2016double}
{Drout} M.~R.,  et~al., 2016, \mn@doi [\apj] {10.3847/0004-637X/821/1/57},
  \href {https://ui.adsabs.harvard.edu/abs/2016ApJ...821...57D} {821, 57}

\bibitem[\protect\citeauthoryear{{Drout}, {G{\"o}tberg}, {Ludwig}, {Groh}, {de
  Mink}, {O{\textquoteright}Grady}  \& {Smith}}{{Drout}
  et~al.}{2023}]{drout2023observed}
{Drout} M.~R.,  {G{\"o}tberg} Y.,  {Ludwig} B.~A.,  {Groh} J.~H.,  {de Mink}
  S.~E.,  {O{\textquoteright}Grady} A.~J.~G.,   {Smith} N.,  2023, \mn@doi
  [Science] {10.1126/science.ade4970}, \href
  {https://ui.adsabs.harvard.edu/abs/2023Sci...382.1287D} {382, 1287}

\bibitem[\protect\citeauthoryear{{Elmhamdi}, {Danziger}, {Cappellaro}, {Della
  Valle}, {Gouiffes}, {Phillips}  \& {Turatto}}{{Elmhamdi}
  et~al.}{2004}]{elmhamdi2004SNib}
{Elmhamdi} A.,  {Danziger} I.~J.,  {Cappellaro} E.,  {Della Valle} M.,
  {Gouiffes} C.,  {Phillips} M.~M.,   {Turatto} M.,  2004, \mn@doi [\aap]
  {10.1051/0004-6361:20041318}, \href
  {https://ui.adsabs.harvard.edu/abs/2004A&A...426..963E} {426, 963}

\bibitem[\protect\citeauthoryear{{Elmhamdi}, {Tsvetkov}, {Danziger}  \&
  {Kordi}}{{Elmhamdi} et~al.}{2011}]{elmhamdi2011photometric}
{Elmhamdi} A.,  {Tsvetkov} D.,  {Danziger} I.~J.,   {Kordi} A.,  2011, \mn@doi
  [\apj] {10.1088/0004-637X/731/2/129}, \href
  {https://ui.adsabs.harvard.edu/abs/2011ApJ...731..129E} {731, 129}

\bibitem[\protect\citeauthoryear{{Ergon} \& {Fransson}}{{Ergon} \&
  {Fransson}}{2022}]{ergon2022spectral}
{Ergon} M.,  {Fransson} C.,  2022, \mn@doi [\aap]
  {10.1051/0004-6361/202243448}, \href
  {https://ui.adsabs.harvard.edu/abs/2022A&A...666A.104E} {666, A104}

\bibitem[\protect\citeauthoryear{{Ergon} et~al.,}{{Ergon}
  et~al.}{2015}]{ergon2015type}
{Ergon} M.,  et~al., 2015, \mn@doi [\aap] {10.1051/0004-6361/201424592}, \href
  {https://ui.adsabs.harvard.edu/abs/2015A&A...580A.142E} {580, A142}

\bibitem[\protect\citeauthoryear{{Ertl}, {Woosley}, {Sukhbold}  \&
  {Janka}}{{Ertl} et~al.}{2020}]{ertl2020explosion}
{Ertl} T.,  {Woosley} S.~E.,  {Sukhbold} T.,   {Janka} H.~T.,  2020, \mn@doi
  [\apj] {10.3847/1538-4357/ab6458}, \href
  {https://ui.adsabs.harvard.edu/abs/2020ApJ...890...51E} {890, 51}

\bibitem[\protect\citeauthoryear{{Fang} \& {Maeda}}{{Fang} \&
  {Maeda}}{2023}]{fang2023inferring}
{Fang} Q.,  {Maeda} K.,  2023, \mn@doi [\apj] {10.3847/1538-4357/acc5e7}, \href
  {https://ui.adsabs.harvard.edu/abs/2023ApJ...949...93F} {949, 93}

\bibitem[\protect\citeauthoryear{{Fang} et~al.,}{{Fang}
  et~al.}{2022}]{fang2022core}
{Fang} Q.,  et~al., 2022, \mn@doi [\apj] {10.3847/1538-4357/ac4f60}, \href
  {https://ui.adsabs.harvard.edu/abs/2022ApJ...928..151F} {928, 151}

\bibitem[\protect\citeauthoryear{{Fang}, {Maeda}, {Kuncarayakti}  \&
  {Nagao}}{{Fang} et~al.}{2023}]{fang2023aspherical}
{Fang} Q.,  {Maeda} K.,  {Kuncarayakti} H.,   {Nagao} T.,  2023, \mn@doi
  [Nature Astronomy] {10.1038/s41550-023-02120-8}, \href
  {https://ui.adsabs.harvard.edu/abs/2023NatAs.tmp..224F} {}

\bibitem[\protect\citeauthoryear{{Fesen} et~al.,}{{Fesen}
  et~al.}{2006}]{fesen2006expansion}
{Fesen} R.~A.,  et~al., 2006, \mn@doi [\apj] {10.1086/504254}, \href
  {https://ui.adsabs.harvard.edu/abs/2006ApJ...645..283F} {645, 283}

\bibitem[\protect\citeauthoryear{{Filippenko} \& {Sargent}}{{Filippenko} \&
  {Sargent}}{1986}]{filippenko1986unique}
{Filippenko} A.~V.,  {Sargent} W.~L.~W.,  1986, \mn@doi [\aj] {10.1086/114051},
  \href {https://ui.adsabs.harvard.edu/abs/1986AJ.....91..691F} {91, 691}

\bibitem[\protect\citeauthoryear{{Filippenko} \& {Sargent}}{{Filippenko} \&
  {Sargent}}{1989}]{filippenko1989spectroscopic}
{Filippenko} A.~V.,  {Sargent} W. L.~W.,  1989, \mn@doi [\apjl]
  {10.1086/185548}, \href
  {https://ui.adsabs.harvard.edu/abs/1989ApJ...345L..43F} {345, L43}

\bibitem[\protect\citeauthoryear{{Folatelli} et~al.,}{{Folatelli}
  et~al.}{2014}]{folatelli2014supernova}
{Folatelli} G.,  et~al., 2014, \mn@doi [\apj] {10.1088/0004-637X/792/1/7},
  \href {https://ui.adsabs.harvard.edu/abs/2014ApJ...792....7F} {792, 7}

\bibitem[\protect\citeauthoryear{{Fransson} \& {Chevalier}}{{Fransson} \&
  {Chevalier}}{1989}]{Fransson1989}
{Fransson} C.,  {Chevalier} R.~A.,  1989, \mn@doi [\apj] {10.1086/167707},
  \href {https://ui.adsabs.harvard.edu/abs/1989ApJ...343..323F} {343, 323}

\bibitem[\protect\citeauthoryear{{Fremling} et~al.,}{{Fremling}
  et~al.}{2018}]{fremling2018oxygen}
{Fremling} C.,  et~al., 2018, \mn@doi [\aap] {10.1051/0004-6361/201731701},
  \href {https://ui.adsabs.harvard.edu/abs/2018A&A...618A..37F} {618, A37}

\bibitem[\protect\citeauthoryear{{Fryxell}, {Mueller}  \& {Arnett}}{{Fryxell}
  et~al.}{1991}]{fryxell1991instabilities}
{Fryxell} B.,  {Mueller} E.,   {Arnett} D.,  1991, \mn@doi [\apj]
  {10.1086/169657}, \href
  {https://ui.adsabs.harvard.edu/abs/1991ApJ...367..619F} {367, 619}

\bibitem[\protect\citeauthoryear{{Gabler}, {Wongwathanarat}  \&
  {Janka}}{{Gabler} et~al.}{2021}]{gabler2021infancy}
{Gabler} M.,  {Wongwathanarat} A.,   {Janka} H.-T.,  2021, \mn@doi [\mnras]
  {10.1093/mnras/stab116}, \href
  {https://ui.adsabs.harvard.edu/abs/2021MNRAS.502.3264G} {502, 3264}

\bibitem[\protect\citeauthoryear{{Gaskell}, {Cappellaro}, {Dinerstein},
  {Garnett}, {Harkness}  \& {Wheeler}}{{Gaskell}
  et~al.}{1986}]{gaskell1986type}
{Gaskell} C.~M.,  {Cappellaro} E.,  {Dinerstein} H.~L.,  {Garnett} D.~R.,
  {Harkness} R.~P.,   {Wheeler} J.~C.,  1986, \mn@doi [\apjl] {10.1086/184709},
  \href {https://ui.adsabs.harvard.edu/abs/1986ApJ...306L..77G} {306, L77}

\bibitem[\protect\citeauthoryear{{Gerardy}, {Fesen}, {Nomoto}, {Maeda},
  {Hoflich}  \& {Wheeler}}{{Gerardy} et~al.}{2002}]{gerardy2002carbon}
{Gerardy} C.~L.,  {Fesen} R.~A.,  {Nomoto} K.,  {Maeda} K.,  {Hoflich} P.,
  {Wheeler} J.~C.,  2002, \mn@doi [\pasj] {10.1093/pasj/54.6.905}, \href
  {https://ui.adsabs.harvard.edu/abs/2002PASJ...54..905G} {54, 905}

\bibitem[\protect\citeauthoryear{{G{\"o}tberg} et~al.,}{{G{\"o}tberg}
  et~al.}{2023}]{gotberg2023stellar}
{G{\"o}tberg} Y.,  et~al., 2023, \mn@doi [\apj] {10.3847/1538-4357/ace5a3},
  \href {https://ui.adsabs.harvard.edu/abs/2023ApJ...959..125G} {959, 125}

\bibitem[\protect\citeauthoryear{{Grefenstette} et~al.,}{{Grefenstette}
  et~al.}{2014}]{grefenstette2014asymmetries}
{Grefenstette} B.~W.,  et~al., 2014, \mn@doi [\nat] {10.1038/nature12997},
  \href {https://ui.adsabs.harvard.edu/abs/2014Natur.506..339G} {506, 339}

\bibitem[\protect\citeauthoryear{{Grefenstette} et~al.,}{{Grefenstette}
  et~al.}{2017}]{grefenstette2017distribution}
{Grefenstette} B.~W.,  et~al., 2017, \mn@doi [\apj]
  {10.3847/1538-4357/834/1/19}, \href
  {https://ui.adsabs.harvard.edu/abs/2017ApJ...834...19G} {834, 19}

\bibitem[\protect\citeauthoryear{{Heger}, {Fryer}, {Woosley}, {Langer}  \&
  {Hartmann}}{{Heger} et~al.}{2003}]{heger2003massive}
{Heger} A.,  {Fryer} C.~L.,  {Woosley} S.~E.,  {Langer} N.,   {Hartmann} D.~H.,
   2003, \mn@doi [\apj] {10.1086/375341}, \href
  {https://ui.adsabs.harvard.edu/abs/2003ApJ...591..288H} {591, 288}

\bibitem[\protect\citeauthoryear{{Hillier} \& {Dessart}}{{Hillier} \&
  {Dessart}}{2012}]{Hillier2012}
{Hillier} D.~J.,  {Dessart} L.,  2012, \mn@doi [\mnras]
  {10.1111/j.1365-2966.2012.21192.x}, \href
  {https://ui.adsabs.harvard.edu/abs/2012MNRAS.424..252H} {424, 252}

\bibitem[\protect\citeauthoryear{{Hobbs}, {Lorimer}, {Lyne}  \&
  {Kramer}}{{Hobbs} et~al.}{2005}]{hobbs2005statistical}
{Hobbs} G.,  {Lorimer} D.~R.,  {Lyne} A.~G.,   {Kramer} M.,  2005, \mn@doi
  [\mnras] {10.1111/j.1365-2966.2005.09087.x}, \href
  {https://ui.adsabs.harvard.edu/abs/2005MNRAS.360..974H} {360, 974}

\bibitem[\protect\citeauthoryear{{Houck} \& {Fransson}}{{Houck} \&
  {Fransson}}{1996}]{Houck1996}
{Houck} J.~C.,  {Fransson} C.,  1996, \mn@doi [\apj] {10.1086/176699}, \href
  {https://ui.adsabs.harvard.edu/abs/1996ApJ...456..811H} {456, 811}

\bibitem[\protect\citeauthoryear{{Hughes}, {Rakowski}, {Burrows}  \&
  {Slane}}{{Hughes} et~al.}{2000}]{hughes2000nucleosynthesis}
{Hughes} J.~P.,  {Rakowski} C.~E.,  {Burrows} D.~N.,   {Slane} P.~O.,  2000,
  \mn@doi [\apjl] {10.1086/312438}, \href
  {https://ui.adsabs.harvard.edu/abs/2000ApJ...528L.109H} {528, L109}

\bibitem[\protect\citeauthoryear{{Hunter} et~al.,}{{Hunter}
  et~al.}{2009}]{hunter2009extensive}
{Hunter} D.~J.,  et~al., 2009, \mn@doi [\aap] {10.1051/0004-6361/200912896},
  \href {https://ui.adsabs.harvard.edu/abs/2009A&A...508..371H} {508, 371}

\bibitem[\protect\citeauthoryear{{Janka} \& {M{\"u}ller}}{{Janka} \&
  {M{\"u}ller}}{1996}]{janka1996neutrino}
{Janka} H.~T.,  {M{\"u}ller} E.,  1996, \aap, \href
  {https://ui.adsabs.harvard.edu/abs/1996A&A...306..167J} {306, 167}

\bibitem[\protect\citeauthoryear{{Jerkstrand}}{{Jerkstrand}}{2017}]{jerkstrand2017spectra}
{Jerkstrand} A.,  2017, in {Alsabti} A.~W.,  {Murdin} P.,  eds, , Handbook of
  Supernovae.
p.~795, \mn@doi{10.1007/978-3-319-21846-5_29}

\bibitem[\protect\citeauthoryear{{Jerkstrand}, {Fransson}  \&
  {Kozma}}{{Jerkstrand} et~al.}{2011}]{jerkstrand201144Ti}
{Jerkstrand} A.,  {Fransson} C.,   {Kozma} C.,  2011, \mn@doi [\aap]
  {10.1051/0004-6361/201015937}, \href
  {https://ui.adsabs.harvard.edu/abs/2011A&A...530A..45J} {530, A45}

\bibitem[\protect\citeauthoryear{{Jerkstrand}, {Fransson}, {Maguire}, {Smartt},
  {Ergon}  \& {Spyromilio}}{{Jerkstrand}
  et~al.}{2012}]{jerkstrand2012progenitor}
{Jerkstrand} A.,  {Fransson} C.,  {Maguire} K.,  {Smartt} S.,  {Ergon} M.,
  {Spyromilio} J.,  2012, \mn@doi [\aap] {10.1051/0004-6361/201219528}, \href
  {https://ui.adsabs.harvard.edu/abs/2012A&A...546A..28J} {546, A28}

\bibitem[\protect\citeauthoryear{{Jerkstrand}, {Ergon}, {Smartt}, {Fransson},
  {Sollerman}, {Taubenberger}, {Bersten}  \& {Spyromilio}}{{Jerkstrand}
  et~al.}{2015}]{jerkstrand2015late}
{Jerkstrand} A.,  {Ergon} M.,  {Smartt} S.~J.,  {Fransson} C.,  {Sollerman} J.,
   {Taubenberger} S.,  {Bersten} M.,   {Spyromilio} J.,  2015, \mn@doi [\aap]
  {10.1051/0004-6361/201423983}, \href
  {https://ui.adsabs.harvard.edu/abs/2015A&A...573A..12J} {573, A12}

\bibitem[\protect\citeauthoryear{{Jerkstrand} et~al.,}{{Jerkstrand}
  et~al.}{2017}]{jerkstrand2017long}
{Jerkstrand} A.,  et~al., 2017, \mn@doi [\apj] {10.3847/1538-4357/835/1/13},
  \href {https://ui.adsabs.harvard.edu/abs/2017ApJ...835...13J} {835, 13}

\bibitem[\protect\citeauthoryear{{Jerkstrand} et~al.,}{{Jerkstrand}
  et~al.}{2020}]{jerkstrand2020properties}
{Jerkstrand} A.,  et~al., 2020, \mn@doi [\mnras] {10.1093/mnras/staa736}, \href
  {https://ui.adsabs.harvard.edu/abs/2020MNRAS.494.2471J} {494, 2471}

\bibitem[\protect\citeauthoryear{{Kageyama} \& {Sato}}{{Kageyama} \&
  {Sato}}{2004}]{kageyama2004yinyang}
{Kageyama} A.,  {Sato} T.,  2004, \mn@doi [Geochemistry, Geophysics,
  Geosystems] {10.1029/2004GC000734}, \href
  {https://ui.adsabs.harvard.edu/abs/2004GGG.....5.9005K} {5, Q09005}

\bibitem[\protect\citeauthoryear{{Kifonidis}, {Plewa}, {Janka}  \&
  {M{\"u}ller}}{{Kifonidis} et~al.}{2003}]{kifonidis2003non}
{Kifonidis} K.,  {Plewa} T.,  {Janka} H.~T.,   {M{\"u}ller} E.,  2003, \mn@doi
  [\aap] {10.1051/0004-6361:20030863}, \href
  {https://ui.adsabs.harvard.edu/abs/2003A&A...408..621K} {408, 621}

\bibitem[\protect\citeauthoryear{{Kifonidis}, {Plewa}, {Scheck}, {Janka}  \&
  {M{\"u}ller}}{{Kifonidis} et~al.}{2006}]{kifonidis2006non}
{Kifonidis} K.,  {Plewa} T.,  {Scheck} L.,  {Janka} H.~T.,   {M{\"u}ller} E.,
  2006, \mn@doi [\aap] {10.1051/0004-6361:20054512}, \href
  {https://ui.adsabs.harvard.edu/abs/2006A&A...453..661K} {453, 661}

\bibitem[\protect\citeauthoryear{{Kozma} \& {Fransson}}{{Kozma} \&
  {Fransson}}{1992}]{kozma1992gamma}
{Kozma} C.,  {Fransson} C.,  1992, \mn@doi [\apj] {10.1086/171311}, \href
  {https://ui.adsabs.harvard.edu/abs/1992ApJ...390..602K} {390, 602}

\bibitem[\protect\citeauthoryear{{Kuncarayakti} et~al.,}{{Kuncarayakti}
  et~al.}{2023}]{Kuncarayakti2023broadlined}
{Kuncarayakti} H.,  et~al., 2023, \mn@doi [\aap] {10.1051/0004-6361/202346526},
  \href {https://ui.adsabs.harvard.edu/abs/2023A&A...678A.209K} {678, A209}

\bibitem[\protect\citeauthoryear{{Larsson} et~al.,}{{Larsson}
  et~al.}{2013}]{larsson2013morphology}
{Larsson} J.,  et~al., 2013, \mn@doi [\apj] {10.1088/0004-637X/768/1/89}, \href
  {https://ui.adsabs.harvard.edu/abs/2013ApJ...768...89L} {768, 89}

\bibitem[\protect\citeauthoryear{{Larsson} et~al.,}{{Larsson}
  et~al.}{2023}]{larsson2023JWST}
{Larsson} J.,  et~al., 2023, \mn@doi [\apjl] {10.3847/2041-8213/acd555}, \href
  {https://ui.adsabs.harvard.edu/abs/2023ApJ...949L..27L} {949, L27}

\bibitem[\protect\citeauthoryear{{Limongi} \& {Chieffi}}{{Limongi} \&
  {Chieffi}}{2003}]{Limongi2003}
{Limongi} M.,  {Chieffi} A.,  2003, \mn@doi [\apj] {10.1086/375703}, \href
  {https://ui.adsabs.harvard.edu/abs/2003ApJ...592..404L} {592, 404}

\bibitem[\protect\citeauthoryear{{Maeda}, {Nakamura}, {Nomoto}, {Mazzali},
  {Patat}  \& {Hachisu}}{{Maeda} et~al.}{2002}]{Maeda2002}
{Maeda} K.,  {Nakamura} T.,  {Nomoto} K.,  {Mazzali} P.~A.,  {Patat} F.,
  {Hachisu} I.,  2002, \mn@doi [\apj] {10.1086/324487}, \href
  {https://ui.adsabs.harvard.edu/abs/2002ApJ...565..405M} {565, 405}

\bibitem[\protect\citeauthoryear{{Maeda}, {Nomoto}, {Mazzali}  \&
  {Deng}}{{Maeda} et~al.}{2006}]{Maeda2006}
{Maeda} K.,  {Nomoto} K.,  {Mazzali} P.~A.,   {Deng} J.,  2006, \mn@doi [\apj]
  {10.1086/500187}, \href
  {https://ui.adsabs.harvard.edu/abs/2006ApJ...640..854M} {640, 854}

\bibitem[\protect\citeauthoryear{{Maeda} et~al.,}{{Maeda}
  et~al.}{2008}]{maeda2008asphericity}
{Maeda} K.,  et~al., 2008, \mn@doi [Science] {10.1126/science.1149437}, \href
  {https://ui.adsabs.harvard.edu/abs/2008Sci...319.1220M} {319, 1220}

\bibitem[\protect\citeauthoryear{{Matheson}, {Filippenko}, {Li}, {Leonard}  \&
  {Shields}}{{Matheson} et~al.}{2001}]{matheson2001optical}
{Matheson} T.,  {Filippenko} A.~V.,  {Li} W.,  {Leonard} D.~C.,   {Shields}
  J.~C.,  2001, \mn@doi [\aj] {10.1086/319390}, \href
  {https://ui.adsabs.harvard.edu/abs/2001AJ....121.1648M} {121, 1648}

\bibitem[\protect\citeauthoryear{{Maurer}, {Mazzali}, {Taubenberger}  \&
  {Hachinger}}{{Maurer} et~al.}{2010}]{maurer2012hydrogen}
{Maurer} I.,  {Mazzali} P.~A.,  {Taubenberger} S.,   {Hachinger} S.,  2010,
  \mn@doi [\mnras] {10.1111/j.1365-2966.2010.17186.x}, \href
  {https://ui.adsabs.harvard.edu/abs/2010MNRAS.409.1441M} {409, 1441}

\bibitem[\protect\citeauthoryear{{Mazzali} \& {Lucy}}{{Mazzali} \&
  {Lucy}}{1998}]{mazzali1998feature}
{Mazzali} P.~A.,  {Lucy} L.~B.,  1998, \mn@doi [\mnras]
  {10.1046/j.1365-8711.1998.01323.x}, \href
  {https://ui.adsabs.harvard.edu/abs/1998MNRAS.295..428M} {295, 428}

\bibitem[\protect\citeauthoryear{{Mazzali}, {Nomoto}, {Patat}  \&
  {Maeda}}{{Mazzali} et~al.}{2001}]{mazzali2001nebular}
{Mazzali} P.~A.,  {Nomoto} K.,  {Patat} F.,   {Maeda} K.,  2001, \mn@doi [\apj]
  {10.1086/322420}, \href
  {https://ui.adsabs.harvard.edu/abs/2001ApJ...559.1047M} {559, 1047}

\bibitem[\protect\citeauthoryear{{Mazzali}, {Deng}, {Maeda}, {Nomoto},
  {Filippenko}  \& {Matheson}}{{Mazzali} et~al.}{2004}]{Mazzali2004}
{Mazzali} P.~A.,  {Deng} J.,  {Maeda} K.,  {Nomoto} K.,  {Filippenko} A.~V.,
  {Matheson} T.,  2004, \mn@doi [\apj] {10.1086/423888}, \href
  {https://ui.adsabs.harvard.edu/abs/2004ApJ...614..858M} {614, 858}

\bibitem[\protect\citeauthoryear{{Mazzali} et~al.,}{{Mazzali}
  et~al.}{2005}]{Mazzali2005}
{Mazzali} P.~A.,  et~al., 2005, \mn@doi [Science] {10.1126/science.1111384},
  \href {https://ui.adsabs.harvard.edu/abs/2005Sci...308.1284M} {308, 1284}

\bibitem[\protect\citeauthoryear{{Mazzali}, {Maurer}, {Valenti}, {Kotak}  \&
  {Hunter}}{{Mazzali} et~al.}{2010}]{mazzali2010type}
{Mazzali} P.~A.,  {Maurer} I.,  {Valenti} S.,  {Kotak} R.,   {Hunter} D.,
  2010, \mn@doi [\mnras] {10.1111/j.1365-2966.2010.17133.x}, \href
  {https://ui.adsabs.harvard.edu/abs/2010MNRAS.408...87M} {408, 87}

\bibitem[\protect\citeauthoryear{{Milisavljevic} et~al.,}{{Milisavljevic}
  et~al.}{2013}]{milisavljevic2013SN2012au}
{Milisavljevic} D.,  et~al., 2013, \mn@doi [\apjl]
  {10.1088/2041-8205/770/2/L38}, \href
  {https://ui.adsabs.harvard.edu/abs/2013ApJ...770L..38M} {770, L38}

\bibitem[\protect\citeauthoryear{{Milisavljevic} et~al.,}{{Milisavljevic}
  et~al.}{2024}]{Mili2024}
{Milisavljevic} D.,  et~al., 2024, \mn@doi [arXiv e-prints]
  {10.48550/arXiv.2401.02477}, \href
  {https://ui.adsabs.harvard.edu/abs/2024arXiv240102477M} {p. arXiv:2401.02477}

\bibitem[\protect\citeauthoryear{{Modjaz}, {Kirshner}, {Blondin}, {Challis}  \&
  {Matheson}}{{Modjaz} et~al.}{2008}]{modjaz2008double}
{Modjaz} M.,  {Kirshner} R.~P.,  {Blondin} S.,  {Challis} P.,   {Matheson} T.,
  2008, \mn@doi [\apjl] {10.1086/593135}, \href
  {https://ui.adsabs.harvard.edu/abs/2008ApJ...687L...9M} {687, L9}

\bibitem[\protect\citeauthoryear{{Modjaz} et~al.,}{{Modjaz}
  et~al.}{2009}]{modjaz2009shock}
{Modjaz} M.,  et~al., 2009, \mn@doi [\apj] {10.1088/0004-637X/702/1/226}, \href
  {https://ui.adsabs.harvard.edu/abs/2009ApJ...702..226M} {702, 226}

\bibitem[\protect\citeauthoryear{{M{\"u}ller}, {Fryxell}  \&
  {Arnett}}{{M{\"u}ller} et~al.}{1991a}]{muller1991high}
{M{\"u}ller} E.,  {Fryxell} B.,   {Arnett} D.,  1991a, in European Southern
  Observatory Conference and Workshop Proceedings. p.~99

\bibitem[\protect\citeauthoryear{{M{\"u}ller}, {Fryxell}  \&
  {Arnett}}{{M{\"u}ller} et~al.}{1991b}]{muller1991instability}
{M{\"u}ller} E.,  {Fryxell} B.,   {Arnett} D.,  1991b, \aap, \href
  {https://ui.adsabs.harvard.edu/abs/1991A&A...251..505M} {251, 505}

\bibitem[\protect\citeauthoryear{{M{\"u}ller}, {Janka}  \&
  {Wongwathanarat}}{{M{\"u}ller} et~al.}{2012}]{muller2012parametrized}
{M{\"u}ller} E.,  {Janka} H.~T.,   {Wongwathanarat} A.,  2012, \mn@doi [\aap]
  {10.1051/0004-6361/201117611}, \href
  {https://ui.adsabs.harvard.edu/abs/2012A&A...537A..63M} {537, A63}

\bibitem[\protect\citeauthoryear{{M{\"u}ller}, {Gay}, {Heger}, {Tauris}  \&
  {Sim}}{{M{\"u}ller} et~al.}{2018}]{muller2018multidimensional}
{M{\"u}ller} B.,  {Gay} D.~W.,  {Heger} A.,  {Tauris} T.~M.,   {Sim} S.~A.,
  2018, \mn@doi [\mnras] {10.1093/mnras/sty1683}, \href
  {https://ui.adsabs.harvard.edu/abs/2018MNRAS.479.3675M} {479, 3675}

\bibitem[\protect\citeauthoryear{{M{\"u}ller} et~al.,}{{M{\"u}ller}
  et~al.}{2019}]{muller2019three}
{M{\"u}ller} B.,  et~al., 2019, \mn@doi [\mnras] {10.1093/mnras/stz216}, \href
  {https://ui.adsabs.harvard.edu/abs/2019MNRAS.484.3307M} {484, 3307}

\bibitem[\protect\citeauthoryear{{Piro} \& {Morozova}}{{Piro} \&
  {Morozova}}{2014}]{piro2014transparent}
{Piro} A.~L.,  {Morozova} V.~S.,  2014, \mn@doi [\apjl]
  {10.1088/2041-8205/792/1/L11}, \href
  {https://ui.adsabs.harvard.edu/abs/2014ApJ...792L..11P} {792, L11}

\bibitem[\protect\citeauthoryear{{Prentice} et~al.,}{{Prentice}
  et~al.}{2019}]{prentice2019investigating}
{Prentice} S.~J.,  et~al., 2019, \mn@doi [\mnras] {10.1093/mnras/sty3399},
  \href {https://ui.adsabs.harvard.edu/abs/2019MNRAS.485.1559P} {485, 1559}

\bibitem[\protect\citeauthoryear{{Ravi} et~al.,}{{Ravi}
  et~al.}{2023}]{ravi2023NIR}
{Ravi} A.~P.,  et~al., 2023, \mn@doi [\apj] {10.3847/1538-4357/accddc}, \href
  {https://ui.adsabs.harvard.edu/abs/2023ApJ...950...14R} {950, 14}

\bibitem[\protect\citeauthoryear{{Reilly} et~al.,}{{Reilly}
  et~al.}{2016}]{reilly2016spectropolarimetry}
{Reilly} E.,  et~al., 2016, \mn@doi [\mnras] {10.1093/mnras/stv3005}, \href
  {https://ui.adsabs.harvard.edu/abs/2016MNRAS.457..288R} {457, 288}

\bibitem[\protect\citeauthoryear{{Rho} et~al.,}{{Rho}
  et~al.}{2021}]{rho2021NIR}
{Rho} J.,  et~al., 2021, \mn@doi [\apj] {10.3847/1538-4357/abd850}, \href
  {https://ui.adsabs.harvard.edu/abs/2021ApJ...908..232R} {908, 232}

\bibitem[\protect\citeauthoryear{{Romani}, {Kandel}, {Filippenko}, {Brink}  \&
  {Zheng}}{{Romani} et~al.}{2021}]{romani2021PSR}
{Romani} R.~W.,  {Kandel} D.,  {Filippenko} A.~V.,  {Brink} T.~G.,   {Zheng}
  W.,  2021, \mn@doi [\apjl] {10.3847/2041-8213/abe2b4}, \href
  {https://ui.adsabs.harvard.edu/abs/2021ApJ...908L..46R} {908, L46}

\bibitem[\protect\citeauthoryear{{Roy} et~al.,}{{Roy}
  et~al.}{2013}]{roy2013sn2007uy}
{Roy} R.,  et~al., 2013, \mn@doi [\mnras] {10.1093/mnras/stt1148}, \href
  {https://ui.adsabs.harvard.edu/abs/2013MNRAS.434.2032R} {434, 2032}

\bibitem[\protect\citeauthoryear{{Rybicki} \& {Lightman}}{{Rybicki} \&
  {Lightman}}{1979}]{rybicki1979radiative}
{Rybicki} G.~B.,  {Lightman} A.~P.,  1979, {Radiative processes in
  astrophysics}

\bibitem[\protect\citeauthoryear{{Sandoval}, {Hix}, {Messer}, {Lentz}  \&
  {Harris}}{{Sandoval} et~al.}{2021}]{sandoval2021three}
{Sandoval} M.~A.,  {Hix} W.~R.,  {Messer} O.~E.~B.,  {Lentz} E.~J.,   {Harris}
  J.~A.,  2021, \mn@doi [\apj] {10.3847/1538-4357/ac1d49}, \href
  {https://ui.adsabs.harvard.edu/abs/2021ApJ...921..113S} {921, 113}

\bibitem[\protect\citeauthoryear{{Sarangi}, {Matsuura}  \&
  {Micelotta}}{{Sarangi} et~al.}{2018}]{sarangi2018dust}
{Sarangi} A.,  {Matsuura} M.,   {Micelotta} E.~R.,  2018, \mn@doi [\ssr]
  {10.1007/s11214-018-0492-7}, \href
  {https://ui.adsabs.harvard.edu/abs/2018SSRv..214...63S} {214, 63}

\bibitem[\protect\citeauthoryear{{Sauer}, {Mazzali}, {Deng}, {Valenti},
  {Nomoto}  \& {Filippenko}}{{Sauer} et~al.}{2006}]{sauer2006properties}
{Sauer} D.~N.,  {Mazzali} P.~A.,  {Deng} J.,  {Valenti} S.,  {Nomoto} K.,
  {Filippenko} A.~V.,  2006, \mn@doi [\mnras]
  {10.1111/j.1365-2966.2006.10438.x}, \href
  {https://ui.adsabs.harvard.edu/abs/2006MNRAS.369.1939S} {369, 1939}

\bibitem[\protect\citeauthoryear{{Scheck}, {Kifonidis}, {Janka}  \&
  {M{\"u}ller}}{{Scheck} et~al.}{2006}]{scheck2006multidimensional}
{Scheck} L.,  {Kifonidis} K.,  {Janka} H.~T.,   {M{\"u}ller} E.,  2006, \mn@doi
  [\aap] {10.1051/0004-6361:20064855}, \href
  {https://ui.adsabs.harvard.edu/abs/2006A&A...457..963S} {457, 963}

\bibitem[\protect\citeauthoryear{{Schulze} et~al.,}{{Schulze}
  et~al.}{2021}]{schulze2021PTFCC}
{Schulze} S.,  et~al., 2021, \mn@doi [\apjs] {10.3847/1538-4365/abff5e}, \href
  {https://ui.adsabs.harvard.edu/abs/2021ApJS..255...29S} {255, 29}

\bibitem[\protect\citeauthoryear{{Shahbandeh} et~al.,}{{Shahbandeh}
  et~al.}{2022}]{shahbandeh2022carnegie}
{Shahbandeh} M.,  et~al., 2022, \mn@doi [\apj] {10.3847/1538-4357/ac4030},
  \href {https://ui.adsabs.harvard.edu/abs/2022ApJ...925..175S} {925, 175}

\bibitem[\protect\citeauthoryear{{Shivvers} et~al.,}{{Shivvers}
  et~al.}{2013}]{shivvers2013nebular}
{Shivvers} I.,  et~al., 2013, \mn@doi [\mnras] {10.1093/mnras/stt1839}, \href
  {https://ui.adsabs.harvard.edu/abs/2013MNRAS.436.3614S} {436, 3614}

\bibitem[\protect\citeauthoryear{{Shivvers} et~al.,}{{Shivvers}
  et~al.}{2019}]{shivvers2019berkeley}
{Shivvers} I.,  et~al., 2019, \mn@doi [\mnras] {10.1093/mnras/sty2719}, \href
  {https://ui.adsabs.harvard.edu/abs/2019MNRAS.482.1545S} {482, 1545}

\bibitem[\protect\citeauthoryear{{Sieverding}, {Kresse}  \&
  {Janka}}{{Sieverding} et~al.}{2023}]{sieverding2023production}
{Sieverding} A.,  {Kresse} D.,   {Janka} H.-T.,  2023, \mn@doi [\apjl]
  {10.3847/2041-8213/ad045b}, \href
  {https://ui.adsabs.harvard.edu/abs/2023ApJ...957L..25S} {957, L25}

\bibitem[\protect\citeauthoryear{{Sollerman}, {Kozma}, {Fransson},
  {Leibundgut}, {Lundqvist}, {Ryde}  \& {Woudt}}{{Sollerman}
  et~al.}{2000}]{Sollerman2000}
{Sollerman} J.,  {Kozma} C.,  {Fransson} C.,  {Leibundgut} B.,  {Lundqvist} P.,
   {Ryde} F.,   {Woudt} P.,  2000, \mn@doi [\apjl] {10.1086/312763}, \href
  {https://ui.adsabs.harvard.edu/abs/2000ApJ...537L.127S} {537, L127}

\bibitem[\protect\citeauthoryear{{Spyromilio}}{{Spyromilio}}{1994}]{spyromilio1994clumping}
{Spyromilio} J.,  1994, \mn@doi [\mnras] {10.1093/mnras/266.1.L61}, \href
  {https://ui.adsabs.harvard.edu/abs/1994MNRAS.266L..61S} {266, L61}

\bibitem[\protect\citeauthoryear{{Stockinger} et~al.,}{{Stockinger}
  et~al.}{2020}]{stockinger2020three}
{Stockinger} G.,  et~al., 2020, \mn@doi [\mnras] {10.1093/mnras/staa1691},
  \href {https://ui.adsabs.harvard.edu/abs/2020MNRAS.496.2039S} {496, 2039}

\bibitem[\protect\citeauthoryear{{Stritzinger} et~al.,}{{Stritzinger}
  et~al.}{2009}]{stritzinger2009herich}
{Stritzinger} M.,  et~al., 2009, \mn@doi [\apj] {10.1088/0004-637X/696/1/713},
  \href {https://ui.adsabs.harvard.edu/abs/2009ApJ...696..713S} {696, 713}

\bibitem[\protect\citeauthoryear{{Tanaka} et~al.,}{{Tanaka}
  et~al.}{2012}]{tanaka2012three}
{Tanaka} M.,  et~al., 2012, \mn@doi [\apj] {10.1088/0004-637X/754/1/63}, \href
  {https://ui.adsabs.harvard.edu/abs/2012ApJ...754...63T} {754, 63}

\bibitem[\protect\citeauthoryear{{Taubenberger} et~al.,}{{Taubenberger}
  et~al.}{2009}]{taubenberger2009nebular}
{Taubenberger} S.,  et~al., 2009, \mn@doi [\mnras]
  {10.1111/j.1365-2966.2009.15003.x}, \href
  {https://ui.adsabs.harvard.edu/abs/2009MNRAS.397..677T} {397, 677}

\bibitem[\protect\citeauthoryear{{Teegarden}}{{Teegarden}}{1991}]{teegarden1991gamma}
{Teegarden} B.~J.,  1991, \mn@doi [Advances in Space Research]
  {10.1016/0273-1177(91)90174-I}, \href
  {https://ui.adsabs.harvard.edu/abs/1991AdSpR..11h.217T} {11, 217}

\bibitem[\protect\citeauthoryear{{Tinyanont} et~al.,}{{Tinyanont}
  et~al.}{2021}]{tinyanont2021infrared}
{Tinyanont} S.,  et~al., 2021, \mn@doi [Nature Astronomy]
  {10.1038/s41550-021-01320-4}, \href
  {https://ui.adsabs.harvard.edu/abs/2021NatAs...5..544T} {5, 544}

\bibitem[\protect\citeauthoryear{{Valenti} et~al.,}{{Valenti}
  et~al.}{2008}]{valenti2008carbon}
{Valenti} S.,  et~al., 2008, \mn@doi [\apjl] {10.1086/527672}, \href
  {https://ui.adsabs.harvard.edu/abs/2008ApJ...673L.155V} {673, L155}

\bibitem[\protect\citeauthoryear{{Valenti} et~al.,}{{Valenti}
  et~al.}{2011}]{valenti2011sn2009jf}
{Valenti} S.,  et~al., 2011, \mn@doi [\mnras]
  {10.1111/j.1365-2966.2011.19262.x}, \href
  {https://ui.adsabs.harvard.edu/abs/2011MNRAS.416.3138V} {416, 3138}

\bibitem[\protect\citeauthoryear{{Williamson}, {Kerzendorf}  \&
  {Modjaz}}{{Williamson} et~al.}{2021}]{williamson2021modeling}
{Williamson} M.,  {Kerzendorf} W.,   {Modjaz} M.,  2021, \mn@doi [\apj]
  {10.3847/1538-4357/abd244}, \href
  {https://ui.adsabs.harvard.edu/abs/2021ApJ...908..150W} {908, 150}

\bibitem[\protect\citeauthoryear{{Wongwathanarat}, {Hammer}  \&
  {M{\"u}ller}}{{Wongwathanarat} et~al.}{2010a}]{wongwathanarat2010yinyang}
{Wongwathanarat} A.,  {Hammer} N.~J.,   {M{\"u}ller} E.,  2010a, \mn@doi [\aap]
  {10.1051/0004-6361/200913435}, \href
  {https://ui.adsabs.harvard.edu/abs/2010A&A...514A..48W} {514, A48}

\bibitem[\protect\citeauthoryear{{Wongwathanarat}, {Janka}  \&
  {M{\"u}ller}}{{Wongwathanarat}
  et~al.}{2010b}]{wongwathanarat2010hydrodynamical}
{Wongwathanarat} A.,  {Janka} H.-T.,   {M{\"u}ller} E.,  2010b, \mn@doi [\apjl]
  {10.1088/2041-8205/725/1/L106}, \href
  {https://ui.adsabs.harvard.edu/abs/2010ApJ...725L.106W} {725, L106}

\bibitem[\protect\citeauthoryear{{Wongwathanarat}, {Janka}  \&
  {M{\"u}ller}}{{Wongwathanarat} et~al.}{2013}]{wongwathanarat2013three}
{Wongwathanarat} A.,  {Janka} H.~T.,   {M{\"u}ller} E.,  2013, \mn@doi [\aap]
  {10.1051/0004-6361/201220636}, \href
  {https://ui.adsabs.harvard.edu/abs/2013A&A...552A.126W} {552, A126}

\bibitem[\protect\citeauthoryear{{Wongwathanarat}, {M{\"u}ller}  \&
  {Janka}}{{Wongwathanarat} et~al.}{2015}]{wongwathanarat2015three}
{Wongwathanarat} A.,  {M{\"u}ller} E.,   {Janka} H.~T.,  2015, \mn@doi [\aap]
  {10.1051/0004-6361/201425025}, \href
  {https://ui.adsabs.harvard.edu/abs/2015A&A...577A..48W} {577, A48}

\bibitem[\protect\citeauthoryear{{Wongwathanarat}, {Janka}, {M{\"u}ller},
  {Pllumbi}  \& {Wanajo}}{{Wongwathanarat}
  et~al.}{2017}]{wongwathanarat2017production}
{Wongwathanarat} A.,  {Janka} H.-T.,  {M{\"u}ller} E.,  {Pllumbi} E.,
  {Wanajo} S.,  2017, \mn@doi [\apj] {10.3847/1538-4357/aa72de}, \href
  {https://ui.adsabs.harvard.edu/abs/2017ApJ...842...13W} {842, 13}

\bibitem[\protect\citeauthoryear{{Woosley}}{{Woosley}}{2019}]{woosley2019evolution}
{Woosley} S.~E.,  2019, \mn@doi [\apj] {10.3847/1538-4357/ab1b41}, \href
  {https://ui.adsabs.harvard.edu/abs/2019ApJ...878...49W} {878, 49}

\bibitem[\protect\citeauthoryear{{Woosley} \& {Weaver}}{{Woosley} \&
  {Weaver}}{1995}]{woosley1995}
{Woosley} S.~E.,  {Weaver} T.~A.,  1995, \mn@doi [\apjs] {10.1086/192237},
  \href {https://ui.adsabs.harvard.edu/abs/1995ApJS..101..181W} {101, 181}

\bibitem[\protect\citeauthoryear{{Woosley}, {Heger}  \& {Weaver}}{{Woosley}
  et~al.}{2002}]{woosley2002evolution}
{Woosley} S.~E.,  {Heger} A.,   {Weaver} T.~A.,  2002, \mn@doi [Reviews of
  Modern Physics] {10.1103/RevModPhys.74.1015}, \href
  {https://ui.adsabs.harvard.edu/abs/2002RvMP...74.1015W} {74, 1015}

\bibitem[\protect\citeauthoryear{{Yaron} \& {Gal-Yam}}{{Yaron} \&
  {Gal-Yam}}{2012}]{wiserep2012}
{Yaron} O.,  {Gal-Yam} A.,  2012, \mn@doi [\pasp] {10.1086/666656}, \href
  {https://ui.adsabs.harvard.edu/abs/2012PASP..124..668Y} {124, 668}

\makeatother
\end{thebibliography}




\appendix

\section{Progenitor data} \label{app:progdata}
\begin{figure}
    \centering
    \includegraphics[width=.98\linewidth]{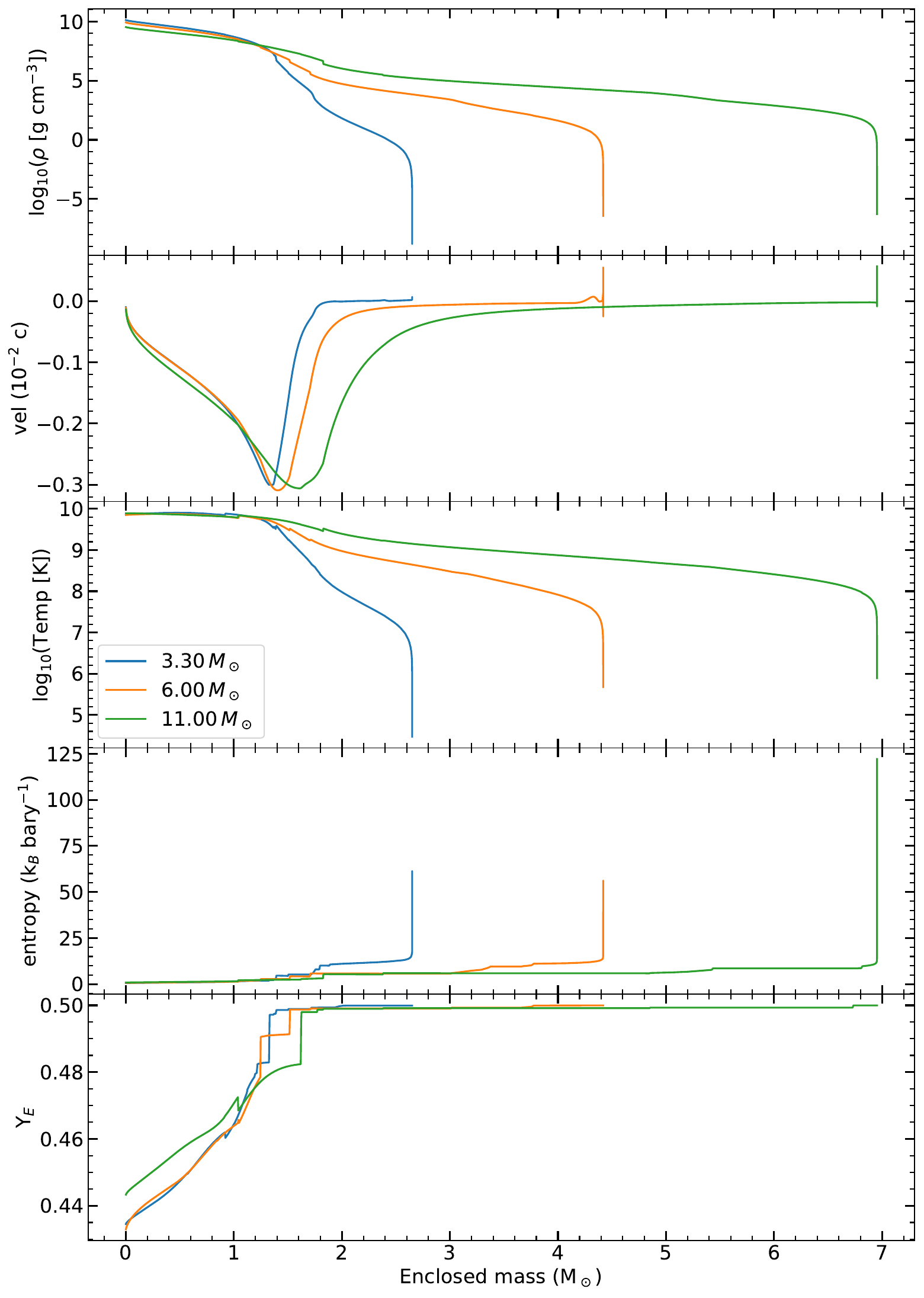}
    \caption{Radial profiles along the mass coordinate for the three progenitor stars; the $3.3\,M_\odot$ He-core in blue, the $6.0\,M_\odot$ He-core in orange and the $11.0\,M_\odot$ in green, all at the onset of iron-core collapse. From top to bottom, the density, velocity, temperature, entropy and electron fraction (Y$_E=n_e/n_\text{baryon}$) are shown. Note that the density and temperature panels are shown on logarithmic scales.}
    \label{fig:appendix-progenitors}
\end{figure}
The structure of the progenitor is a critical detail in the modelling of the core-collapse and thus critical for our late-time modelling as these core-collapse models form the input for \texttt{ExTraSS}. In Figure \ref{fig:appendix-progenitors} some additional physical details are shown for the interior structure of the three progenitor models (each of the $3.3\,M_\odot$, $6.0\,M_\odot$ and $11.0\,M_\odot$ models is based of these same progenitor stars which are exploded with a different energy). The profiles for the $3.3\,M_\odot$ star were previously also shown in the Appendix of \citet{vanbaal2023modelling}.

\section{\texttt{P-HotB} details $\&$ more 3D renderings} \label{app:BonusRenderings}
\begin{figure*}
    \centering
    \includegraphics[width=.32\linewidth]{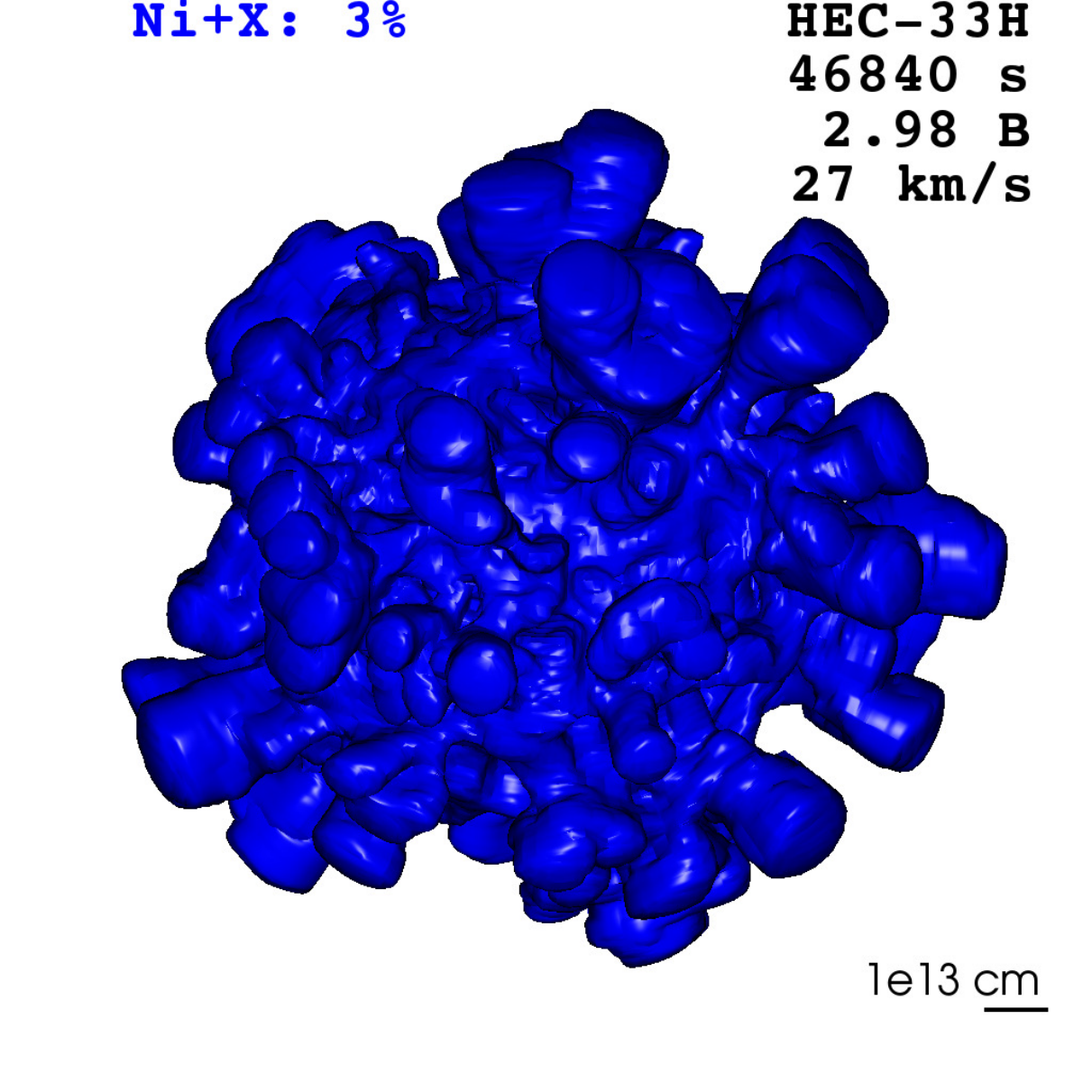}   
    \includegraphics[width=.32\linewidth]{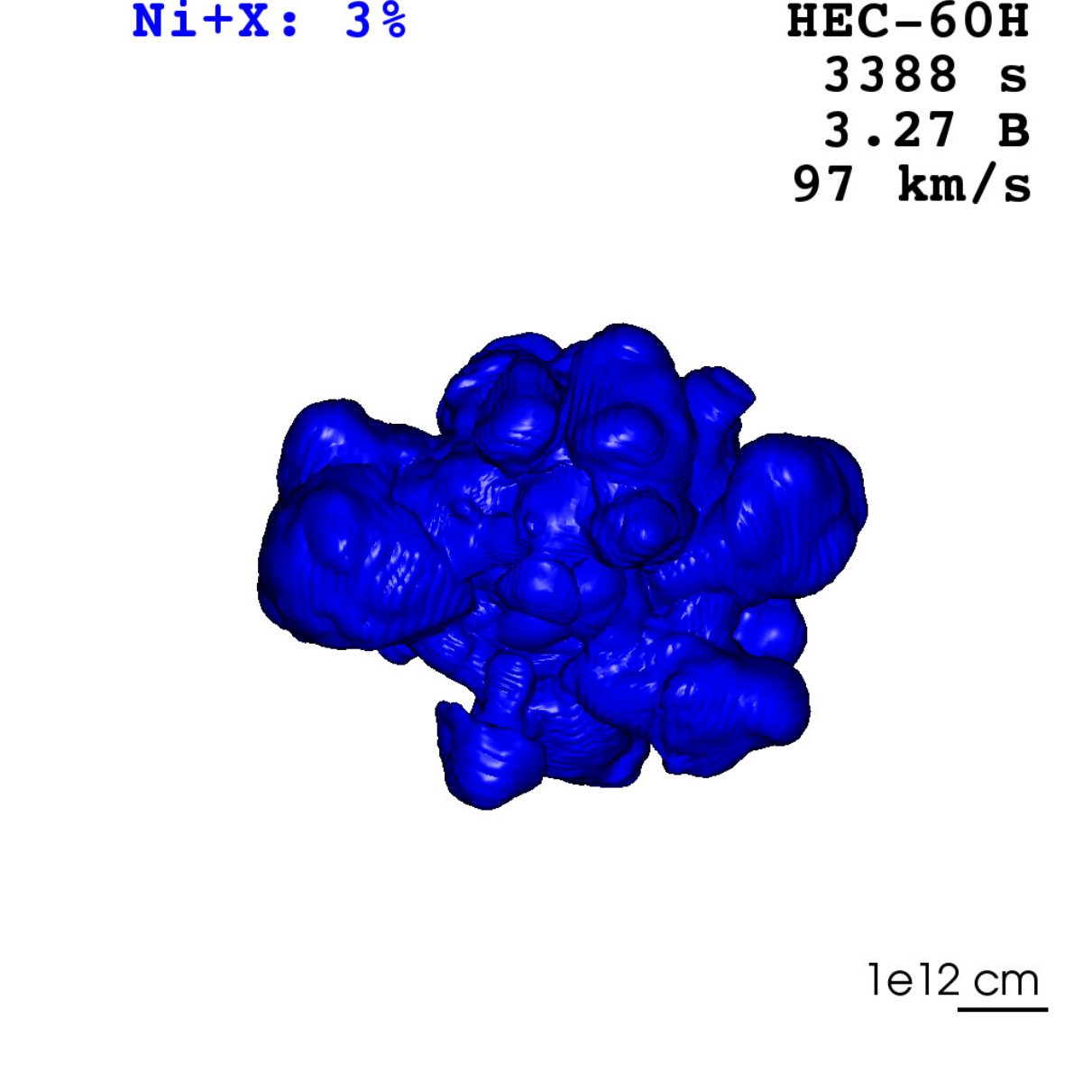}   
    \includegraphics[width=.32\linewidth]{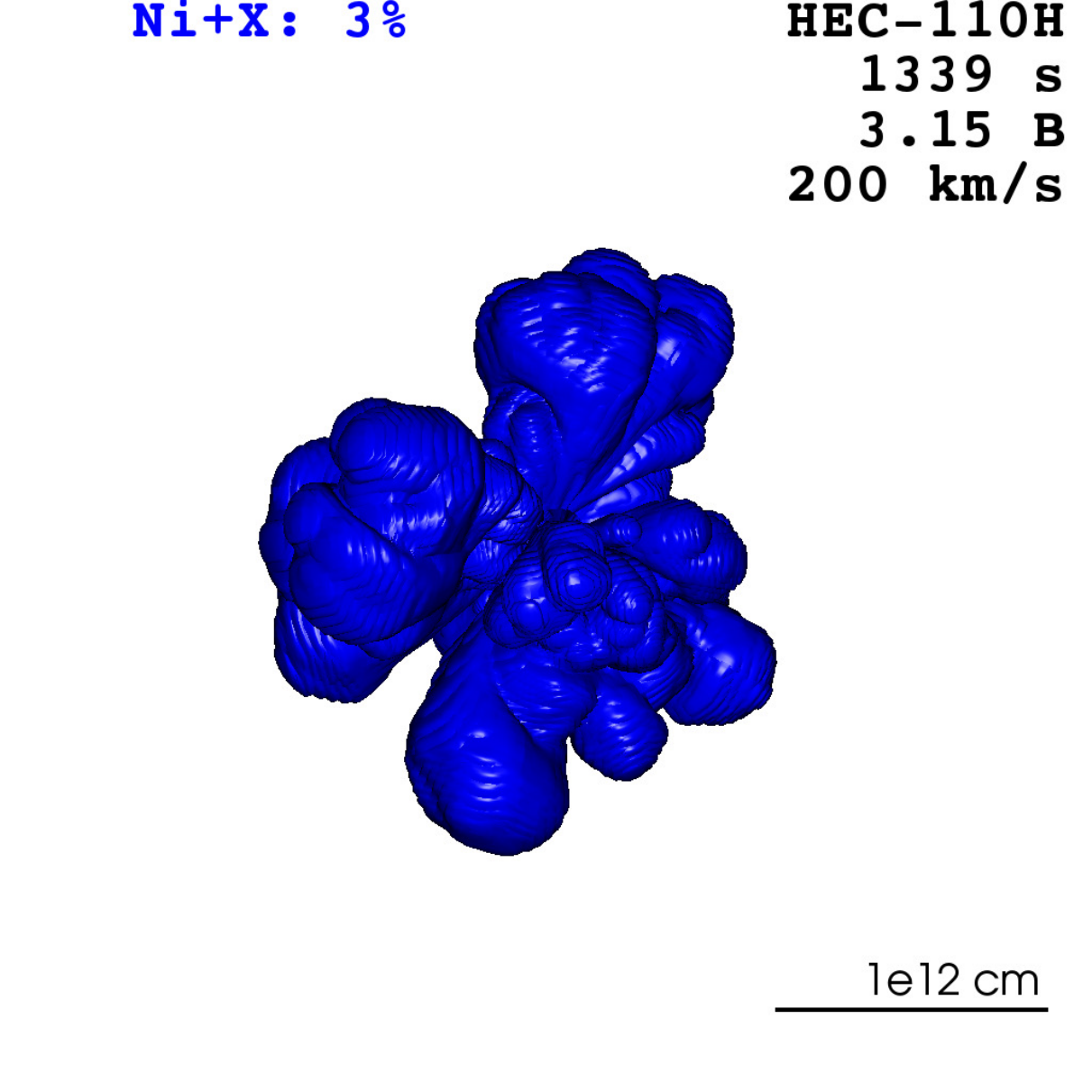}   
    \includegraphics[width=.32\linewidth]{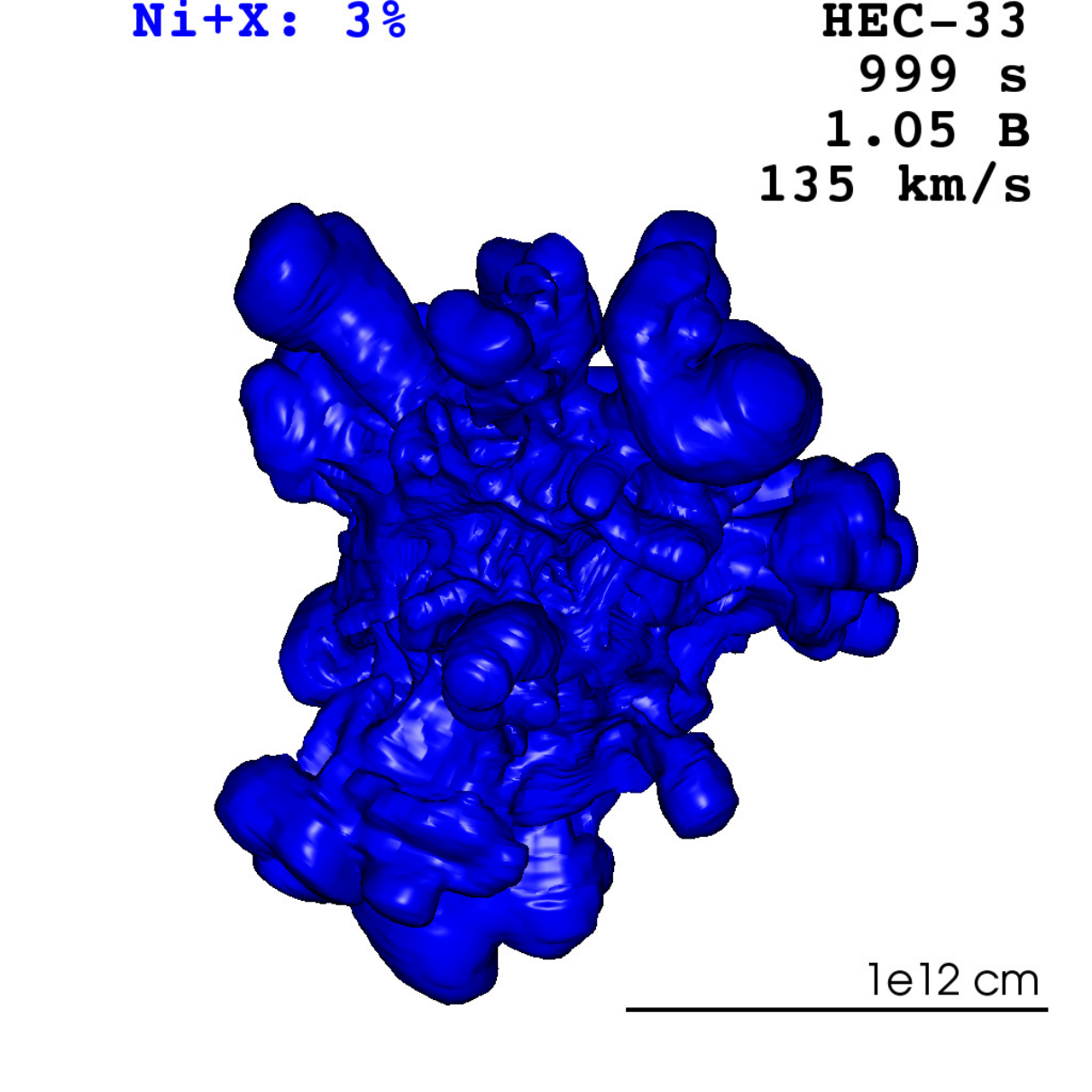}   
    \includegraphics[width=.32\linewidth]{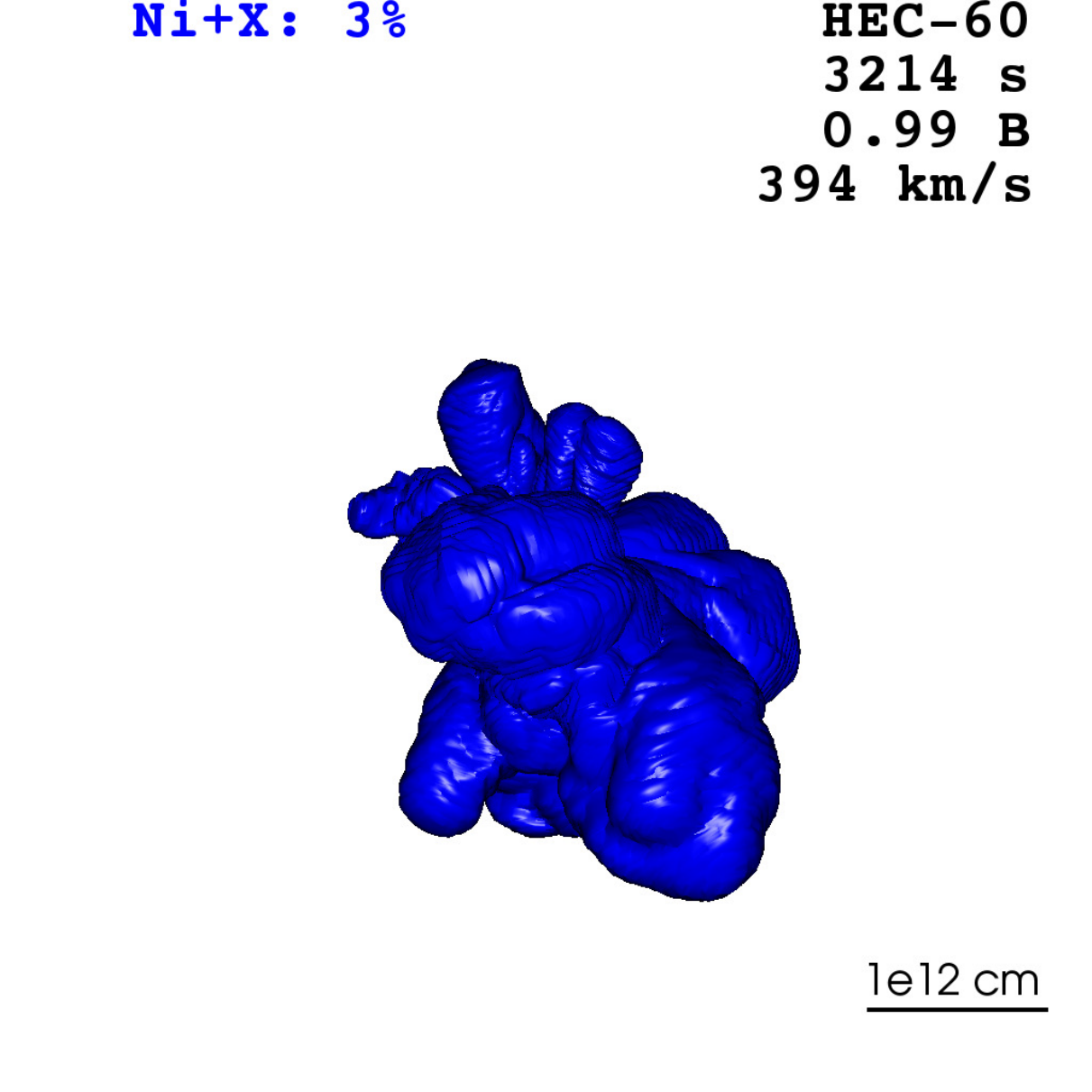}   
    \includegraphics[width=.32\linewidth]{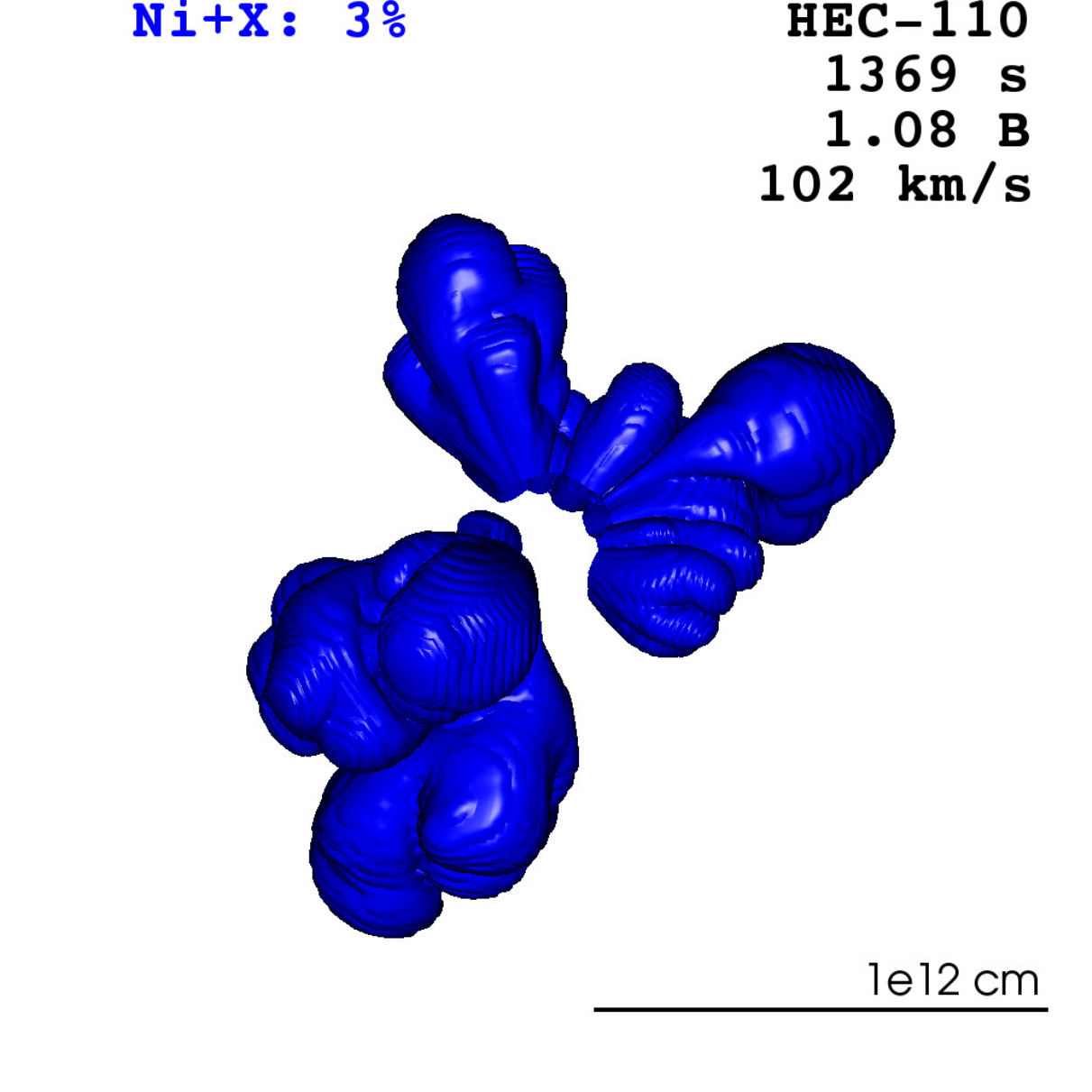}   
    \includegraphics[width=.32\linewidth]{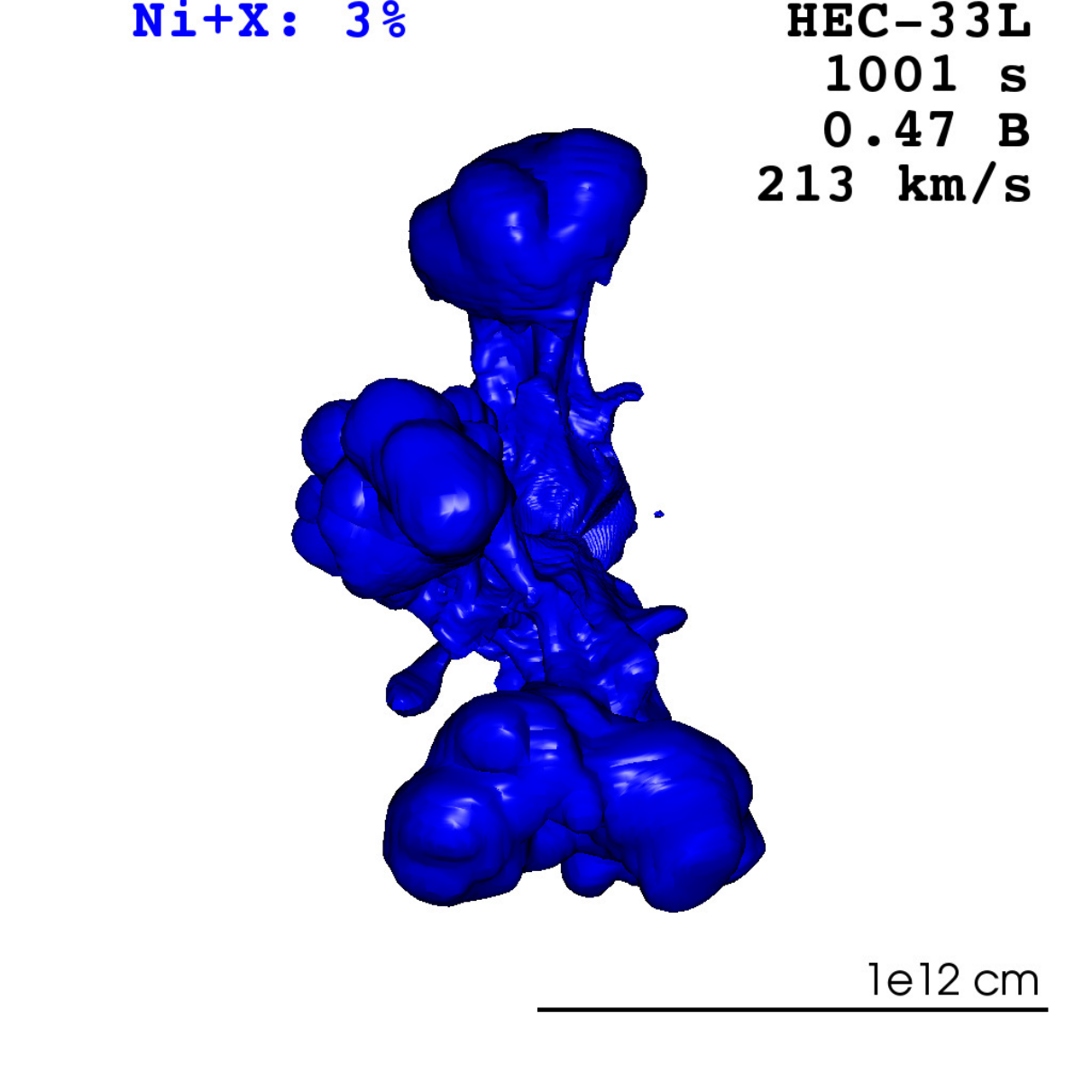}   
    \includegraphics[width=.32\linewidth]{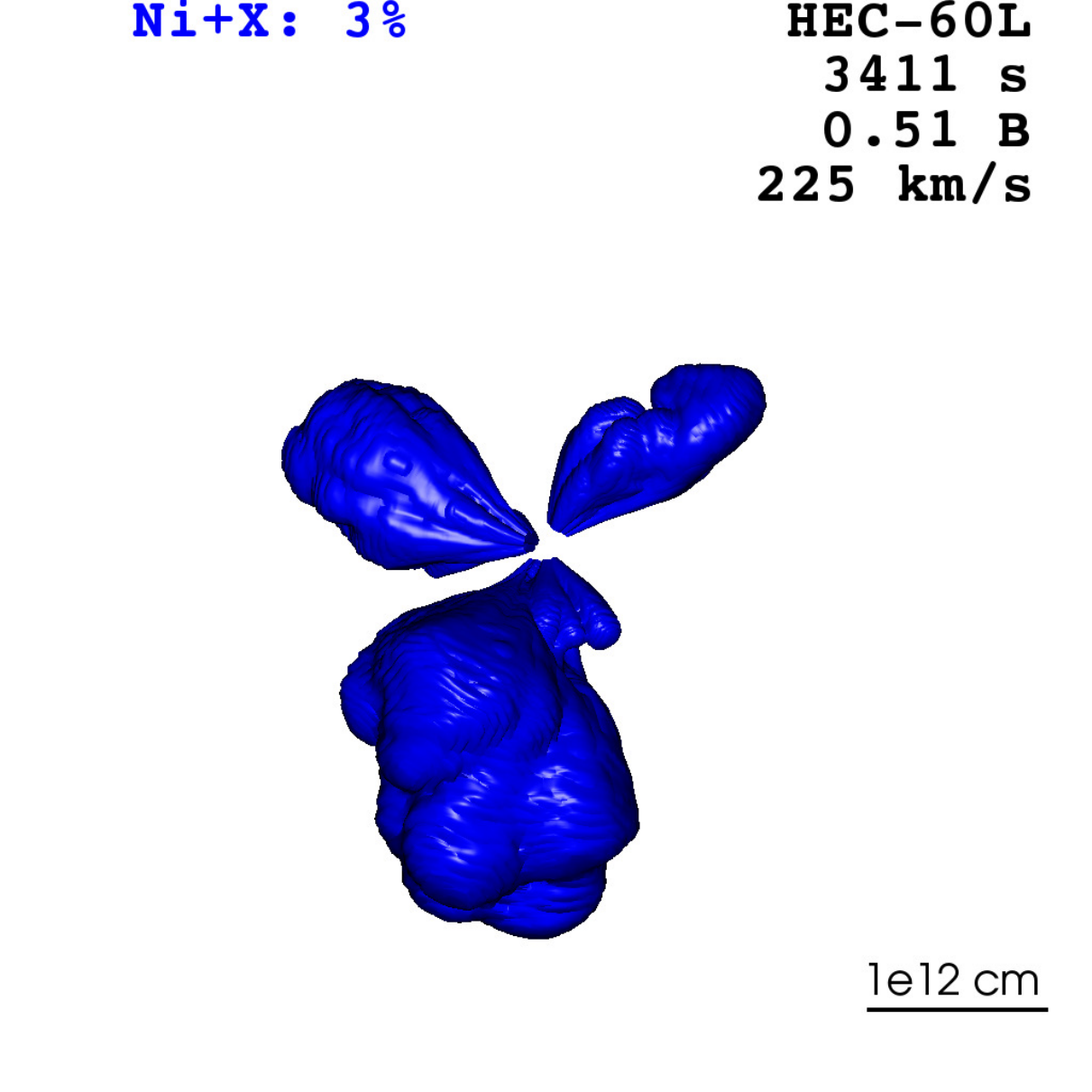}   
    \includegraphics[width=.32\linewidth]{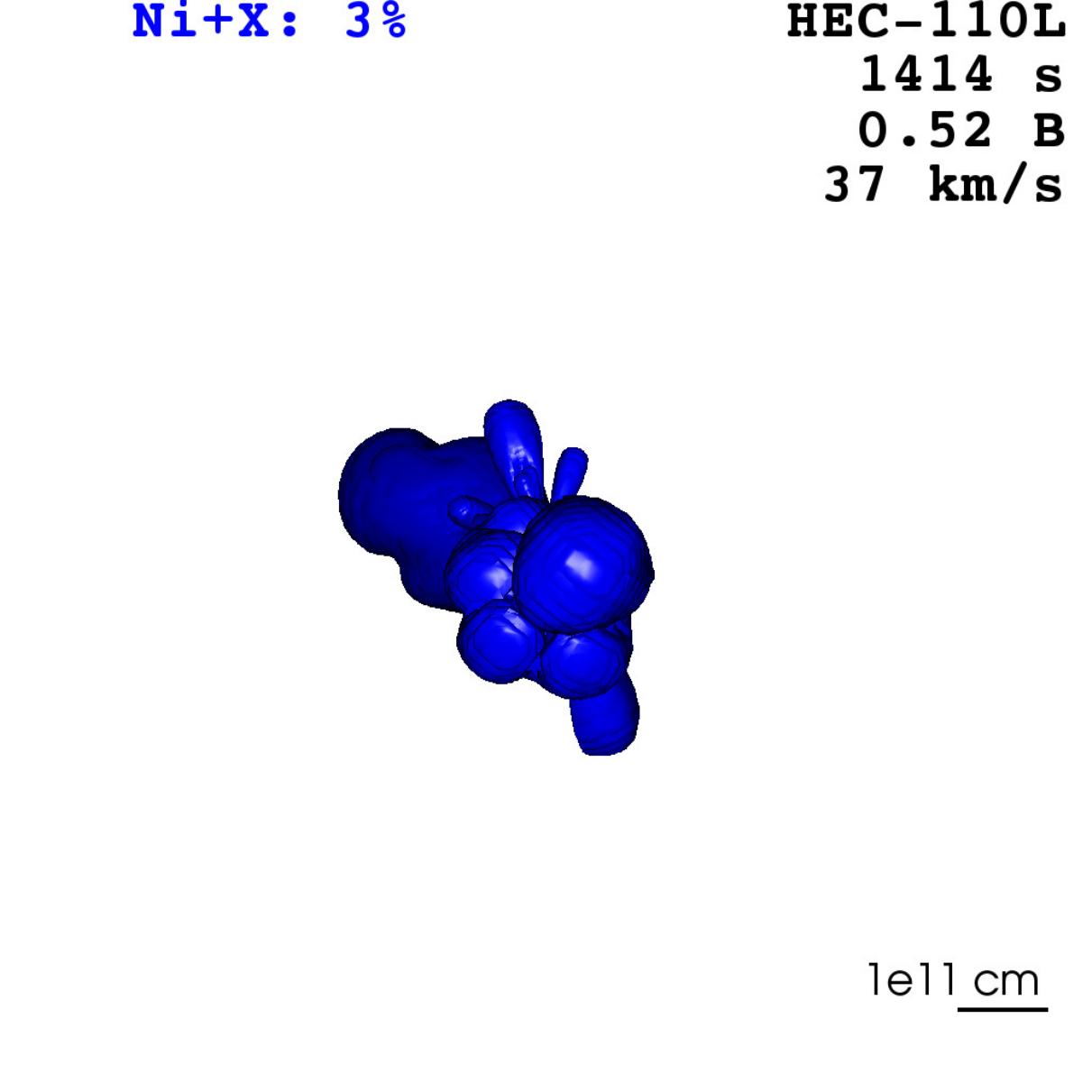}   
    \caption{The 3D $^{56}$Ni-only renderings of all explosion models, as shown in Figure \ref{fig:3D-Rendering}. The top-right corner in each panel denotes the final \texttt{P-HotB} time, explosion energy and compact object kick velocity. The orientations are the same as in Figure \ref{fig:3D-Rendering}.}
    \label{fig:3D-Rendering_NiOnly}
\end{figure*}
\begin{figure*}
    \centering
    \includegraphics[width=.32\linewidth]{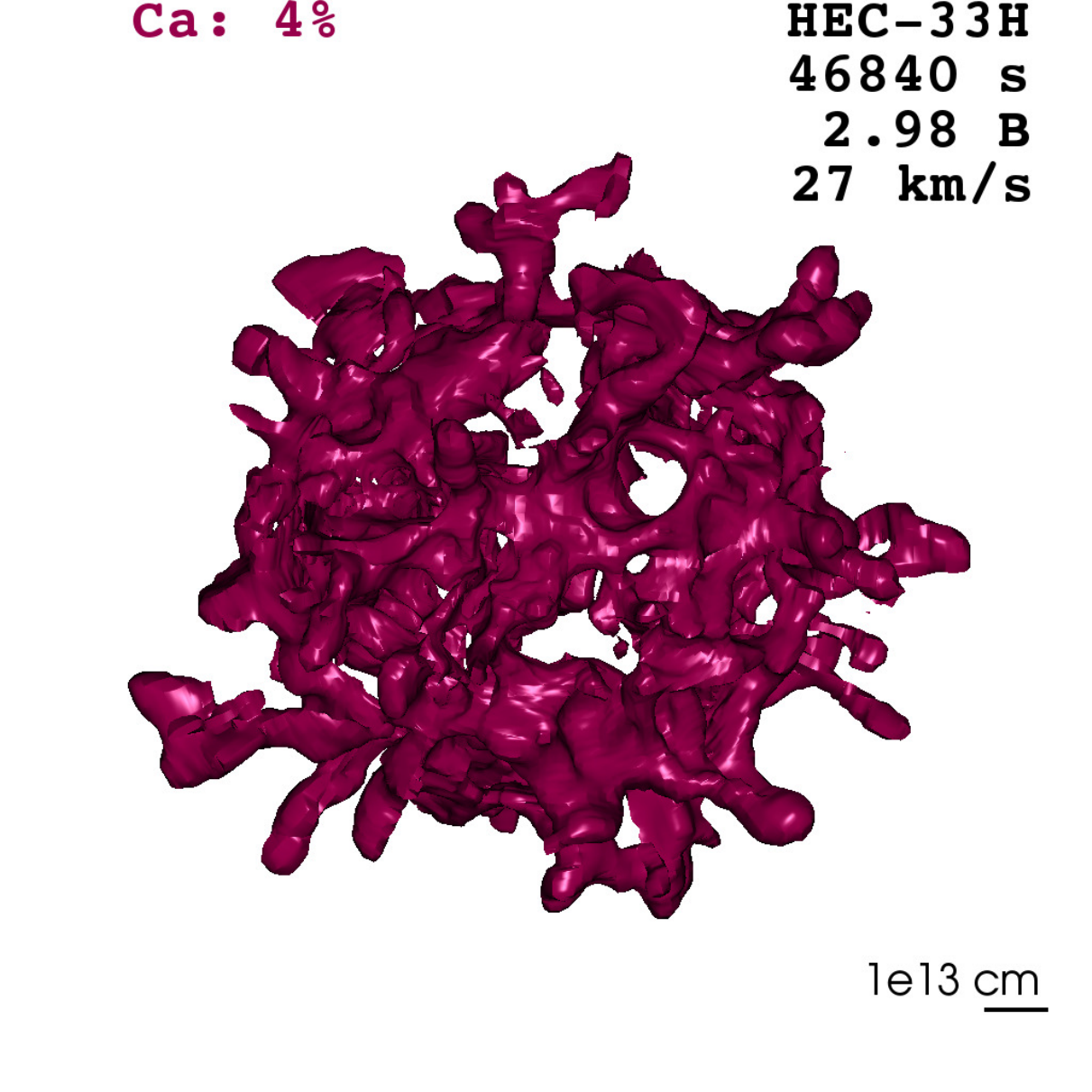}   
    \includegraphics[width=.32\linewidth]{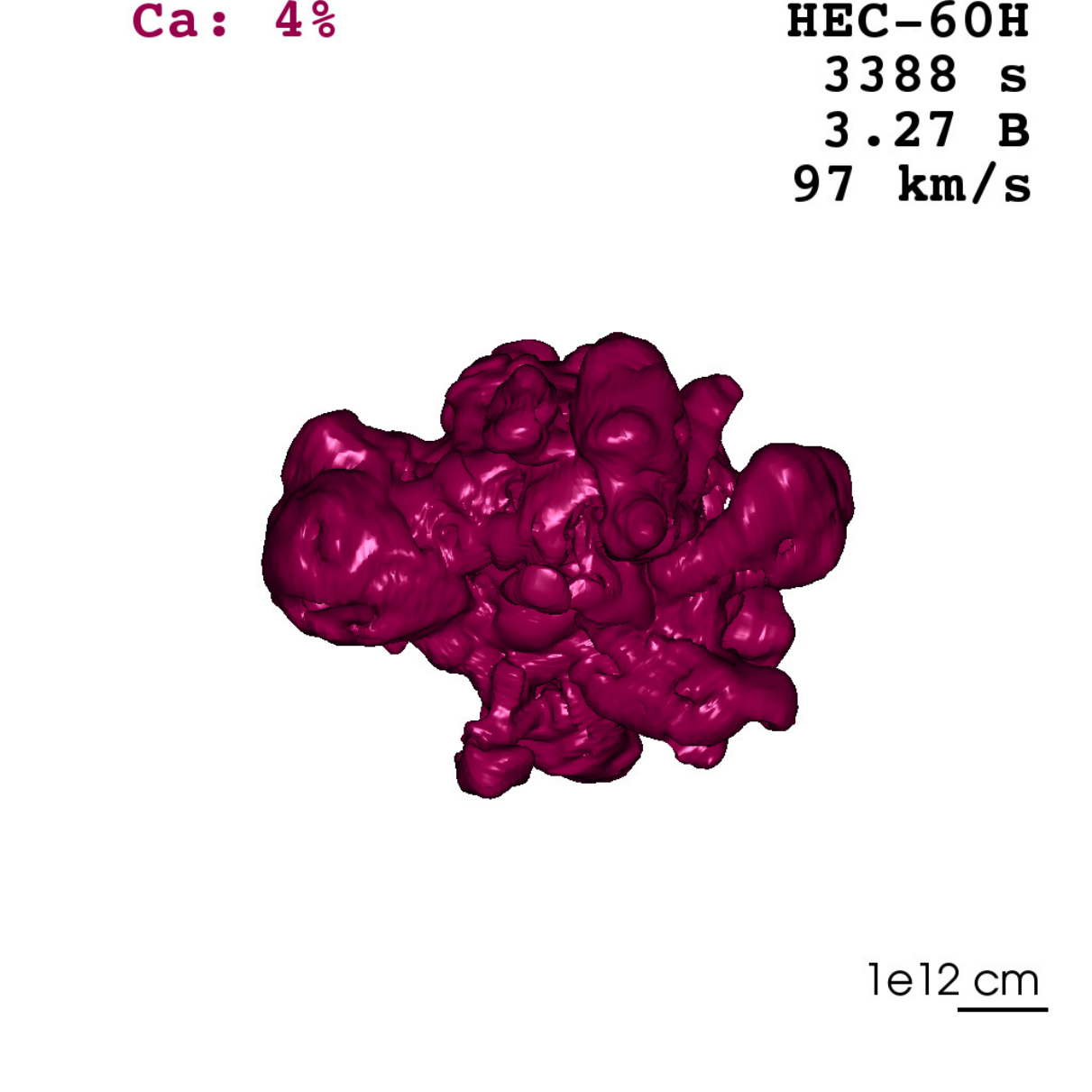}   
    \includegraphics[width=.32\linewidth]{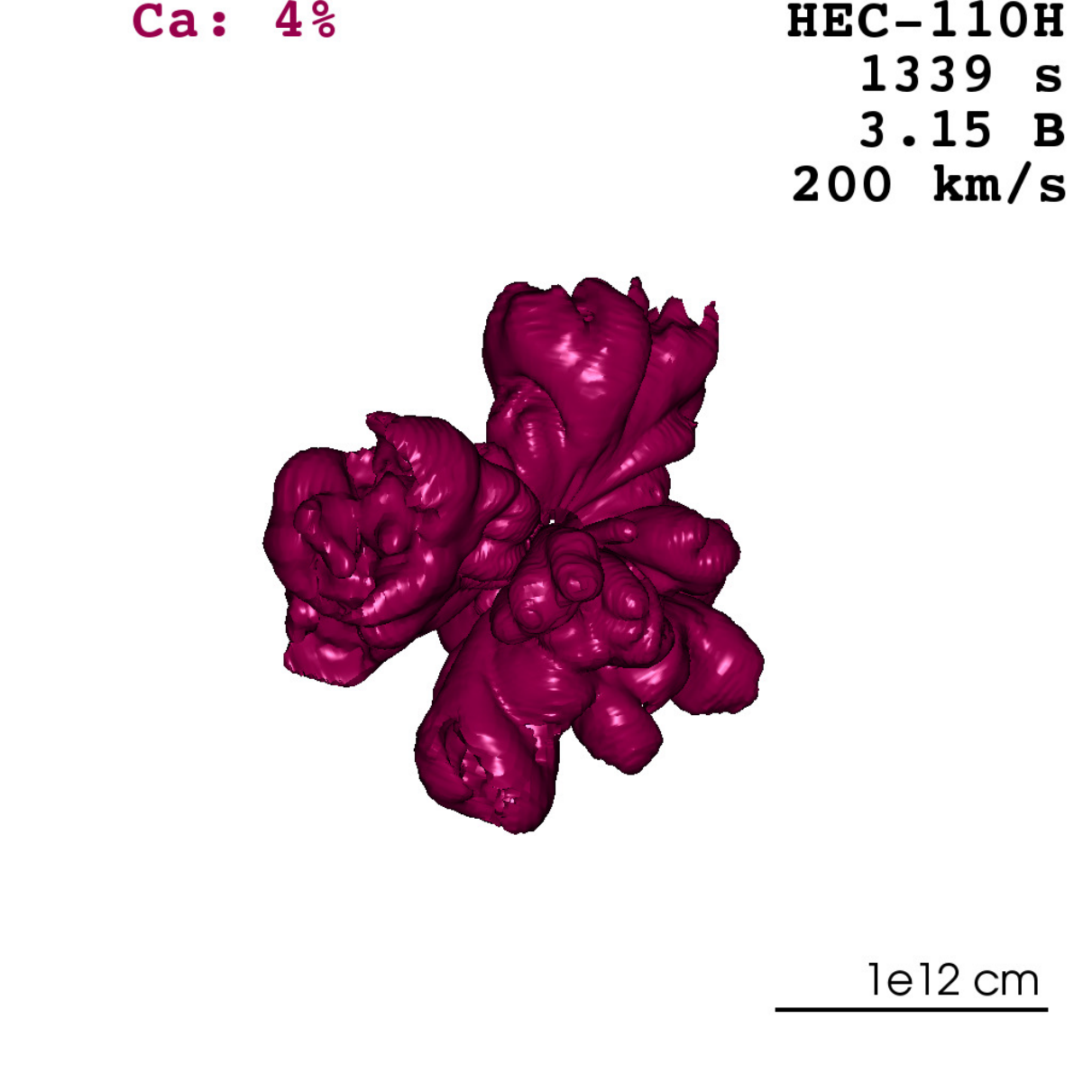}   
    \includegraphics[width=.32\linewidth]{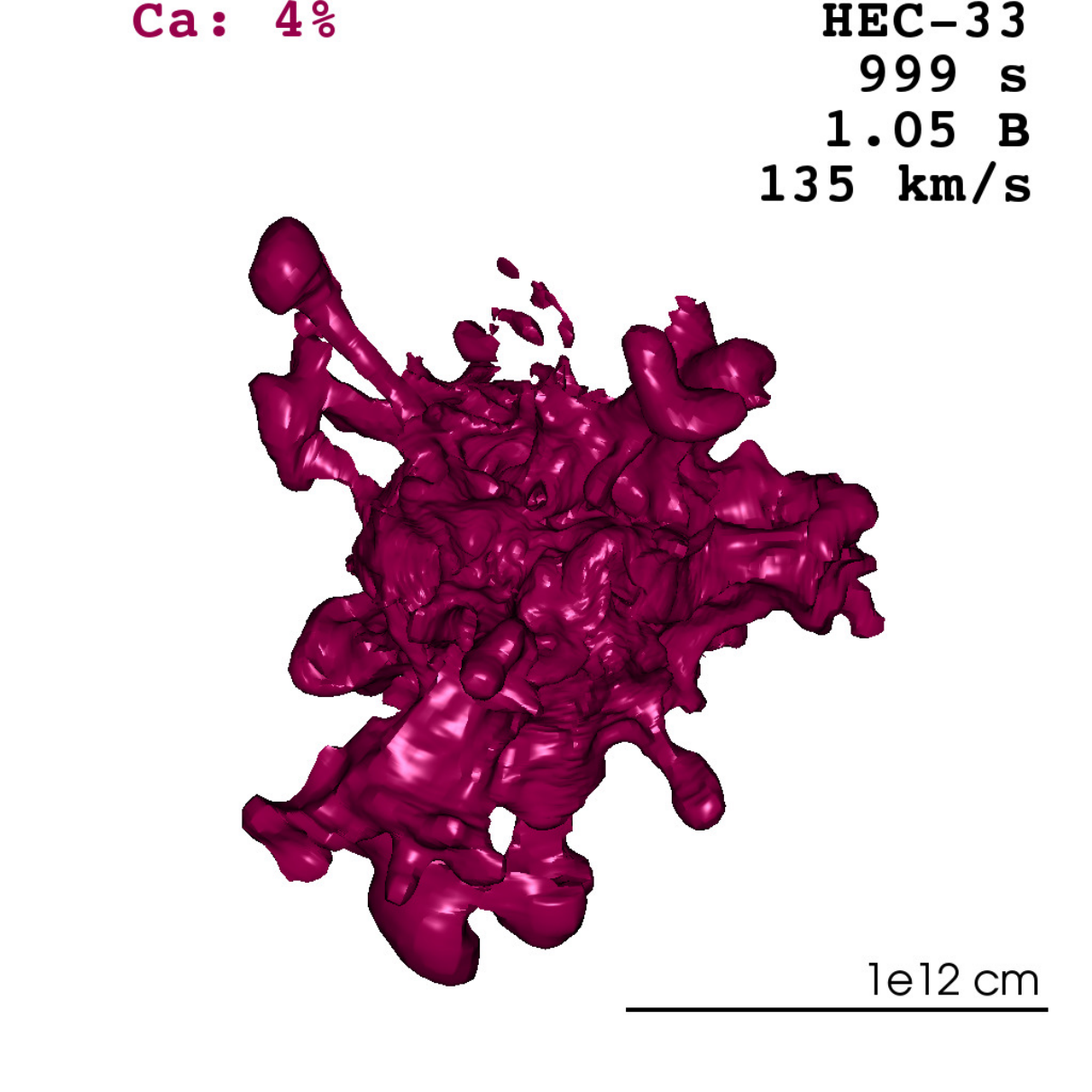}   
    \includegraphics[width=.32\linewidth]{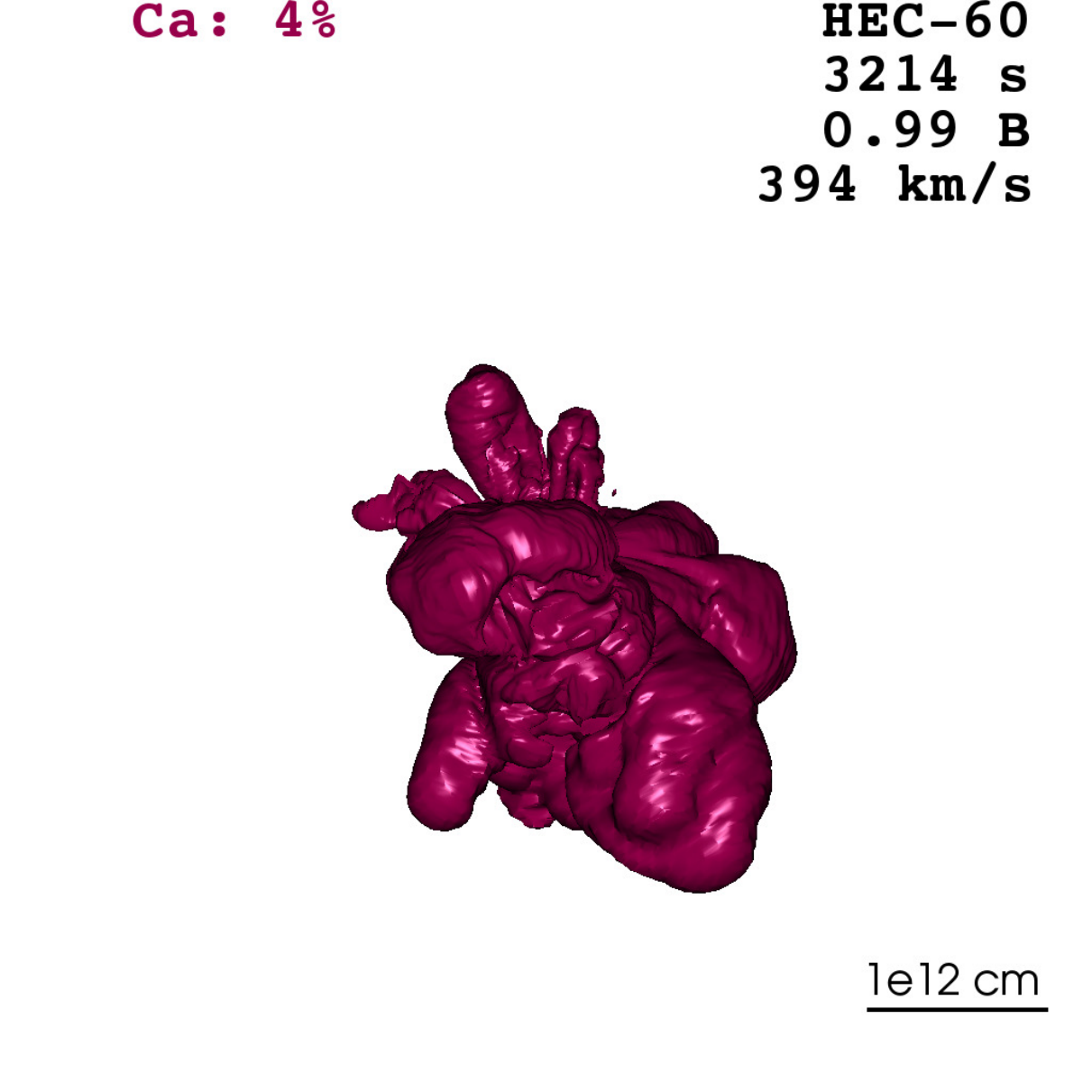}   
    \includegraphics[width=.32\linewidth]{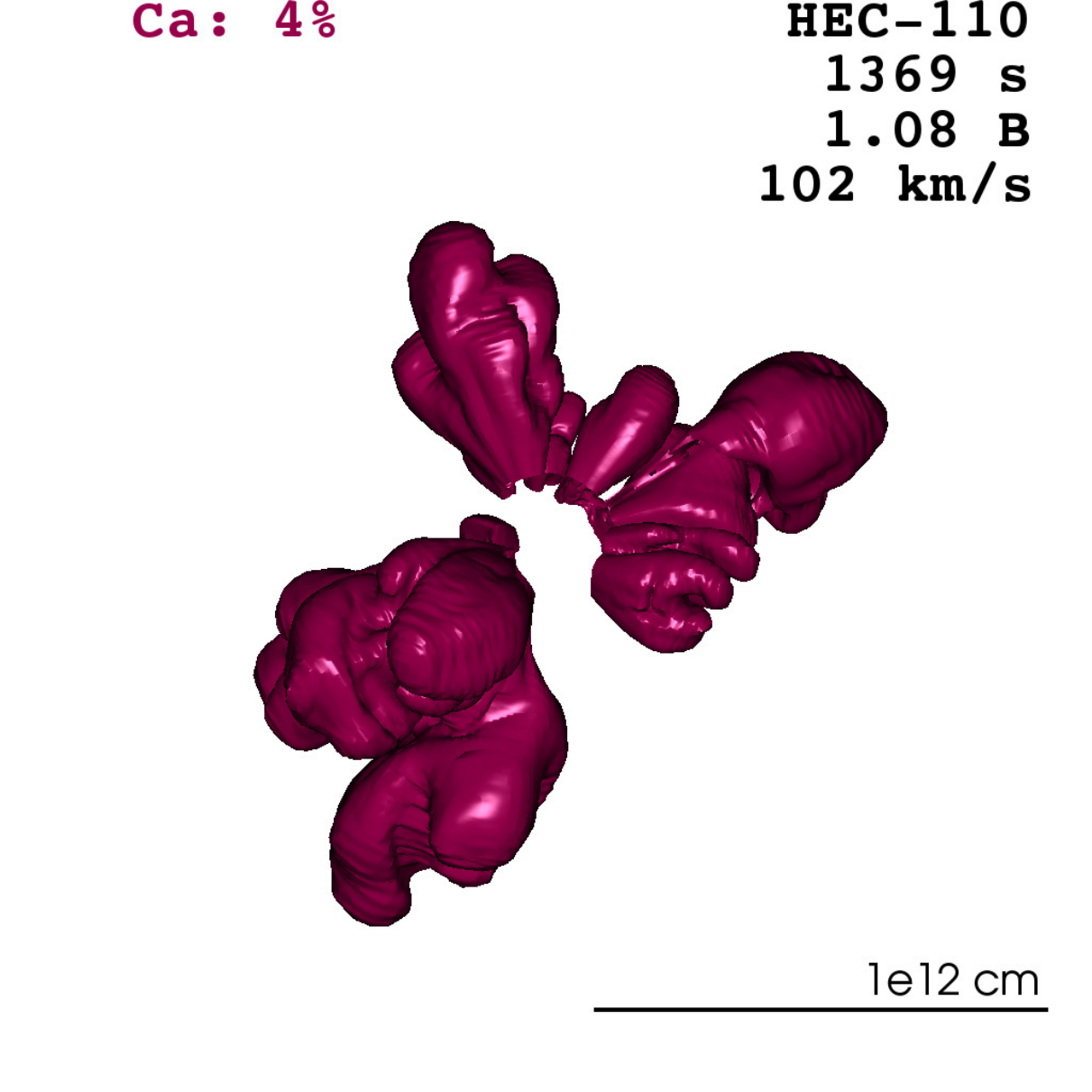}   
    \includegraphics[width=.32\linewidth]{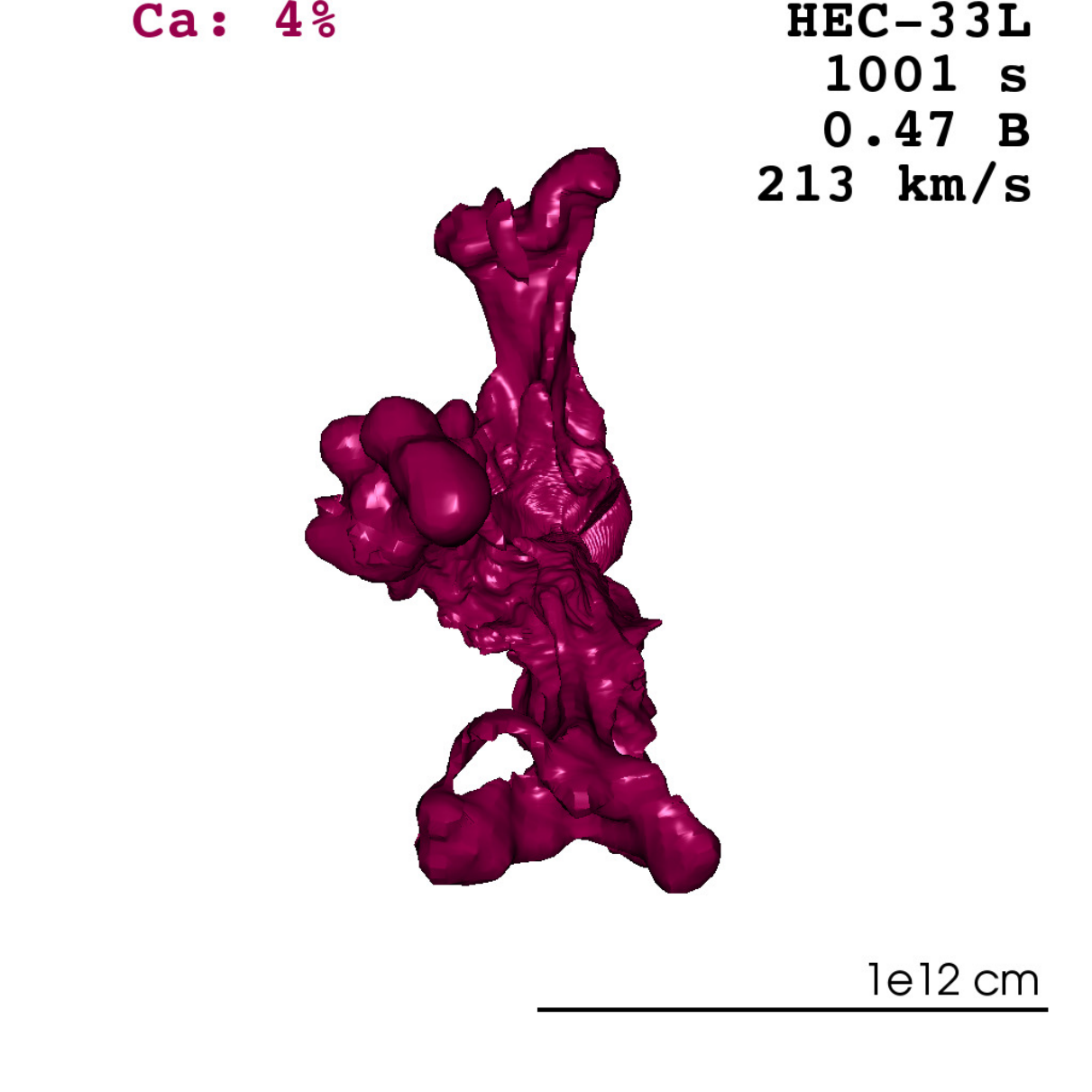}   
    \includegraphics[width=.32\linewidth]{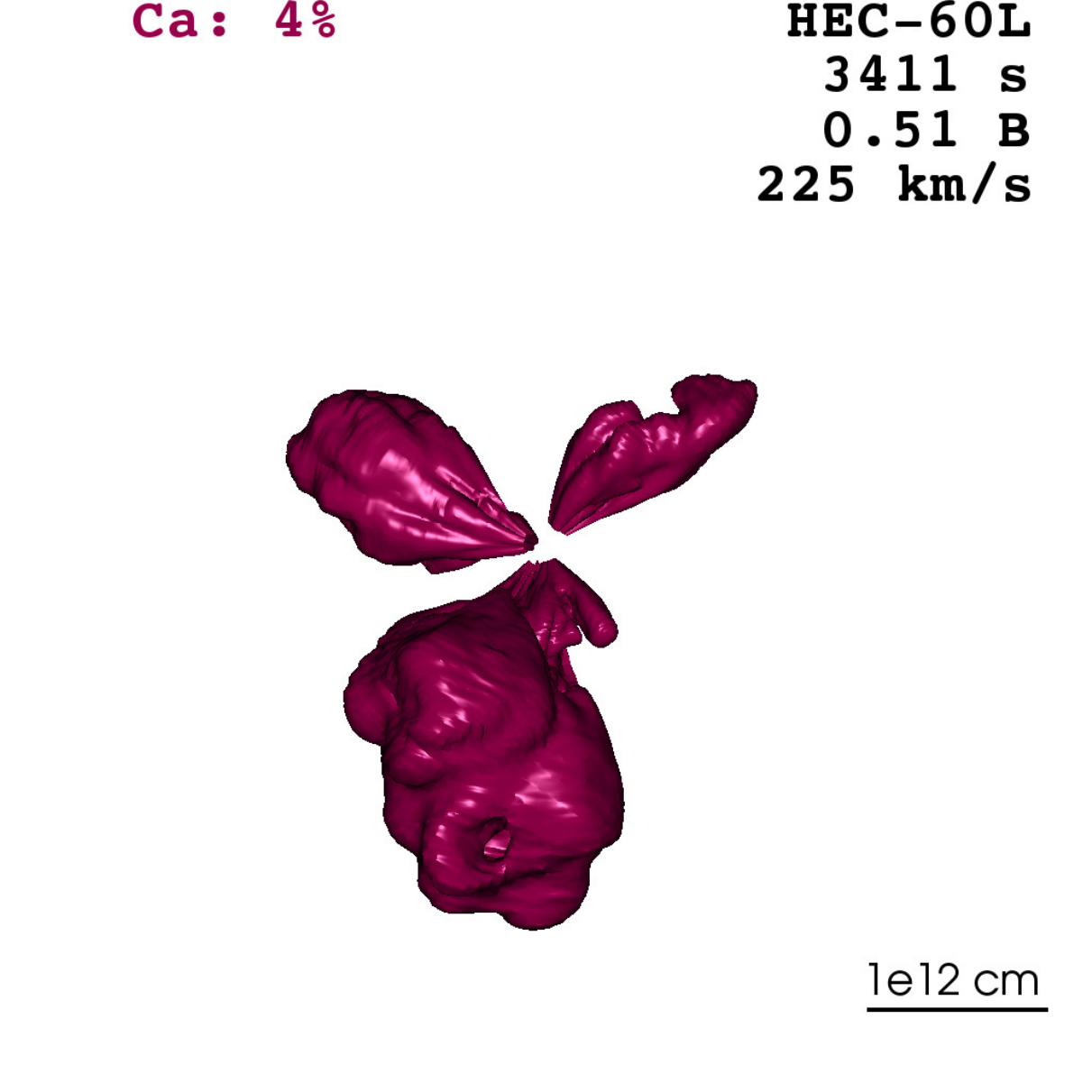}   
    \includegraphics[width=.32\linewidth]{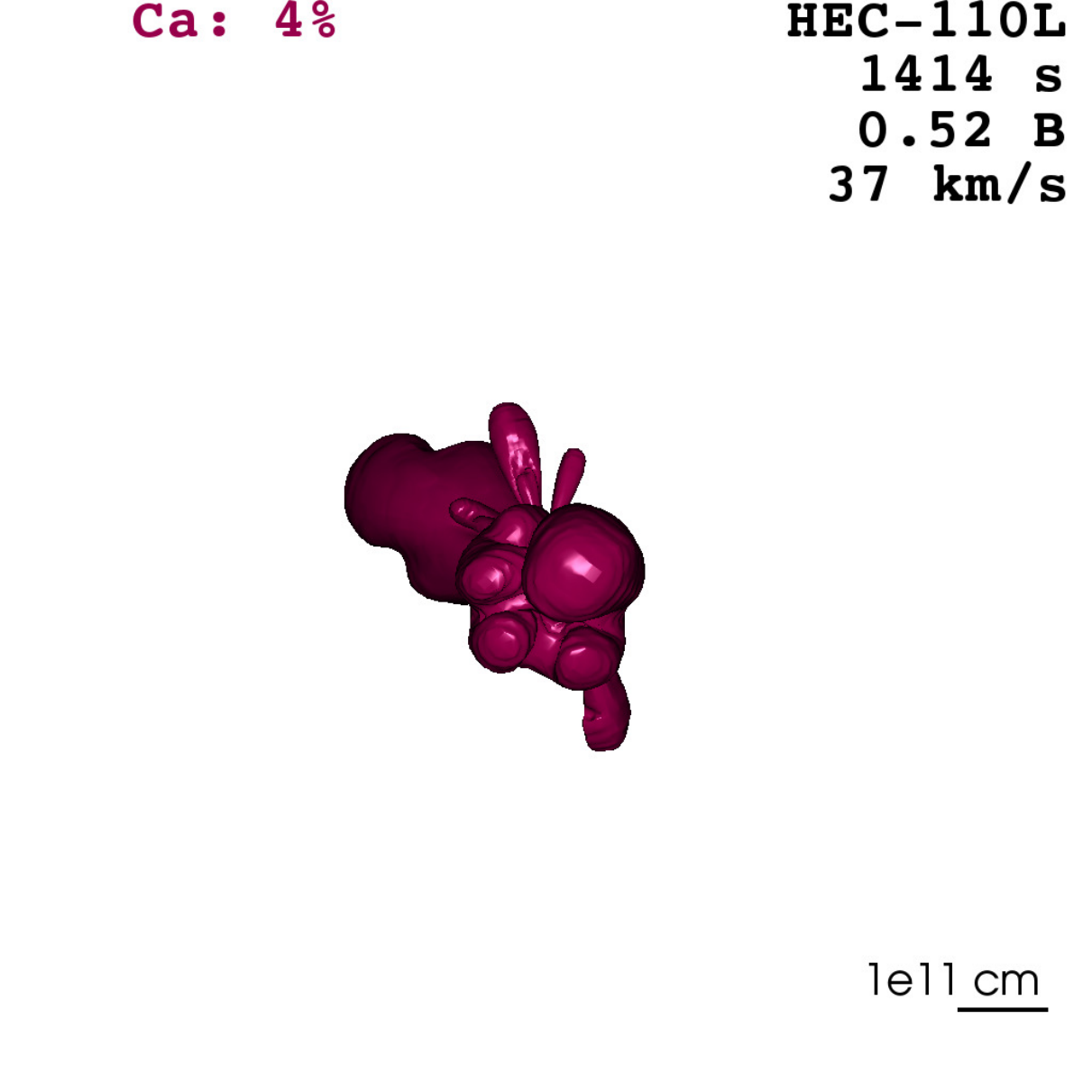}   
    \caption{The 3D $^{40}$Ca-only renderings of all explosion models, complementary to Figure \ref{fig:3D-Rendering}. The top-right corner in each panel denotes the final \texttt{P-HotB} time, explosion energy and compact object kick velocity. The orientations are the same as in Figure \ref{fig:3D-Rendering}.}
    \label{fig:3D-Rendering_CaOnly}
\end{figure*}

In this appendix we list additional 3D renderings from our explosion models, to further highlight asymmetries inside the ejecta, and we expand in more detail on \texttt{P-HotB}, the code used to explode our progenitor stars. In Figure \ref{fig:3D-Rendering_NiOnly} the renderings from the $3\%$ isosurfaces of $^{56}$Ni are shown, under the same orientation as in Figure \ref{fig:3D-Rendering}; i.e. the neutron star movement is directly upwards in each panel. In Figure \ref{fig:3D-Rendering_CaOnly} the 3D renderings of the $4\%$ isosurfaces of $^{40}$Ca are shown for all models, under the same orientation as in Figure \ref{fig:3D-Rendering}.

Neutrino-driven SN explosions were initiated by neutrino energy deposition in the layer behind the stalled core-bounce shock. For this purpose a gray approximation of the neutrino transport in the subnuclear near-surface layers of the proto-neutron star (PNS) was used, while the central, high-density core of the PNS was excluded from the axis-free computational polar grid (Yin-Yang grid; \citealt{kageyama2004yinyang,wongwathanarat2010yinyang}). Instead, an inner grid boundary was placed at a time-dependent radius connected to a Lagrangian mass shell, whose inward motion was chosen to describe the slow contraction behavior of the cooling PNS (for details, see \citealt{scheck2006multidimensional,arcones2007nucleosynthesis}). The three different explosion energies ($0.5\,$B, $1\,$B and $3\,$B) were obtained by changing the time-dependent neutrino luminosities at the inner grid boundary, which allows us to achieve explosion energies close to the desired values. The progenitor models from \citet{woosley2019evolution} are 1D, and the symmetry in these is broken roughly $10\,$ms after the core bounce. This is done by applying small-amplitude perturbations (randomized cell-by-cell), which become the seeds that grow into the post-shock instabilities in the neutrino-heating layer \citep[see also][]{wongwathanarat2010hydrodynamical,wongwathanarat2013three}.

About one second after the explosion was initiated by neutrino heating, the simulations were mapped to a larger spatial grid covering the entire star and using an adequate description of the low-density equation of state. For the subsequent long-time runs over hours to days, the time-stepping constraints connected to the high densities and neutrino treatment in the near-surface layers of the PNS were circumvented by replacing this central region by an inner grid boundary relocated to $500\,$km. A time-dependent inflow of neutrino-heated wind matter was prescribed as an inner boundary condition for several seconds (i.e. for the typical duration of the neutrino cooling of the hot PNS until neutrino transparency), before the boundary was opened by setting a free outflow condition that was stepwise moved to larger radii \citep[for details, see][]{wongwathanarat2013three,wongwathanarat2015three}.

During our simulations, nucleosynthesis in the expanding neutrino-heated and shock-heated SN ejecta was approximated by a small $\alpha-$network. It consists of the 13 $\alpha-$nuclei ($^4$He, $^{12}$C, $^{16}$O, $^{20}$Ne, $^{24}$Mg, $^{28}$Si, $^{32}$S, $^{36}$Ar, $^{40}$Ca, $^{44}$Ti, $^{48}$Cr, $^{52}$Fe and $^{56}$Ni) plus an additional 'tracer nucleus' X, to which we channel the flow resulting from the reaction $^{52}$Fe$(\alpha,\gamma)^{56}$Ni in the case of the electron fraction $Y_e=n_e/n_b$ (electron density divided by baryon density) drops below 0.490 and $^{56}$Ni is not the dominant nucleus synthesized in the iron group. The tracer species X thus tracks the production of neutron-rich (non-alpha) nuclei in ejecta that freeze out from nuclear statistical equilibrium (NSE) under the condition of neutron excess. It allows us to distinguish neutron-rich material from $^{56}$Ni, whose yield would ltherwise be overestimated. The 14-species network was solved in all grid cells with temperatures between $10^8-8\times10^9\,$K. When NSE freezes out in ejecta that cool down from above $8\times10^9\,$K, the network is started from pure $\alpha$ composition, which is compatible with the high-temperature NSE composition in the absence of free neutrons and protons, in particular since $Y_e$ has values close to 0.5 in all of the ejecta. As the hot ejecta expand and their temperature decreases these $\alpha-$particles recombine to produce heavier $\alpha-$nuclei considered in our network, among them $^{56}$Ni. Once the temperature drops below $10^8\,$K, all nuclear reactions are switched off, as they become too slow to change the nuclear composition on the explosion timescale. Nuclear burning in shock-heated matter is also described by our $\alpha-$network. For details of the treatment of nuclear composition and equation of state during the neutrino-heating simulations and the subsequent long-time evolution calculations, see \citet{kifonidis2003non} and \citet{wongwathanarat2013three,wongwathanarat2015three}.

\section{Physical conditions} \label{app:PhysicalCon}
\begin{figure*}
    \centering
    \includegraphics[width=.32\linewidth]{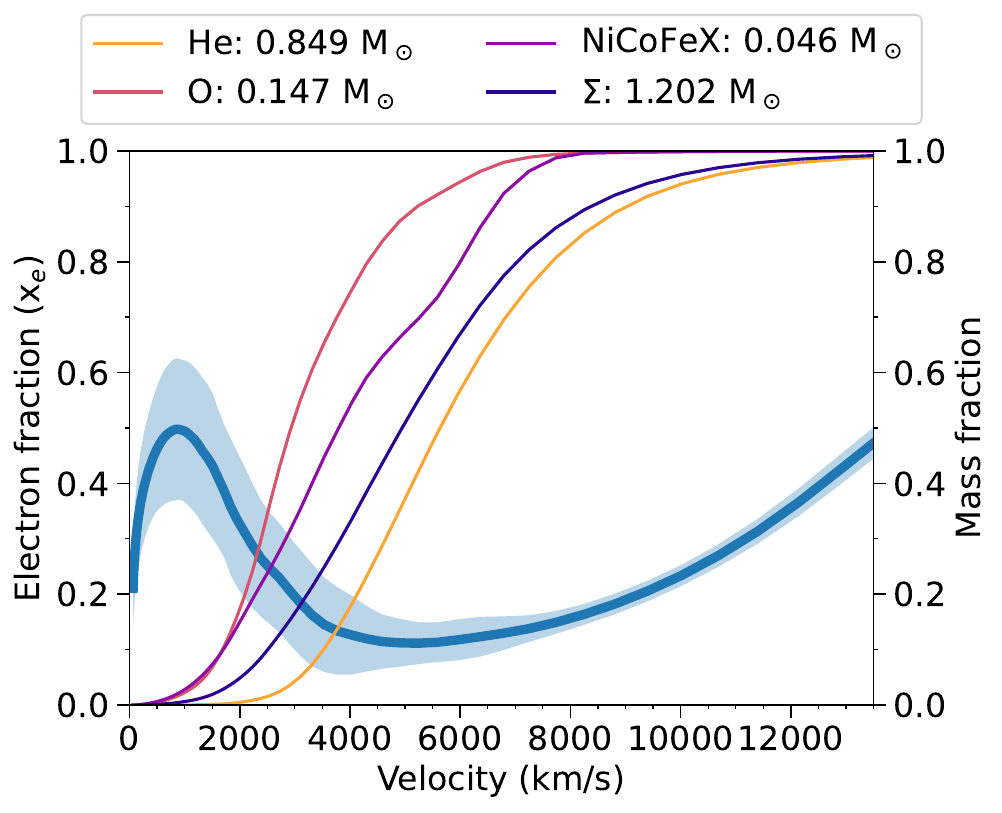}
    \includegraphics[width=.32\linewidth]{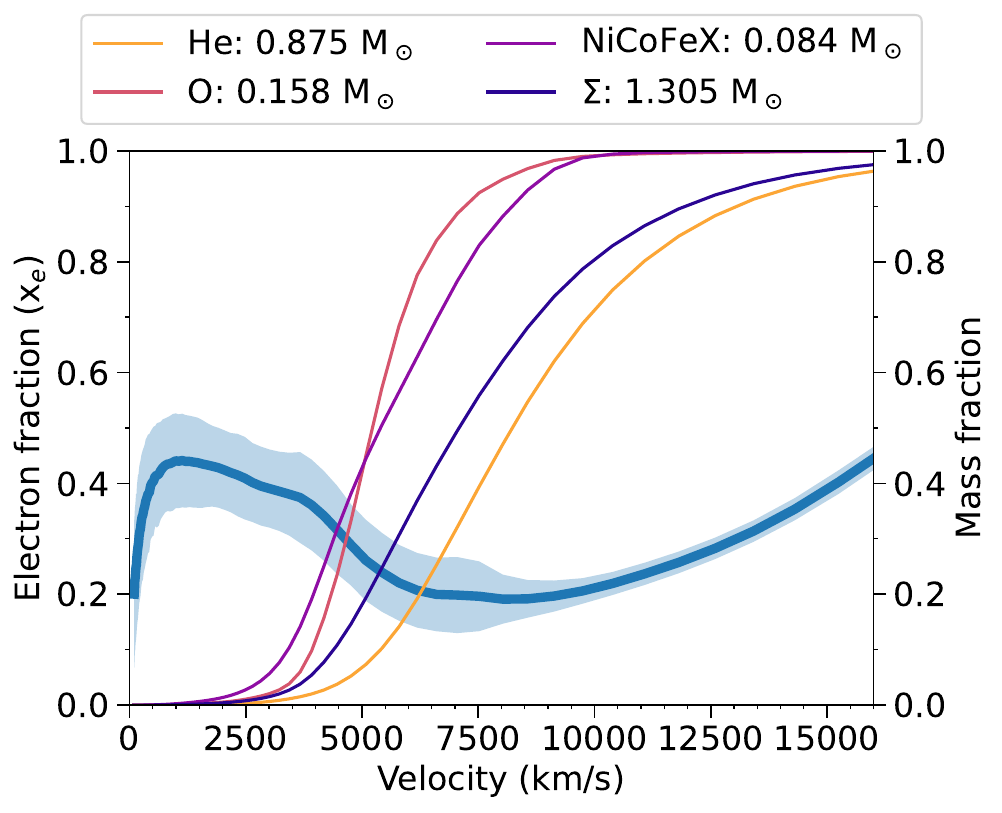}
    \includegraphics[width=.32\linewidth]{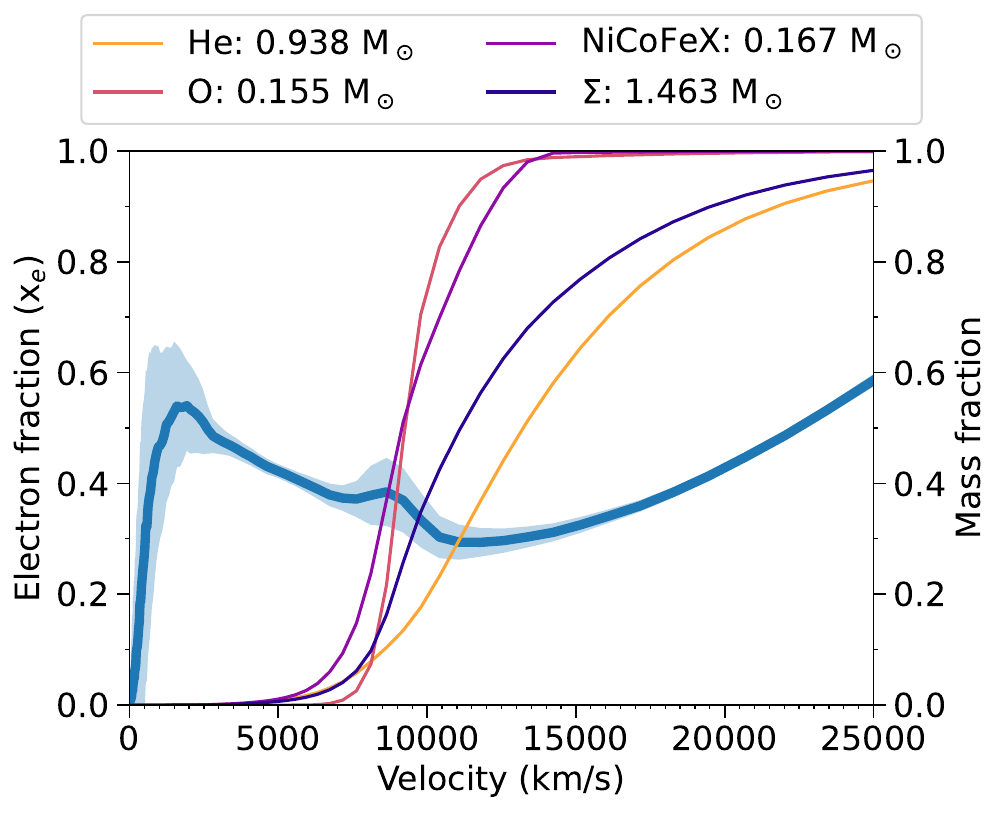}
    \includegraphics[width=.32\linewidth]{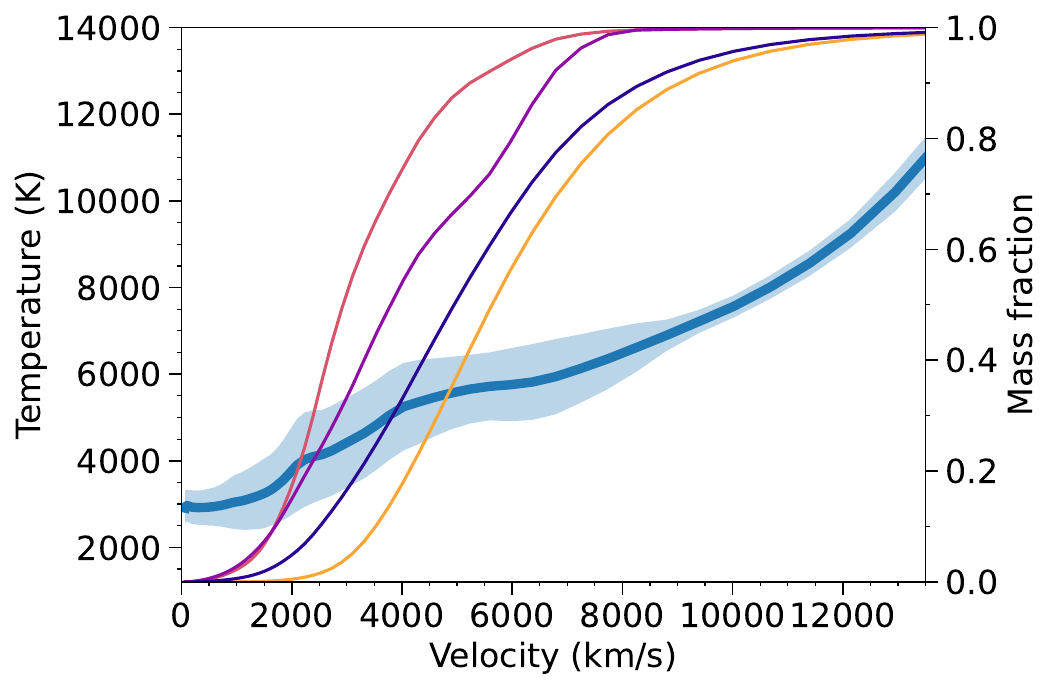}
    \includegraphics[width=.32\linewidth]{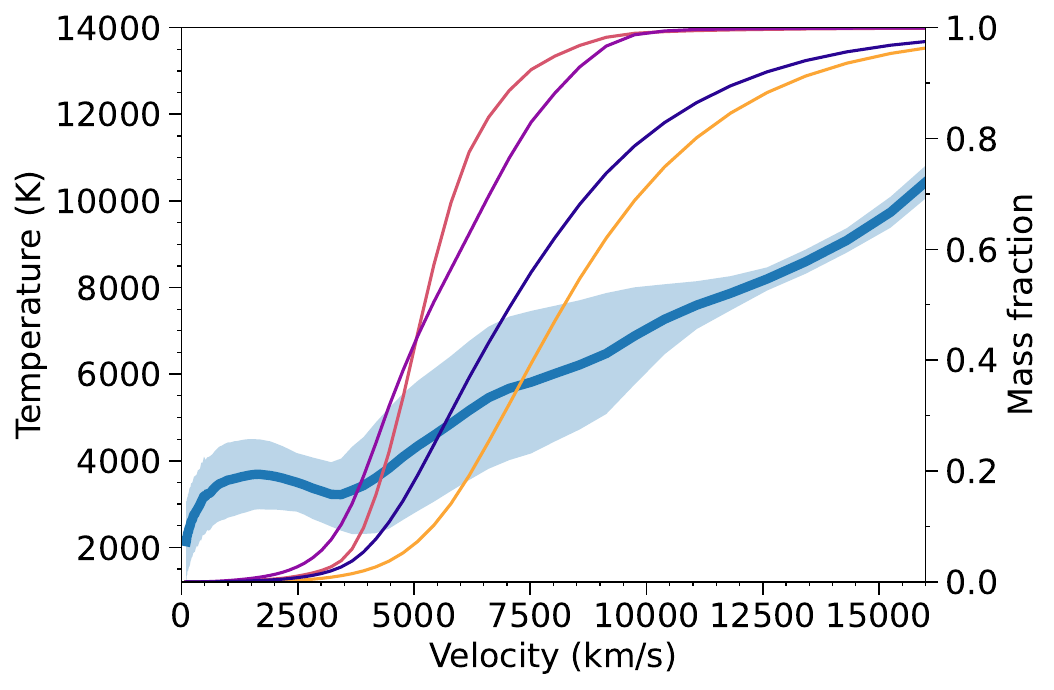}
    \includegraphics[width=.32\linewidth]{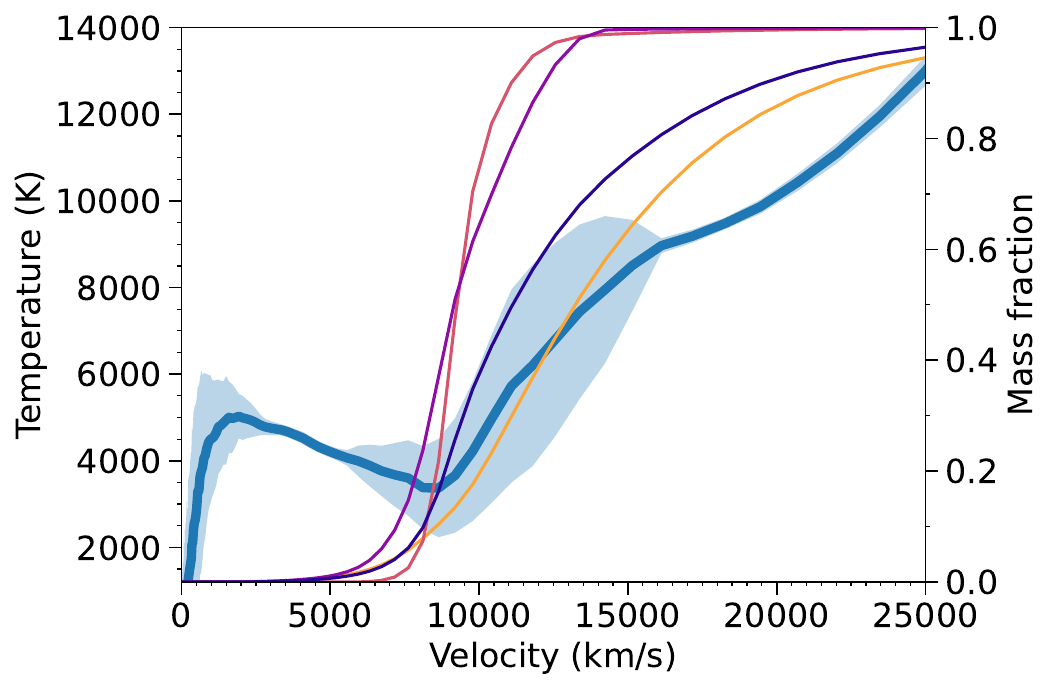}
    \caption{A display of the electron fractions ($x_e$, top row) and temperatures (bottom row) in blue together with a shaded $1\sigma$ region throughout the ejecta for the HEC-33 models (low, medium, high from left to right), together with the He, O NiCoFeX and overall ejecta curves (masses denoted at the top of each panel). Note the different velocity scales on the x-axis between the different explosion energies (different columns). In all models it can be observed that the highest velocity ejecta, which are (almost) pure He (see Figure \ref{fig:composition_mixing_hec33}) are symmetrical. The temperatures for the medium and high energy models display a minor bump at the innermost ejecta, while the low energy model does not have this although it does have a stronger bump for $x_e$ in that region. Overall the temperature curves are more monotone for lower explosion energies.}
    \label{fig:Txecombi-HEC-33}
\end{figure*}

\begin{figure*}
    \centering
    \includegraphics[width=.32\linewidth]{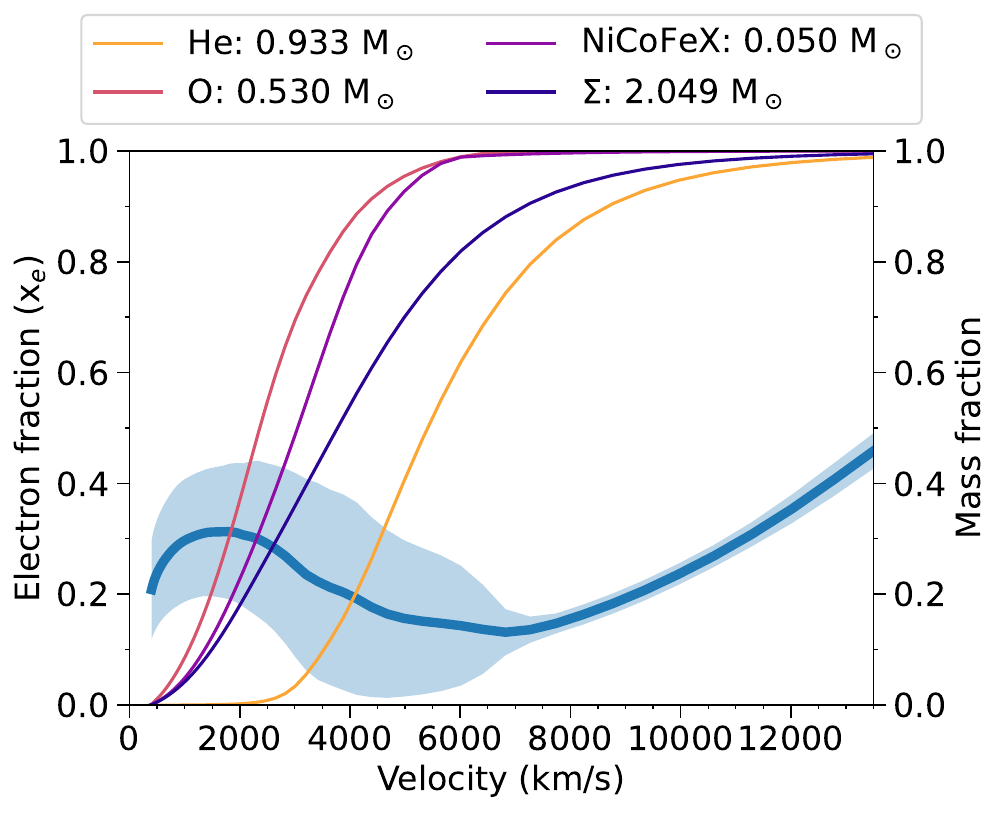}
    \includegraphics[width=.32\linewidth]{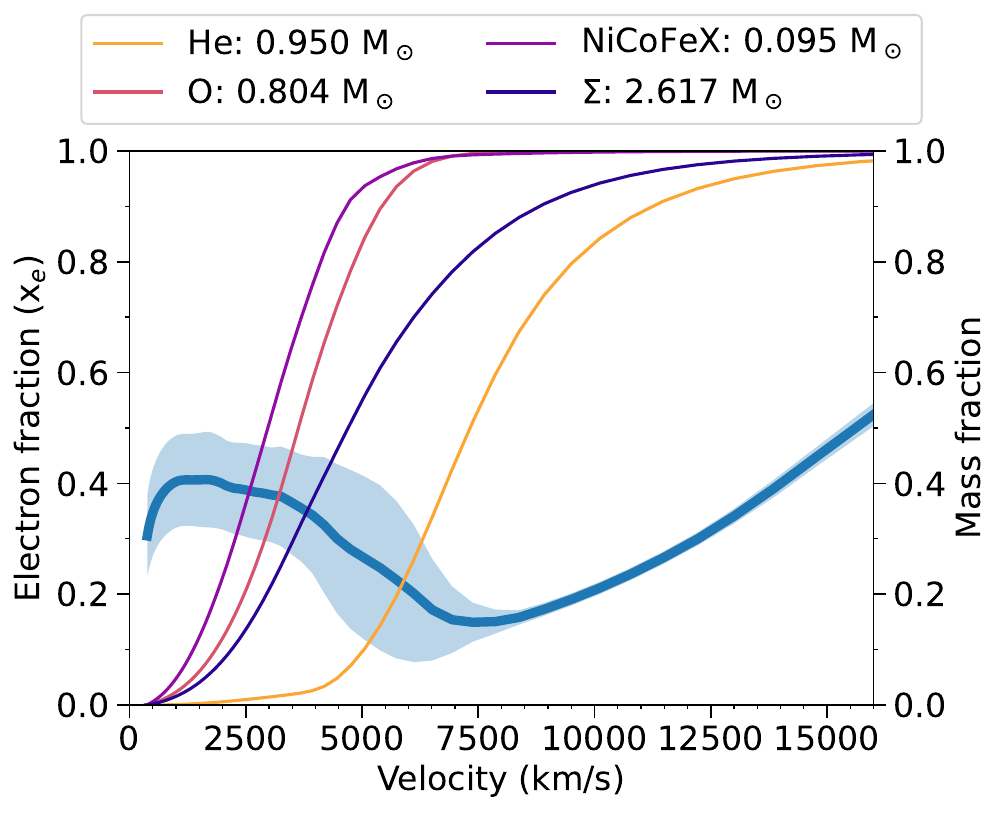}
    \includegraphics[width=.32\linewidth]{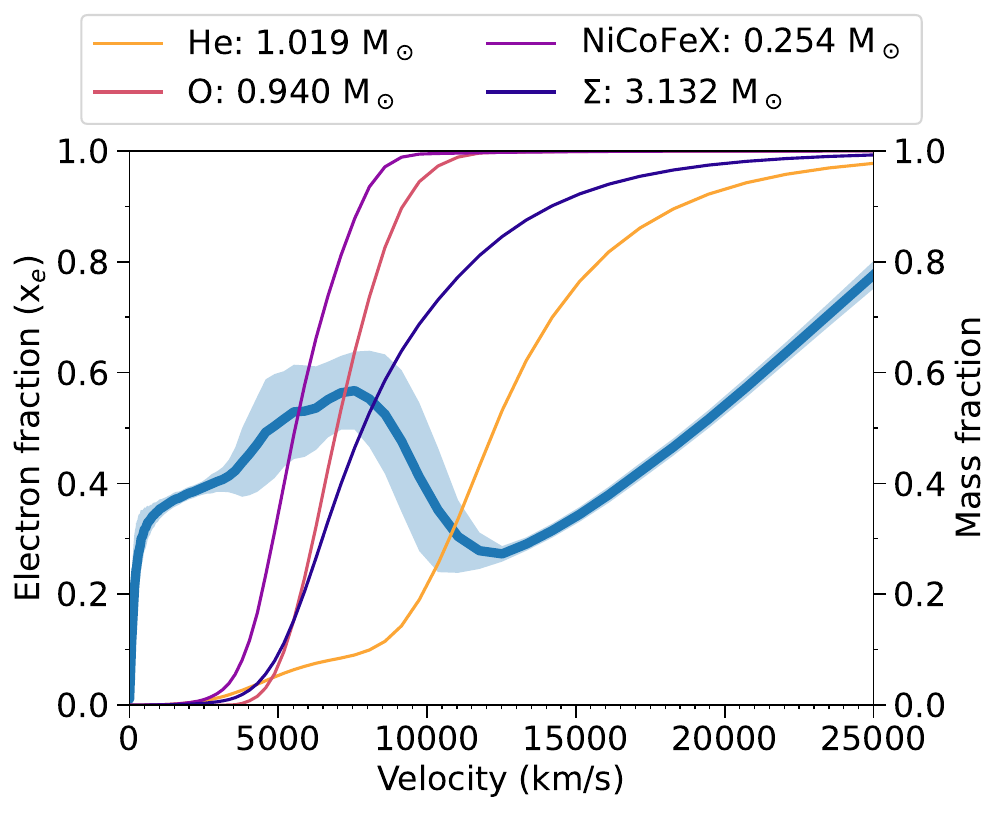}
    \includegraphics[width=.32\linewidth]{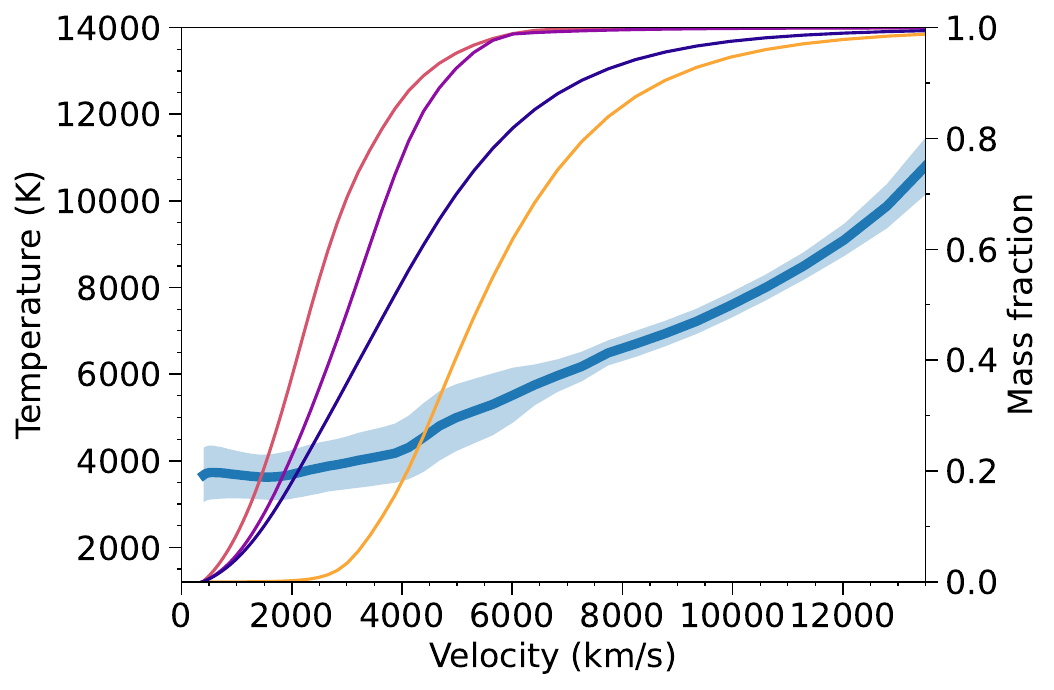}
    \includegraphics[width=.32\linewidth]{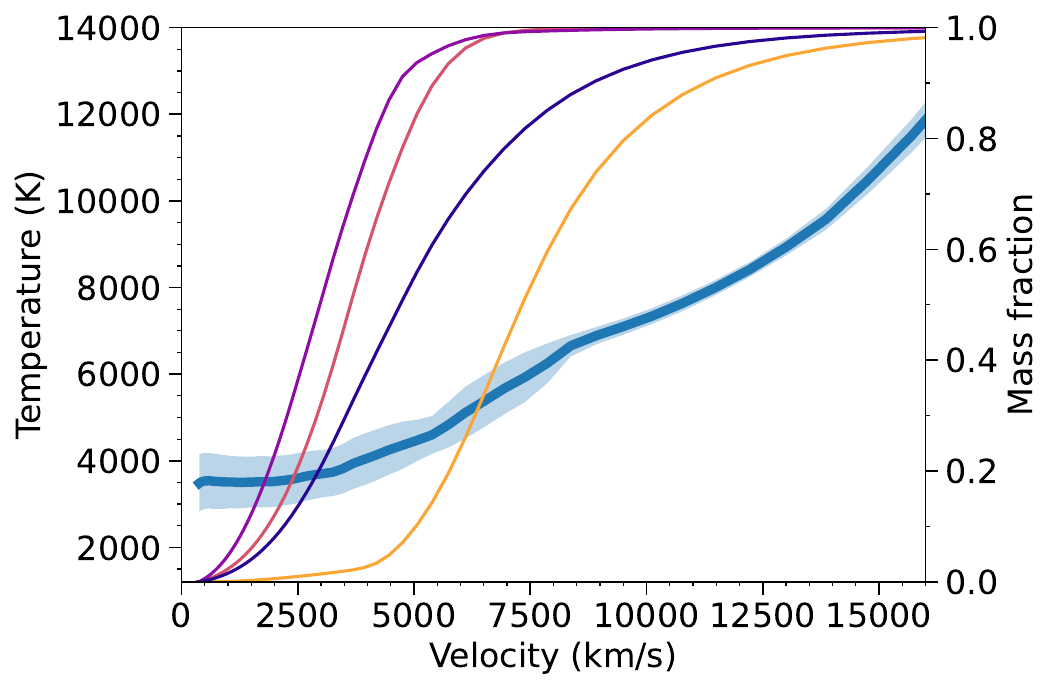}
    \includegraphics[width=.32\linewidth]{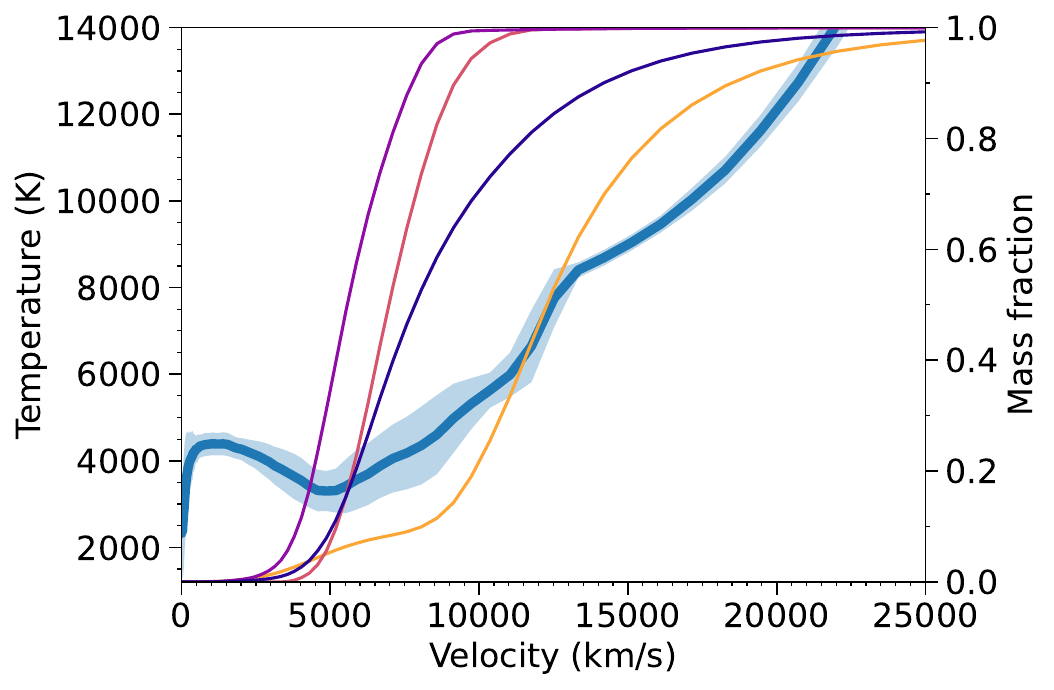}
    \caption{The same as in Figure \ref{fig:Txecombi-HEC-33} but for the HEC-60 models (low, medium, high from left to right; $x_e$ on top and temperature below). Note the different velocity scales on the x-axis. The low and medium models are shaped similarly to the HEC-33 models, although the temperature spreads are smaller. The high energy model are roughly symmetrical for both the innermost ejecta ($v\leqslant3000\,\text{km}\,\text{s}^{-1}$) and outermost ejecta ($v \geqslant12000\,\text{km}\,\text{s}^{-1}$). The intermediate region has a rise in $x_e$ where Ni is present, and a drop where it is not, while the temperature curve does not display such a strong reaction to Ni-presence.}
    \label{fig:Txecombi-HEC-60}
\end{figure*}

\begin{figure*}
    \centering
    \includegraphics[width=.32\linewidth]{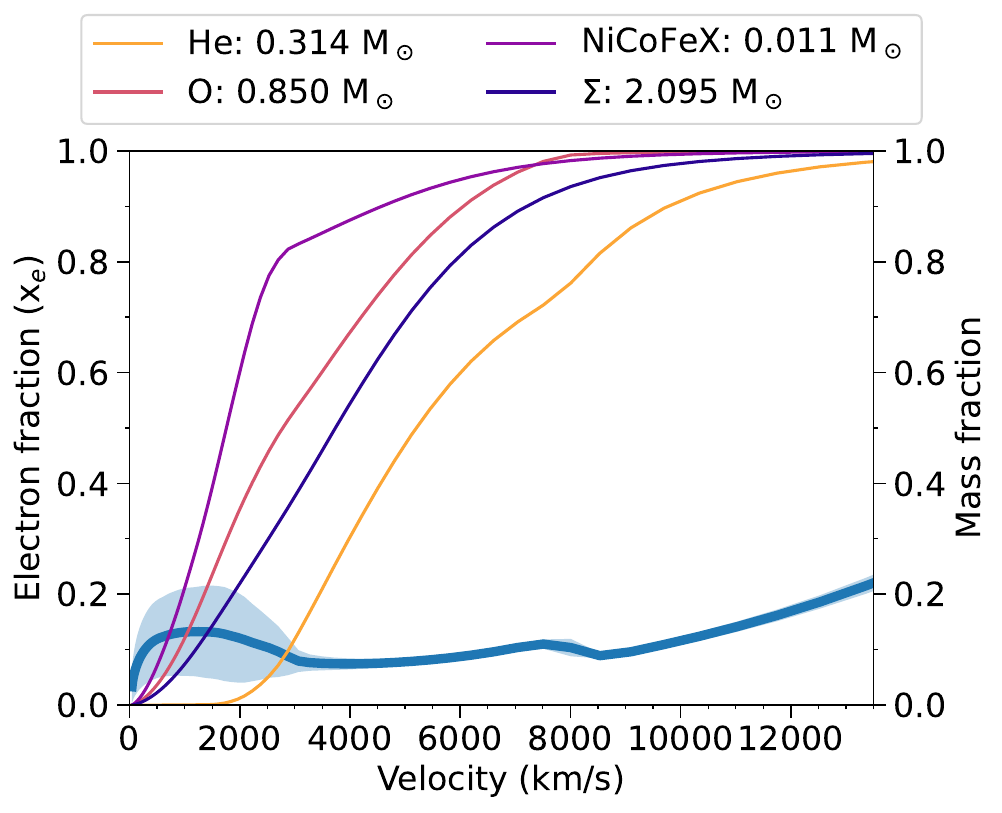}
    \includegraphics[width=.32\linewidth]{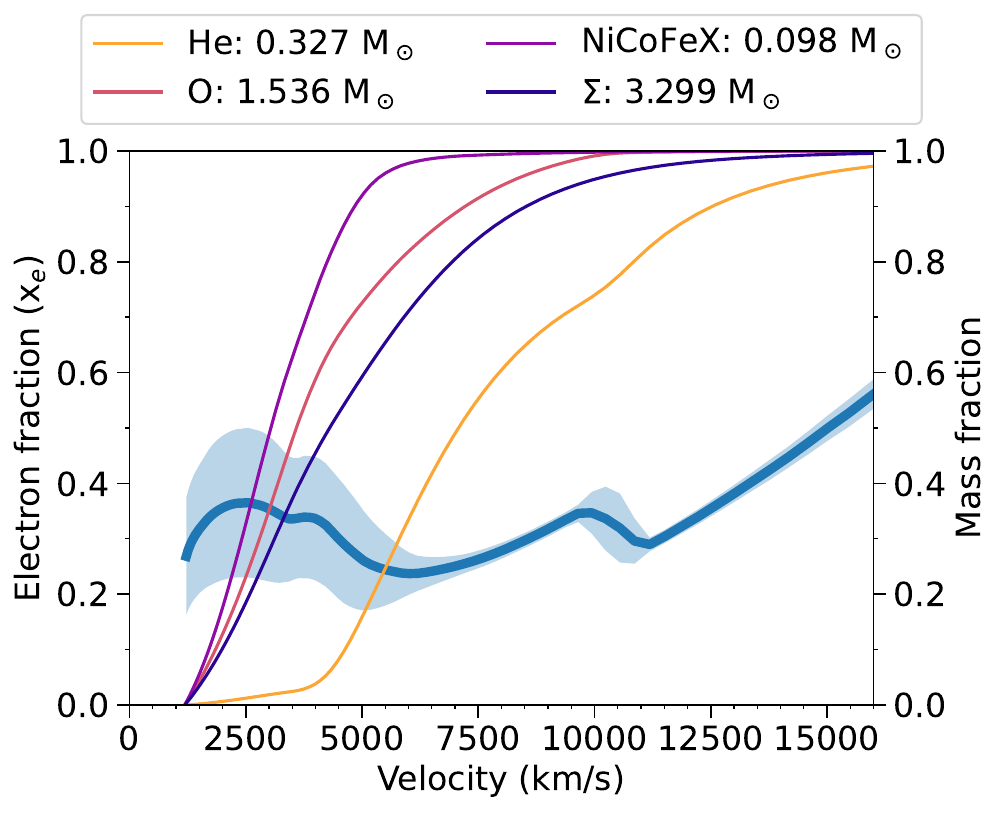}
    \includegraphics[width=.32\linewidth]{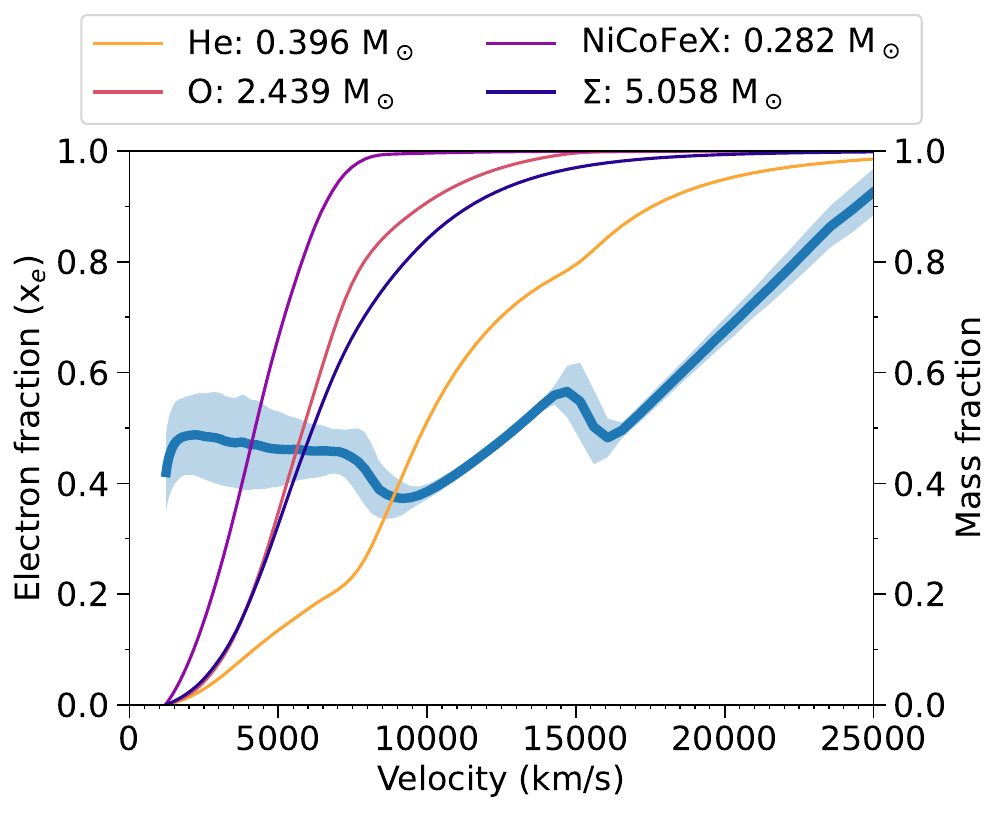}
    \includegraphics[width=.32\linewidth]{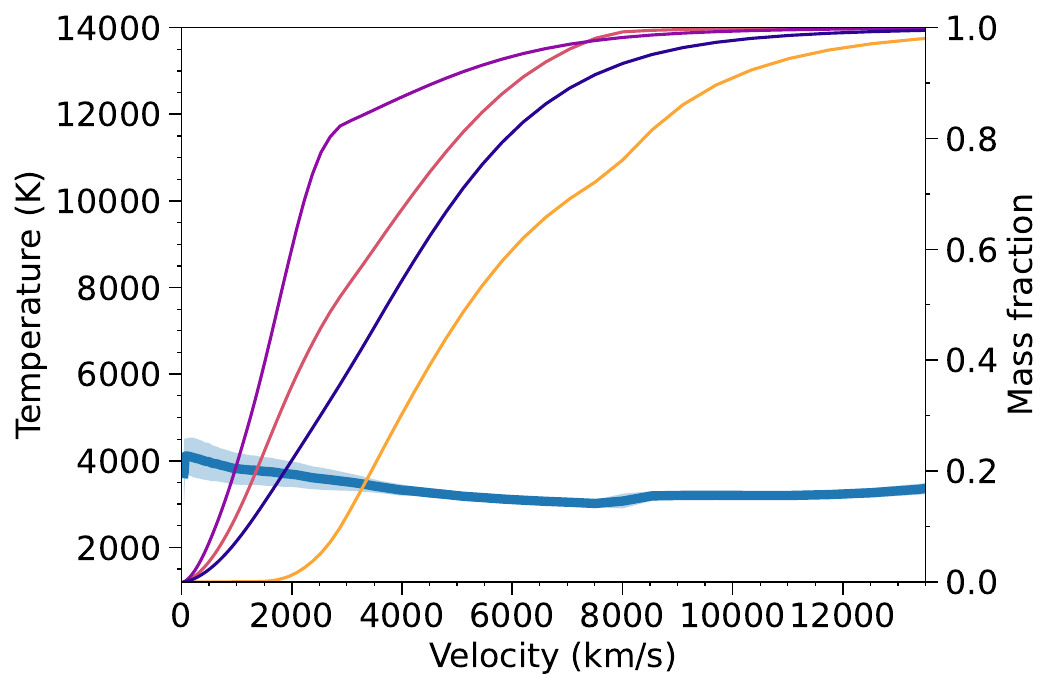}
    \includegraphics[width=.32\linewidth]{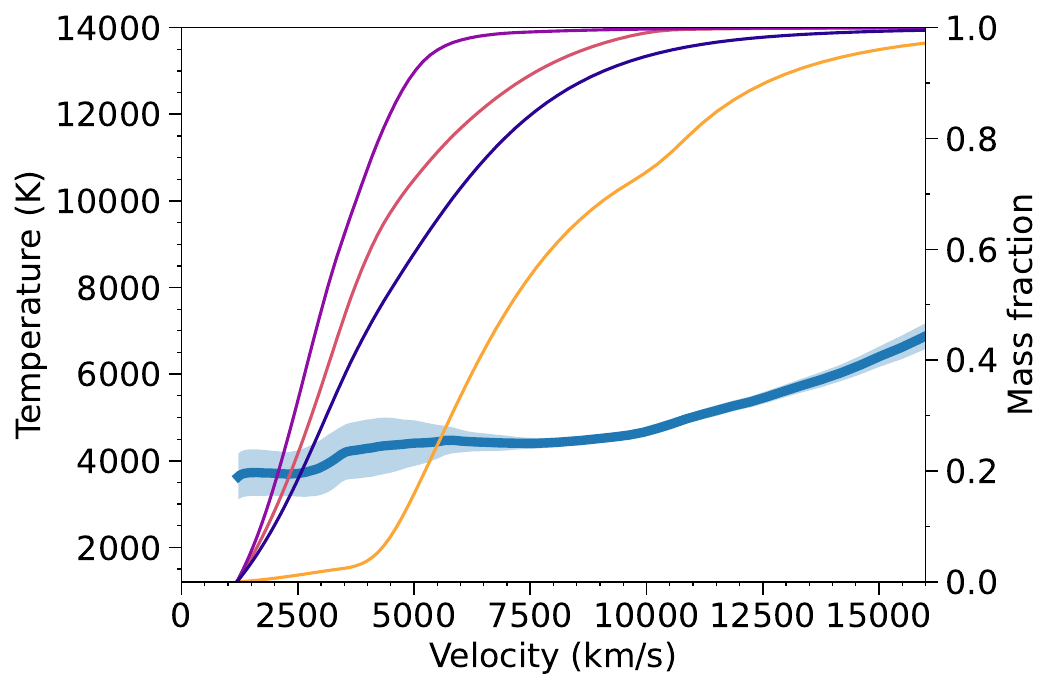}
    \includegraphics[width=.32\linewidth]{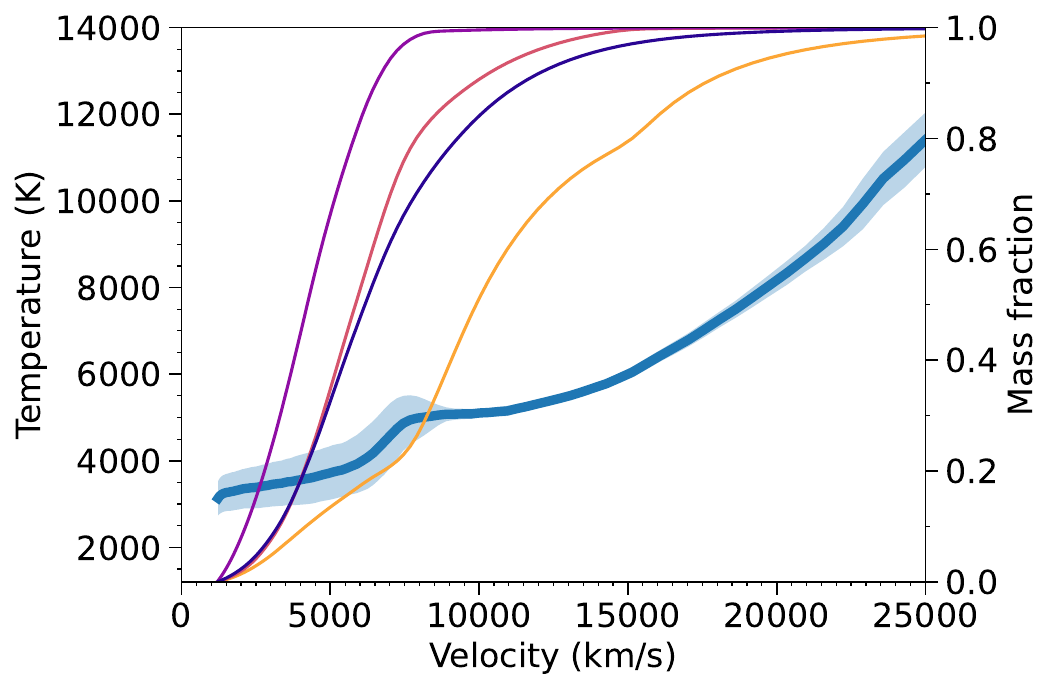}
    \caption{The same as in Figure \ref{fig:Txecombi-HEC-33} but for the HEC-110 models (low, medium, high from left to right; $x_e$ on top and temperature below). Note the different velocity scales on the x-axis. For the low energy model, both the $x_e$ and temperature curve are much flatter than for any other model. For the medium model, the temperature curve is still fairly flat, while the $x_e$ curve shows a minor drop at $v\sim10000\,\text{km}\,\text{s}^{-1}$. In the high energy model this drop in $x_e$ appears at $v\sim15000\,\text{km}\,\text{s}^{-1}$ and a small bump in the temperature curve can be noted as well at this velocity.}
    \label{fig:Txecombi-HEC-110}
\end{figure*}
In Figures \ref{fig:Txecombi-HEC-33}, \ref{fig:Txecombi-HEC-60} and \ref{fig:Txecombi-HEC-110} the angle-averaged free electron fractions ($x_e$, which is also the ionization fraction, top row, including a shaded $1\sigma$ standard deviation) and temperatures (bottom row, also including a shaded $1\sigma$ standard deviation) are given for the $3.3\,M_\odot$, $6.0\,M_\odot$ and $11.0\,M_\odot$ models respectively (in each Figure, from left-to-right, the $0.5\,$B, $1.0\,$B and $3.0\,$B explosions), together with the curves for the helium, oxygen, 'iron-group' elements (labelled as NiCoFeX, similar to \citealt{gabler2021infancy}) as well as the overall ejecta distribution ($\Sigma$) across the velocity space. Note the different velocity scales between the different models; they are chosen such that (generally) the $^{56}$Ni-containing ejecta encompass about half of the velocity range.

For the HEC-33 models, it can be seen that $x_e$ rises to around 0.5 for the inner ejecta but is lower for the bulk of the ejecta, before rising again to $\sim0.45$ for the low and medium models, and $\sim0.6$ for the high energy model in the envelope. In the high energy model it can also be seen that a small bump in standard deviation occurs where the iron-group elements are present. This bump can also clearly be seen in the temperature curve for this model (bottom right panel in Figure \ref{fig:Txecombi-HEC-33}), where the angle-averaged T increases rapidly from $\sim4000\,$K to $\sim9500\,$K. For the low energy model, the temperature increases throughout the ejecta with the innermost ejecta being the coldest at $\sim3000\,$K, and the outer envelope reaching $\sim10000\,$K. The medium energy model reaches similar envelope temperatures but has a small bump for the inner ejecta, which start colder but increase quite quickly to $\sim4000\,$K. Compared to \citet{vanbaal2023modelling}, for the medium-energy $3.3\,M_\odot$ model, both the $x_e$ and T values around this inner ejecta bump (for $v\leqslant2500\,\text{km}\,\text{s}^{-1}$) have gone up a little bit due to the composition change originating with the $^{56}$Ni $\rightarrow$ $^{56}$Co $\rightarrow$ $^{56}$Fe which is now included properly in the code. 

For the HEC-60 models (Figure \ref{fig:Txecombi-HEC-60}) more variety for the $x_e$ values can be observed. The low energy model here stays below $x_e\leqslant0.35$ until the He-dominated envelope, while the medium energy model stays below $x_e\leqslant0.4$; both models display some variation in the metal-rich ejecta but trend towards narrow standard deviations for the He-dominated envelope. For the high energy model, a different pattern emerges, as the inner regions ($v\leqslant4000\,\text{km}\,\text{s}^{-1}$) are very light on ejecta and thus display very little deviation and rise quicly to $x_e\approx0.4$. Between $4000\leqslant v \leqslant8000\,\text{km}\,\text{s}^{-1}$ a further increase to $x_e\approx0.55$ occurs, but then $x_e$ decreases again to $\sim0.3$ when the iron-group elements become scarce in the ejecta. For the outer, helium-dominated ejecta a rise to $\sim0.75$ occurs. The temperature curves for the low and medium energy models are quite monotonic, with the innermost ejecta at $\sim3500\,$K and the outermost ejecta at $\sim11000\,$K. The high energy model displays more variation, with a temperature bump for the innermost, low-mass ejecta ($v\leqslant4000\,\text{km}\,\text{s}^{-1}$) and a hotter outer ejecta of $\gtrsim14000\,$K.

The HEC-110 models (Figure \ref{fig:Txecombi-HEC-110}) display some interesting similarities between them. For the $x_e$ curves in each model, three regions can be identified: an inner, NiCoFeX-rich component with some spread in the electron fractions, an inner envelope composed of (predominantly) He, O and C (see also Figure \ref{fig:composition_mixing_hec11}) and lastly an outer envelope dominated by He, with a small drop in $x_e$ in the transition region in between. Comparing these different models a large difference in values for $x_e$ can however be seen, with the low energy model never going above $0.2$ and the high energy model (almost) never going below $0.4$; the medium energy model mostly lies in these intermediate values. The temperature evolution for these models can also be divided in the same three regions, with the inner NiCoFeX-rich region reaching $\sim4000\,$K for each model. However, for the low energy model the temperature never increases even in the outer envelope, displaying a unique temperature curve among the nine models. The medium energy model does show an increase in temperature in the outer envelope but only reaches $\sim6000\,$K, while the high energy model gets up to $\sim11000\,$K in the outer envelope.

\section{Observational data} \label{app:observationdata}  
In Table \ref{tab:obsSN_data} we give the full list of our observations which we used in Section \ref{sec:obscomp}. For each SN we note the host galaxy, the redshift we used in our redshift corrections (taken from \href{https://www.wiserep.org}{WiSeREP} $-$ \citealt{wiserep2012}), the date of the observation and our dating label. The elements column lists which elements we were able to (solidly) identify in each spectrum and which was used. References to the spectra's original papers are given.
\begin{table*}
	\centering
	\caption{List of SNe included in this work. The elements column indicates which features were used from a particular observation; [C I] is used to refer to the blended [C I]/Ca II NIR triplet feature. }
	\label{tab:obsSN_data}
    \setlength\tabcolsep{4pt}
    \begin{tabular*}{.98\linewidth}{@{\extracolsep{\fill}}l ccc ccr} 
        \hline
        \,Name & Host & Redshift & Date & Phase & Elements & Reference \\
        \hline
		1983N & NGC 5236 & 0.0017  & 1984-03-01 & 226 & Mg I] & \citealt{gaskell1986type} \\ 
        1985F & NGC 4618 & 0.0018  & 1985-03-19 & 280 & Mg I], [O I], [Ca II], [C I] & \citealt{filippenko1986unique} \\  
        1990I & NGC 4960A & 0.0097 & 1990-12-21 & 237 & [O I] & \citealt{elmhamdi2004SNib} \\
              &          &         & 1991-01-11 & 257 & [O I] & \citealt{elmhamdi2004SNib} \\
        1990U & NGC 7479 & 0.0079  & 1990-11-23 & 134 & [O I], [Ca II], [C I] & \citealt{taubenberger2009nebular} \\ 
              &          &         & 1990-12-12 & 153 & [O I], [Ca II], [C I] & \citealt{matheson2001optical} \\ 
              &          &         & 1990-12-20 & 161 & [O I] & \citealt{taubenberger2009nebular} \\ 
              &          &         & 1991-01-06 & 178 & [O I], [Ca II], [C I] & \citealt{matheson2001optical} \\ 
        1996aq& NGC 5584 & 0.0056  & 1997-02-11 & 176 & [O I], [Ca II], [C I] & \citealt{taubenberger2009nebular} \\
              &          &         & 1997-04-02 & 226 & Mg I], [O I], [Ca II] & \citealt{taubenberger2009nebular} \\ 
              &          &         & 1997-05-14 & 268 & Mg I], [O I], [Ca II], [C I] & \citealt{taubenberger2009nebular}\\
        1999dn& NGC 7714 & 0.0094  & 2000-09-01 & 375 & [O I] & \citealt{taubenberger2009nebular} \\ 
        2004ao& UGC 10862& 0.0056  & 2004-07-21 & 134 & Mg I], [O I], [Ca II] & \citealt{modjaz2008double} \\
              &          &         & 2004-08-16 & 160 & Mg I], [O I], [Ca II], [C I] & \citealt{shivvers2019berkeley}\\
              &          &         & 2004-09-09 & 184 & Mg I], [O I], [Ca II] & \citealt{modjaz2008double}\\
              &          &         & 2004-09-14 & 189 & Mg I], [O I] & \citealt{modjaz2008double}\\
              &          &         & 2004-09-22 & 197 & Mg I], [O I]  & \citealt{modjaz2008double}\\
              &          &         & 2004-09-24 & 199 & [O I], [Ca II] & \citealt{shivvers2019berkeley}\\
              &          &         & 2004-11-14 & 250 & Mg I], [O I], [Ca II], [C I] & \citealt{elmhamdi2011photometric}\\
        2004dk& NGC 6118 & 0.0052  & 2005-05-11 & 263 & Mg I], [O I], [Ca II], [C I] & \citealt{modjaz2008double} \\
              &          &         & 2005-07-09 & 322 & Mg I], [O I], [Ca II] & \citealt{modjaz2008double} \\
        2004gq& NGC 1832 & 0.0065  & 2005-10-08 & 291 & [O I], [Ca II] & \citealt{modjaz2008double} \\
        2004gt& NGC 4038 & 0.0055  & 2005-05-24 & 160 & [O I], [Ca II], [C I] & \citealt{taubenberger2009nebular}\\
        2004gv& NGC 856  & 0.02    & 2005-10-08 & 279 & [O I] & \citealt{modjaz2008double}\\
        2006gi& NGC 3147 & 0.0094  & 2007-02-10 & 145 & Mg I], [O I], [Ca II], [C I] & \citealt{taubenberger2009nebular}\\
        2006ld& UGC 348  & 0.014   & 2007-07-17 & 280 & [O I], [Ca II] & \citealt{taubenberger2009nebular}\\
              &          &         & 2007-08-06 & 300 & [O I], [Ca II] & \citealt{taubenberger2009nebular} \\
              &          &         & 2007-08-20 & 314 & [O I], [Ca II] & \citealt{taubenberger2009nebular} \\
        2007C & NGC 4981 & 0.0056  & 2007-05-17 & 131 & Mg I], [O I], [Ca II], [C I] & \citealt{taubenberger2009nebular}\\
              &          &         & 2007-06-20 & 165 & Mg I], [O I], [Ca II], [C I] & \citealt{taubenberger2009nebular}\\
        2007Y & NGC 1187 & 0.0047  & 2007-09-22 & 200 & Mg I], [O I], [Ca II] & \citealt{stritzinger2009herich}\\
              &          &         & 2007-10-21 & 229 & Mg I], [O I], [Ca II], [C I] & \citealt{stritzinger2009herich}\\
              &          &         & 2007-11-30 & 269 & Mg I], [O I], [Ca II], [C I] & \citealt{stritzinger2009herich}\\
        2007uy& NGC 2270 & 0.007   & 2008-06-06 & 141 & [O I], [Ca II] & \citealt{roy2013sn2007uy}\\
        2008D & NGC 2770 & 0.0065  & 2008-06-07 & 140 & [O I], [Ca II], [C I] & \citealt{modjaz2009shock}\\
        2009jf& NGC 7479 & 0.008   & 2010-06-19 & 245 & Mg I], [O I], [Ca II] & \citealt{valenti2011sn2009jf}\\
              &          &         & 2010-07-08 & 264 & Mg I], [O I], [Ca II] & \citealt{valenti2011sn2009jf}\\
              &          &         & 2010-10-04 & 352 & Mg I], [O I], [Ca II] & \citealt{valenti2011sn2009jf}\\
              &          &         & 2010-10-11 & 359 & Mg I], [O I], [Ca II] & \citealt{valenti2011sn2009jf}\\
              &          &         & 2010-10-12 & 360 & [O I], [Ca II] & \citealt{shivvers2019berkeley}\\
              &          &         & 2010-11-01 & 380 & [O I] & \citealt{schulze2021PTFCC}\\
        2011dh& M51      & 0.002   & 2011-12-18 & 202 & Mg I], [O I], [Ca II], [C I] & \citealt{shivvers2013nebular}\\
              &          &         & 2012-02-23 & 268 & Mg I], [O I], [Ca II], [C I] & \citealt{shivvers2013nebular}\\
              &          &         & 2012-03-18 & 292 & Mg I], [O I], [Ca II], [C I] & \citealt{ergon2015type}\\
              &          &         & 2012-05-24 & 359 & Mg I], [O I], [Ca II], [C I] & \citealt{ergon2015type}\\
              &          &         & 2012-07-19 & 416 & Mg I], [O I], [Ca II], [C I] & \citealt{ergon2015type}\\
        2012au& NGC 4790 & 0.0045  & 2012-12-19 & 284 & Mg I], [O I], [Ca II] & \citealt{milisavljevic2013SN2012au}\\
              &          &         & 2013-02-05 & 332 & Mg I], [O I], [Ca II] & \citealt{milisavljevic2013SN2012au}\\
        2014C & NGC 7331 & 0.0027  & 2014-08-25 & 221 & [O I], [Ca II], [C I] & \citealt{shivvers2019berkeley}\\
              &          &         & 2014-09-03 & 230 & [O I], [Ca II] & \citealt{shivvers2019berkeley}\\
              &          &         & 2014-09-28 & 255 & [O I], [Ca II], [C I] & \citealt{shivvers2019berkeley}\\
              &          &         & 2014-10-02 & 259 & [O I], [Ca II] & \citealt{shivvers2019berkeley}\\
        2014ei&MCG-01-13-50& 0.014 & 2015-03-27 & 142 & [O I], [Ca II], [C I] & \citealt{shivvers2019berkeley}\\
        2015Q & NGC 3888 & 0.0079  & 2015-12-16 & 190 & [Ca II], [C I] & \citealt{shivvers2019berkeley}\\
              &          &         & 2016-01-07 & 212 & [O I], [Ca II], [C I] & \citealt{shivvers2019berkeley}\\
              &          &         & 2016-02-09 & 245 & Mg I], [O I], [Ca II], [C I] & \citealt{shivvers2019berkeley}\\
              &          &         & 2016-05-07 & 332 & Mg I], [O I], [C I] & \citealt{shivvers2019berkeley}\\
        2017bgu&2MASX J16555976+4233370 & 0.008 & 2017-09-04 & 183 & [O I] & \citealt{prentice2019investigating}\\
  \hline
	\end{tabular*}
\end{table*}

In order to calculate the line properties from the observational spectra, we apply the following routine. First we correct the spectra for the redshift, and the wavelength range around the feature is selected. Next we apply a background subtraction, by taking the average flux of the seven bins at both ends of the feature and drawing a straight line between these, which serves as our correcting factor\footnote{Not every spectrum needs this subtraction, but we apply it to all spectral features to be impartial. The background subtraction is important for cases where the feature might be clearly visible but sit on an obviously non-zero background, which would impact the integral calculations.}. Every feature is checked by eye to make sure that this process has not resulted in strange artefacts or slanted profiles. If the feature passes this check, we determine the $v_\text{shift}$ and $v_\text{width}$ integrals as we do for our models.

Some of our spectra clearly display narrow features, and for this subset we choose a smaller wavelength range around the feature than the default, to avoid errors being introduced in the background subtraction process. This selection of narrow line spectra is also performed by eye.


\bsp	
\label{lastpage}
\end{document}